\documentclass[traditabstract]{aa} 
\usepackage{amsmath}
\usepackage{graphicx}
\usepackage{txfonts} %Conflict with amsmath (so should be defined afterwards)$M$
\usepackage[usenames,dvipsnames]{color}  %SdM: allows fancy color names
\usepackage{longtable,lscape}
\usepackage{natbib}
\usepackage{numprint}
\bibpunct{(}{)}{;}{a}{}{,} % to follow the A&A style
\newcommand{\Msun}{\,\mbox{${M}_{\odot}$}}
\newcommand{\Rsun}{\,\mbox{${R}_{\odot}$}}
\newcommand{\Lsun}{\,\mbox{${L}_{\odot}$}}
\begin{document}

	\title{Theoretical uncertainties of the type Ia supernova rate}
   \author{J.S.W. Claeys\inst{1}
          \and
          O.R. Pols\inst{1}
          \and
          R.G. Izzard\inst{2}
          \and
          J. Vink\inst{3}          
          \and
          F.W.M. Verbunt\inst{1}
          }
   \institute{Department of Astrophysics/IMAPP, Radboud University Nijmegen, P.O. Box 9010, 6500 GL Nijmegen, The Netherlands;
\and   
   Argelander Institute for Astronomy, University of Bonn, Auf dem H\"ugel 71, D-53121 Bonn, Germany;
\and  
   Anton Pannekoek Institute/GRAPPA, University of Amsterdam, PO Box 94249, 1090 GE Amsterdam, The Netherlands;\\
                            \email{jsclaeys.claeys@gmail.com, O.Pols@astro.ru.nl}
             }

   \date{Received 20 September 2013; accepted 12 January 2014}
   
\abstract{It is thought that type Ia supernovae (SNe Ia) are explosions of carbon-oxygen white dwarfs (CO WDs). Two main evolutionary channels are proposed for the WD to reach the critical density required for a thermonuclear explosion: the single degenerate scenario (SD), in which a CO WD accretes from a non-degenerate companion, and the double degenerate scenario (DD), in which two CO WDs merge. However, it remains difficult to reproduce the observed SN~Ia rate with these two scenarios.

With a binary population synthesis code we study the main evolutionary channels that lead to SNe~Ia and we calculate the SN~Ia rates and the associated delay time distributions. We find that the DD channel is the dominant formation channel for the longest delay times. The SD channel with helium-rich donors is the dominant channel at the shortest delay times. Our standard model rate is a factor five lower than the observed rate in galaxy clusters.

We investigate the influence of ill-constrained aspects of single- and binary-star evolution and uncertain initial binary distributions on the rate of type Ia~SNe. These distributions, as well as uncertainties in both helium star evolution and common envelope evolution, have the greatest influence on our calculated rates. Inefficient common envelope evolution increases the relative number of SD explosions such that for $\alpha_{\rm ce} = 0.2$ they dominate the SN~Ia rate. Our highest rate is a factor three less than the galaxy-cluster SN~Ia rate, but compatible with the rate determined in a field-galaxy dominated sample. If we assume unlimited accretion onto WDs, to maximize the number of SD explosions, our rate is compatible with the observed galaxy-cluster rate.}


\maketitle

\section{Introduction}
Type Ia supernovae (SNe~Ia) are important astrophysical phenomena. On the one hand they drive galactic chemical evolution as the primary source of iron, on the other hand they are widely used as cosmological distance indicators \citep{Phillips93,Riess96,Riess98,Perlmutter99} because of their homogeneous light curves. Even so, the exact progenitor evolution remains uncertain. It is generally accepted that SNe~Ia are thermonuclear explosions of carbon-oxygen white dwarfs \citep[CO WDs;][]{Nomoto82, Bloom12}. The explosion can be triggered when the CO WD reaches a critical density, which is reached when the mass approaches the Chandrasekhar mass ($M_{\rm Ch}$ = 1.4\Msun, for non-rotating WDs). Single stars form a CO WD with a mass up to about 1.2\Msun~\cite[]{Weidemann00}. To explain how the WD then reaches $M_{\rm Ch}$~two main channels are proposed: the single degenerate channel \cite[SD; ][]{WhelanIben73,Nomoto82}, in which the WD accretes material from a non-degenerate companion; and the double degenerate channel \cite[DD; ][]{Webbink84,IbenTutukov84}, in which two CO WDs merge.

Both channels cannot fully explain the observed properties of SNe~Ia \cite[][]{Howell11}. WDs in the SD channel burn accreted material into carbon and oxygen, and therefore these systems should be observed as supersoft X-ray sources \cite[SSXS,][]{Vandenheuvel92}, not enough of which are observed to explain the number of SNe~Ia \cite[]{Gilfanov10, DiStefano10}. However, \cite{Nielsen13} argue that only a small amount of circumstellar mass loss is able to obscure these sources, making it difficult or even impossible for observers to detect them as SSXS. Moreover, during the supernova explosion some of the material from the donor star is expected to be mixed with the ejecta, which has never been conclusively observed \citep{Leonard07,GarciaBadenes11}. In some cases NaD absorption lines have been observed, which can be interpreted as circumstellar material from the donor star \citep{Patat07, Sternberg11}. Furthermore, the donor star is expected to survive the supernova explosion and to have a high space velocity. In the case of the 400 year old Tycho supernova remnant, we would expect to observe this surviving star. Nevertheless, no such object has been unambiguously identified \citep{RuizLapuente04,Kerzendorf09,Schaefer12}. Some SNe and supernova remnants show evidence of interaction with circumstellar material which links them to the SD channel \citep{Hamuy03, Chiotellis12, Dilday12}. However, a variation on the DD channel, in which the merger occurs during or shortly after the CE phase \cite[]{Hamuy03, Chevalier12, Soker13}, cannot be excluded as the progenitor channel for these SNe. Additionally, SN 2011fe, a SN~Ia which exploded in a nearby galaxy, showed no evidence of interaction with circumstellar material, in radio, X-ray and optical \cite[e.g.][]{Margutti12, Chomiuk12, Patat13} and pre-explosion images exclude most types of donors stars of the SD channel \cite[]{Li11}.

These difficulties in reconciling the observational evidence with the predictions of the SD channel are avoided in the DD channel, but this channel has its own difficulties. In particular, \cite{NomotoKondo91} show that the merger product evolves towards an accretion-induced collapse (AIC) rather than towards a normal SN~Ia. Only recently, some studies indicate that the merger can lead to a SN~Ia explosion under certain circumstances \citep{Pakmor10, Pakmor11,Pakmor13, Shen13, Dan13}. Finally, binary population synthesis studies show that neither channel reproduces the observed SN~Ia rate \cite[e.g.][]{Ruiter09, Mennekens10, Toonen12, Bours13, Chen12}.

We investigate with a binary population synthesis (BPS) code the progenitor evolution towards SNe~Ia through the canonical Chandrasekhar mass channels: the SD and the DD channels. Our code makes it possible to study large stellar populations, to calculate SN~Ia rates and to compare these with the observed rate of SNe~Ia. We analyse not only the different evolutionary channels, but also --in the case of the SD channel-- determine general properties of the donor star at the moment of the explosion, and --in the case of the DD channel-- of the merger product. The SN~Ia rate has been studied by several groups through binary population synthesis, however they used only a standard model, without investigating in great detail the uncertainties in their model, or they investigated one uncertainty in binary evolution, such as the common envelope evolution \citep[e.g.][]{YungelsonLivio00, Ruiter09, Wang09, Wang10, Mennekens10, Toonen12}. We perform a study of the main parameters that play a role in the evolution towards type~Ia~SNe which are ill-constrained. We study the effects on the predicted rate and on the progenitor evolution towards SNe~Ia.

In Sect.\,\ref{sec:StandAssum}, the binary population synthesis code and our model are described. In Sect.\,\ref{Sect::BinEvol}, the general progenitor evolution as it follows from our standard model is outlined. The resulting SN~Ia rate is discussed in Sect.\,\ref{sec:DTD}, while Sect.\,\ref{sec:inflPar}~and~\ref{sec:inflDistr} treat the effects of varying the parameters on the progenitor evolution and the total rate. Finally, Sect.\,\ref{sec::disc} discusses the validity of the Chandrasekhar model.

\section{Binary population synthesis\label{sec:StandAssum}}
We employ the rapid binary evolution code \texttt{binary\_c/nucsyn} based on the work of \citet{Hurley00,Hurley02} with updates described in \citet{Izzard04,Izzard06,Izzard09} and below. We discuss the key assumptions of our model of single and binary star evolution. In Table\,\ref{ParTable} we list the assumptions in our standard model and give an overview of possible other assumptions, and the dependence of the theoretical SN~Ia rate on them is discussed in Sects.\,\ref{sec:inflPar}~and~\ref{sec:inflDistr}. This code serves as the basis of our binary population synthesis (BPS) calculations.

\begin{table*} 
\caption{Physical parameters of single and binary evolution in our model and distribution functions of initial binary parameters for the standard model and their variations (used in Sects.\,\ref{sec:inflPar}\,and\,\ref{sec:inflDistr}).}
\label{ParTable}
\begin{center}
%\begin{tabular}{l|c|c|c}
\begin{tabular}{lccc}
\hline \noalign{\smallskip} 
 Name parameter & Reference & Standard model & Variation \\
\hline \noalign{\smallskip} %\hline 
$\eta$ (R75) & Eq.\,\ref{Reimers} & 0.5 & 5\\
TP-AGB wind & Eq.\,\ref{eq::VW}\,-\,\ref{eq::P0} & HPT00 & KLP02, R75 ($\eta = 1$) , B95, VL05\\
$\alpha_{\rm BH}$ & Eq.\,\ref{eq::BH} & 1.5 & 5\\
$B_{\rm wind}$ & Eq.\,\ref{TE88} & 0.0 & $10^3$\\
$Q_{\rm crit,HG}$\tablefootmark{a} & Table\,\ref{TableQ} & 0.25 & 0.5\\
$Q_{\rm crit,He-stars}$ & Table\,\ref{TableQ} & $Q_{\rm crit,He-HG}$ = $Q_{\rm crit,HG}$ = 0.25 & $Q_{\rm crit,He-HG} = Q_{\rm crit,He-GB}$ = 1.28\\
$\alpha_{\rm ce}$ & Eq.\,\ref{eq::alphace} & 1.0 & 0.2-10\\
CE accretion\tablefootmark{b} & Eq.\,\ref{eq::alphace} & 0.0 & 0.05\Msun\\
$\lambda_{\rm ce}$ & Eq.\,\ref{eq::lambdace}\,\& Appendix\,\ref{app::lambda} & variable & 1 \\
$\lambda_{\rm ion}$ & Eq.\,\ref{eq::lambdace}\,\& Appendix\,\ref{app::lambda} & 0.0 & 0.5\\
$\sigma$ & Eq.\,\ref{MaxMaccr} & 10 & 1, 20, $\infty$\tablefootmark{c} \\
$\gamma_{\rm RLOF}$ & Eq.\,\ref{eq::jorb} & $M_{\rm d}$/$M_{\rm a}$ & 2, $M_{\rm a}$/$M_{\rm d}$\\
$\gamma_{\rm wind}$ & Eq.\,\ref{eq::jorb} & $M_{\rm a}$/$M_{\rm d}$ & 2\\
$\eta_{\rm H}$, $\eta_{\rm He}$ & Eq.\,\ref{eq::etaH}\,\&\,\ref{eq::etaHe} & HKN96 & 1\\
$\psi(M_{\rm 1,i})$ & Sect.\,\ref{Sect::BPS} & KTG93 & S98, K01, C03, B03\\
$\phi(q_{\rm i})$ $\propto q_{\rm i}^{x}$ & Sect.\,\ref{Sect::BPS} & $x=0$  & $-1 \leq x \leq 1$ \\
\hline
\end{tabular}
\end{center}
\tablefoot{\tablefoottext{a}{Only in the case of mass transfer with a non-degenerate accretor.}
\tablefoottext{b}{Only when the companion is a MS star.}
\tablefoottext{c}{The symbol $\infty$ implies conservative mass transfer to all types of stars.}
\tablebib{B03 = \cite{Bell03}, B95 = \cite{Bloecker95}, C03 = \cite{Chabrier03}, HKN96 = \cite{Hachisu96}, HPT00 = \cite{Hurley00}, HPT02 = \cite{Hurley02}, K01 = \cite{Kroupa01}, KLP02 = \cite{Karakas02}, KTG93 = \cite{Kroupa93}, R75 = \cite{Reimers75}, S98 = \citep{Scalo98}, VL05 = \cite{VanLoon05}.}}
\end{table*}

\subsection{Single star evolution \label{sec:SSE}}
Our single star models are analytic fits to detailed stellar models, described in~\cite{Hurley00}, with updates of the thermally pulsating asymptotic giant branch (TP-AGB) which are fits to the models of~\cite{Karakas02}, as described in \cite{Izzard06}. In the present study we adopt a metallicity $Z$ of 0.02. We assume that only a CO WD can explode as a SN~Ia based on the work of \cite{NomotoKondo91} who found that an ONe WD that reaches $M_{\rm Ch}$ undergoes AIC. CO WDs are formed only in low and intermediate mass binary systems, with a primary mass up to 10\Msun, therefore we limit our discussion to this mass range.

\subsubsection{TP-AGB models\label{sec:TPAGB}}
\begin{figure}
\begin{center}
\includegraphics[width=9cm]{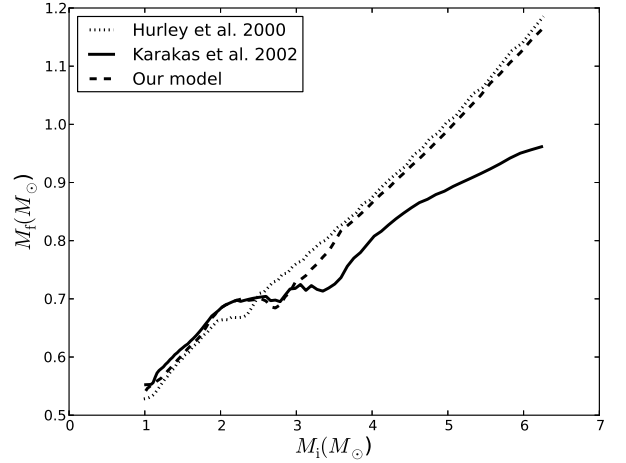}
\caption{Initial ($M_{\rm i}$) versus final ($M_{\rm f}$)  mass of single stars that become CO WDs for different TPAGB models (Sect.\,\ref{sec:TPAGB}). The dotted line shows the results of the \cite{Hurley00}-models, the full line shows the results for the fits to the models of \cite{Karakas02}, and the dashed line shows our model.}
\label{fig:MiMf}
\end{center}
\end{figure}
The evolution of the core, luminosity and radius of the TP-AGB star versus time are based on the models of \cite{Karakas02}, with the prescriptions described in \cite{Izzard04,Izzard06}. Because the core masses are determined without taking overshooting during previous evolution phases into account, a fit is made to take overshooting into consideration, based on the core masses calculated by \cite{Hurley02} during the early-AGB (E-AGB). Up to and including the first thermal pulse, the core mass is calculated by the formulae described by \cite{Hurley02}. A smooth function is built in to guarantee a continuous transition of the radius between the E-AGB and the TP-AGB. CO WDs form from initial stars with masses up to 6.2\Msun, which corresponds to a maximum CO WD of about 1.15\Msun\ (Fig.\,\ref{fig:MiMf}).

\subsubsection{Helium star evolution\label{sec:HeStar}}
In our model, if a hydrogen-rich star is stripped of its outer layers after the main-sequence (MS) but before the TP-AGB, and the helium core is not degenerate,  a helium star is formed. If the exposed core is degenerate a He~WD is made. The prescriptions to describe the evolution of naked helium stars are discussed in \cite{Hurley00}. However, because of its importance for the formation of type Ia~SNe we emphasize some details.

A distinction is made between three phases of helium star evolution: He-MS, the equivalent of the MS, when helium burns in the centre of the helium star; He-HG, the equivalent of the Hertzsprung gap (HG), when He-burning moves to a shell around the He-depleted core; and He-GB, the equivalent of the giant branch (GB), when the  helium star has a deep convective envelope. If a non-degenerate helium core is exposed during the E-AGB, a star on the He-HG forms, otherwise a He-MS star forms.

A He-MS star becomes a He-WD when its mass is less than 0.32\Msun, the minimum mass necessary to burn helium. The boundary between a helium star forming a CO WD or a ONe WD is determined by the mass of the star at the onset of the He-HG. The mass of the star on the He-HG should exceed 1.6\Msun\ in order to form a dense enough core to burn carbon and form an ONe-core, which is based on detailed models \cite[carbon ignites off-centre when $M_{\rm CO,core}\gtrsim 1.08$\Msun,][]{Pols98}. The evolution of a He-HG star with $M < 1.6\Msun$ leads to the formation of a CO WD with a mass greater than 1.2\Msun~unless its envelope is lost by wind or binary mass transfer.

\subsubsection{Wind mass loss\label{sec:wind}}
We adopt for both low- and intermediate-mass stars the prescription based on \cite{Reimers75} during the HG and beyond, multiplied by a factor $\eta$, 
\begin{equation}
\label{Reimers}
\dot{M}_{\rm R} = \eta \cdot 4.0 \cdot 10^{-13} \, \frac{R}{\Rsun} \frac{L}{\Lsun} \frac{\Msun}{M} \, \Msun \, \rm yr^{-1}\,,
\end{equation}
where $M$ is the mass, $R$ the radius and $L$ the luminosity of the star, and $\eta = 0.5$ in our standard model (Table \ref{ParTable}). During the TP-AGB we use the prescription based on \cite{VassiliadisWood93},
\begin{equation}
\label{eq::VW}
\log \left(\frac{\dot{M}_{\rm VW}}{\Msun\,{\rm yr^{-1}}}\right) = -11.4 + 0.0125 \left(\frac{P_0}{d}-100 \max\left[\frac{M}{\Msun}-2.5,0.0\right]\right),
\end{equation}
where $P_0$ is the Mira pulsation period in days ($d$), given by
\begin{equation}
\label{eq::P0}
\log \left(\frac{P_0}{d}\right) = {\min}\left(3.3, -2.7 - 0.9 \log\left[\frac{M}{\Msun}\right] + 1.94 \log \left[\frac{R}{\Rsun}\right]\right)\,.
\end{equation}
The wind is limited by the steady superwind rate, $\dot{M}_{\rm SW} = 1.36 \, (L/\Lsun)$  $ M_{\odot} \, \rm yr^{-1}$.

In our model the wind of helium stars is either Reimers-like (Eq.\,\ref{Reimers}) or Wolf-Rayet-(WR) like \citep{Hamann95,Hamann98},
\begin{equation}
\dot{M}_{\rm WR} = 10^{-13} \,  (L/\Lsun)^{1.5} \Msun\,{\rm yr^{-1}}\,,
\end{equation}
depending on which of the two is stronger,
\begin{equation}
\label{eq::Mhestar}
\dot{M}_{\rm He-star} = {\max} (\dot{M}_{\rm WR}, \dot{M}_{\rm R})\,.
\end{equation}

\subsection{Binary star evolution \label{sec:BSE}}
Binary evolution can significantly impact the evolution of the individual stars in a binary system. Binary evolution is primarily determined by the initial semi-major axis ($a$) and the masses of the two stars. In the widest systems ($a \gtrsim 10^5\Rsun$) the stars do not interact, but evolve as though they were single. In closer systems ($a \lesssim 10^5\Rsun$) interaction with the wind of the companion star can alter the evolution. In even closer systems ($a \lesssim 3\cdot 10^3\Rsun$) Roche-lobe overflow (RLOF) or common envelope (CE) evolution often has dramatic consequences for the further evolution of the two stars \cite[]{Hurley02}. These different interactions are discussed below.

We assume the initial eccentricity ($e_{\rm i}$) is zero, based on the work of \cite{Hurley02} who show that the evolution of close binary populations is almost independent of the initial eccentricity. We use the subscripts d and a for the donor and accreting star, respectively, and i for the initial characteristics of the star on the zero-age main-sequence (ZAMS). We use the subscripts $1$ and $2$ for the initially more massive star, the primary, and initially less massive star, the secondary, respectively. The mass ratio $M_{\rm a}/M_{\rm d}$ is denoted by $Q$, while the initial mass ratio $q_{\rm i}$ is $M_{\rm 2,i}/M_{\rm 1,i}$.

\subsubsection{Interaction with a stellar wind\label{sec:IntWind}}
Mass lost in the form of a stellar wind is accreted by the companion at a rate that depends on the mass loss rate, the velocity of the wind and the distance to the companion star. To describe this process and to calculate the rate $\dot{M}_{\rm a}$ at which material is accreted by the companion star, we use the Bondi-Hoyle prescription \citep{BondiHoyle44,Hurley02}, more specifically,
\begin{equation}
\label{eq::BH}
\dot{M}_{\rm a} = \min \left(0.8 \dot{M}_{\rm d}, -1 \left[\frac{G M_{\rm a}}{v_{\rm d}^2}\right]^2 \frac{\alpha_{\rm BH}}{2a^2} \frac{1}{(1+v_{\rm rel}^2)^{3/2}} \dot{M}_{\rm d}\right),
\end{equation}
with $\dot{M}_{\rm d}$ the mass loss rate from the donor star, $e$ the eccentricity, $a$ the semi-major axis, $v_{\rm d}$ the wind velocity of the donor star which we assume equals 0.25 times the escape velocity as proposed by \cite{Hurley02}, and $v_{\rm rel} = |v_{\rm orb}/v_{\rm d}|$, with $v_{\rm orb}$ the orbital velocity. The Bondi-Hoyle accretion efficiency parameter, $\alpha_{\rm BH}$, is set to 3/2 in our standard model (Table\,\ref{ParTable}). When both stars lose mass through a wind they are treated independently, no interaction of the two winds is assumed.

Some observed RS CVn systems show that the less evolved star is the more massive of the binary before RLOF occurs, therefore it has been suggested that wind mass loss is tidally enhanced by a close companion \cite[]{ToutEggleton88}. It is uncertain if this phenomenon occurs for other types of binaries. To approximate this effect the following formula is implemented \cite[]{ToutEggleton88},

\begin{equation}
\label{TE88}
\dot{M} = \dot{M}_{\rm R} \left( 1+ B_{\rm wind} \cdot  \max \left[\frac{1}{2},\frac{R}{R_{\rm L}} \right]^6 \right),
\end{equation}
where $R_{\rm L}$ is the Roche radius of the star \cite[]{Eggleton83}. In our standard model $B_{\rm wind}=0$ (Table\,\ref{ParTable}).

\subsubsection{Stability of RLOF\label{sec:RLOF}} 

\begin{table} 
\caption{The critical mass ratio $Q_{\rm crit}$ for stable RLOF for different types of donor stars, in the case of a non-degenerate and a degenerate accretor.}
\label{TableQ}
\begin{center}
%\begin{tabular}{l|c|c}
\begin{tabular}{lcc}
\hline \noalign{\smallskip} 
Type of donor & Non-degenerate & Degenerate \\
 & accretor\tablefootmark{a,b} & accretor\tablefootmark{a} \\
\hline \noalign{\smallskip} %\hline 
MS, $M>0.7$\Msun\tablefootmark{c} & 0.625\tablefootmark{d} (MS accretor) & -- \\
MS, $M<0.7$\Msun\tablefootmark{c} & 1.44 & 1.0 \\
HG & 0.25 & 0.21 \\
GB, AGB\tablefootmark{e} & $2.13/\left[1.67-x+2\left(\frac{M_{\rm c,d}}{M_{\rm d}}\right)^5\right]$ & 0.87 \\
He-HG & 0.25 & 0.21 \\
He-GB & 1.28 & 0.87 \\
WD & -- & 1.6 \\
\hline
\end{tabular}
\tablefoot{\tablefoottext{a}{The symbol -- indicates that no value is defined.}
\tablefoottext{b}{Based on \cite{Hurley02}.}
\tablefoottext{c}{We distinguish between MS stars ($M > 0.7$\Msun) and low-mass MS stars ($M < 0.7$\Msun), because the latter are almost completely convective, and react differently to mass loss and gain.}
\tablefoottext{d}{Based on \cite{DeMink07}.}
\tablefoottext{e}{At solar metallicity $x$ $\approx$ 0.3 \cite[]{Hurley00,Hurley02}.}}
\end{center}
\end{table}

Whether or not mass transfer is stable depends on: 1) the reaction of the donor, because a star with a convective envelope responds differently to mass loss than a star with a radiative envelope; 2) the evolution of the orbit, which itself depends on the accretion efficiency ($\beta = \dot{M}_{\rm a}/\dot{M}_{\rm d}$) and the mass ratio $Q = M_{\rm a}/M_{\rm d}$; and 3) the reaction of the accreting star. In our model the criterion to determine whether mass transfer is stable is given by a critical mass ratio $Q_{\rm crit}$, which depends on the types of donor and accreting stars (Table\,\ref{TableQ}). During RLOF the mass ratio $Q$ of the binary system is compared with $Q_{\rm crit}$: if $Q < Q_{\rm crit}$ mass transfer is dynamically unstable and a CE phase follows, otherwise RLOF is stable. In Table\,\ref{TableQ} we list the values of the critical mass ratios for the different phases of the evolution of a star during accretion onto either a non-degenerate or a degenerate star.

RLOF from a MS donor is generally stable, because a radiative MS star shrinks in reaction to mass loss. When material is transferred to a MS companion, RLOF is stable only for certain mass ratios because a radiative accreting star expands and can fill its own Roche lobe. If the mass ratio is less than 0.625 the system evolves into a contact system and we assume the two MS stars merge~\cite[][]{DeMink07}. The stability criteria for the other types of donor stars transferring mass to a non-degenerate companion are calculated according to \cite{Hurley02}.

To calculate $Q_{\rm crit}$ of WDs, we assume that during RLOF barely anything is accreted onto the WD, $\beta=0.01$, and the specific angular momentum  of the ejected material is that of the orbit of the accreting star \cite[][]{Hachisu96}. Because of the low accretion efficiencies of WDs (Appendix\,\ref{app::efficiencyWD}), the critical mass ratio for stable mass transfer decreases. Our adopted critical mass ratios for mass transfer from non-degenerate donor stars onto a WD are calculated with the formulae from \citet[][Chiotellis priv. comm.]{Soberman97}.

\subsubsection{Common envelope evolution\label{sec:CE}}
If the donor star is an evolved star and the mass ratio of the Roche-lobe overflowing system is less than its critical value (Table\,\ref{TableQ}), the system evolves into a CE. We use the $\alpha$-prescription to describe this complex phase \cite[]{Webbink84}, in which the binding energy ($E_{\rm bind}$) of the CE is compared with the orbital energy ($E_{\rm orb}$) of the system to calculate the amount the orbit should shrink in order to lose the envelope of the donor star. More specifically, $\alpha_{\rm ce}$ is defined by,
\begin{equation}
\label{eq::alphace}
{E}_{\rm bind,i} = \alpha_{\rm ce} (E_{\rm orb,f} - E_{\rm orb,i}),
\end{equation}
where the subscripts i and f correspond to the states before and after the CE phase, respectively, and where
\begin{equation}
\label{eq::lambdace}
{E}_{\rm bind,i} = -G \left(\frac{M_{\rm d} M_{\rm env,d}}{\lambda_{\rm ce} R_{\rm d}} \right),
\end{equation}
and $M_{\rm d}$, $M_{\rm env,d}$ and $R_{\rm d}$ are the mass, the envelope mass, and the radius of the donor, respectively, and $G$ is the gravitational constant.

The CE efficiency parameter, $\alpha_{\rm ce}$, describes how efficiently the energy is transferred from the orbit to the envelope. Its value is expected to be between 0 and 1, although it could be larger than 1 if another energy source is available, such as nuclear energy. In our standard model we take $\alpha_{\rm ce}$ = 1 (see Table\,\ref{ParTable}). The parameter $\lambda_{\rm ce}$ depends on the relative mass distribution of the envelope, and is not straightforward to define \cite[][]{Ivanova11}. According to \cite{Ivanova11} its value is close to 1 for low-mass red giants and therefore in many studies taken as such. In our model $\lambda_{\rm ce}$ is variable and is dependent on the type of star, its mass and luminosity. We use a prescription based on \citet[][our Appendix\,\ref{app::lambda}]{DewiTauris00}, which gives a value of $\lambda_{\rm ce}$ between 0.25 and 0.75 for HG stars, between 1.0 and 2.0 for GB and AGB stars and $\lambda_{\rm ce}$ = 0.5 for helium stars. Because of the short timescale of the CE phase \cite[e.g.][]{Passy12}, we assume in our standard model that no mass is accreted by the companion star (Table\,\ref{ParTable}).

Additional energy sources can boost the envelope loss, e.g.\,the ionization energy of the envelope. The extended envelope can become cool enough that recombination of hydrogen occurs in the outer layers. In our model the fraction of this energy which is used is expressed by $\lambda_{\rm ion}$, which is between 0 and 1 (Appendix\,\ref{app::lambda}). In our standard model this effect is not considered and $\lambda_{\rm ion}$ = 0 (Table\,\ref{ParTable}).

\subsubsection{Stable RLOF\label{sec:stableRLOF}}
When mass transfer is stable ($Q \, > \, Q_{\rm crit}$, Table\,\ref{TableQ}), the mass transfer rate is calculated as a function of the ratio of the stellar radius of the donor $R_{\rm d}$ and the Roche radius $R_{\rm L}$ \cite[based on ][]{WhyteEggleton80},
\begin{equation}
\label{H02}
\dot{M} = f \cdot 3.0\cdot 10^{-6} \, \left(\log\left[\frac{R_{\rm d}}{R_{\rm L}}\right]\right)^3 \, \left({\min}\left[\frac{M_{\rm d}}{\Msun},5.0\right]\right)^2 \Msun \, \rm yr^{-1}\,.
\end{equation}
The last, mass-dependent factor in the equation is defined by \cite{Hurley02} for stability reasons. In \citet{Hurley02} the dimensionless factor $f$ is 1, however this underestimates the rate when mass transfer proceeds on the thermal timescale. In our model $f$ is not a constant, but depends on the stability of mass transfer. If $f$ is large, the radius stays close to the Roche radius and mass transfer is self-regulating. However, in our model numerical instabilities arise when $ \log R_{\rm d}/R_{\rm L} \lesssim 10^{-3}$ and the envelope of the donor star is small. Therefore, a function is defined which forces the radius to follow the Roche radius more closely during thermal-timescale mass transfer ($f$ = 1000) and more loosely during nuclear timescale mass transfer ($f$=1). A smooth transition is implemented between the two extreme values of $f$. To test the stability of the function to calculate the mass transfer rate, a binary grid was run of $50^3$ binary systems with \texttt{binary\_c/nucsyn}, with initial primary masses between 2 and 10\Msun, secondary masses between 0.5 and 10\Msun~and initial separations between 3 and $10^4$\Rsun. The mass transfer rates of systems evolving through the SD channel with a hydrogen-rich donor were used to optimize this function for computational stability. The resulting function is given by
\begin{equation}
\label{calcF}
f = \begin{cases}
1000 &  Q < 1 \text{ and} \\
\displaystyle \max \left(1, \frac{1000}{Q}\cdot \exp\left[-\frac{1}{2} \left\{\frac{\ln Q}{0.15}\right\}^2  \right]\right) &  Q > 1. \\
\end{cases}
\end{equation}

Because mass transfer is dynamically stable, the mass transfer rate is capped at the thermal timescale mass transfer rate $\dot{M}_{\rm KH}$, given by
\begin{equation}
\label{KH}
\dot{M}_{\rm KH,d} = \frac{M_{\rm d}}{\tau_{\rm KH}} \Msun\,{\rm yr^{-1}},
\end{equation}
with $\tau_{\rm KH}$ the thermal timescale of the donor star,
\begin{equation}
\label{tKH} 
\tau_{\rm KH} = 1.0 \cdot 10^7 \frac{{M M'}}{\Msun^2} \frac{\Rsun}{R} \frac{\Lsun}{L} \rm yr\,,
\end{equation}
and $M'$ either the mass of the star, if the star is on the MS or He-MS, or the envelope mass of the star otherwise.

In order to test the mass transfer rate calculated with our BPS code, as a next step a set of binary systems and their mass transfer rates were calculated with the our BPS code and with a detailed binary stellar evolution code \cite[STARS,][]{Eggleton71,Eggleton06, Pols95,Glebbeek08} and compared. We simulated a grid with primary masses between 2 and 6\Msun, secondary masses varied between 0.7 and 3.5\Msun~and orbital periods varied between 2 and 4 days (to check binary systems with both MS and HG donors) and a more massive binary system of 12 and 7\Msun~at orbital periods of 3.5 and 4 days. We find that the resulting maximum mass transfer rate computed with the BPS code is up to about a factor three (for MS donors) and a factor five (for HG donors) larger than the maximum from the detailed stellar evolution code, with the duration of thermal timescale mass transfer correspondingly shorter. Additionally, we find similar durations of the entire mass transfer phase calculated with both codes.

In addition, the accretor adjusts its structure to the accreted mass. If the mass transfer rate is higher than the thermal timescale rate of the accretor,  the accreting star is brought out of thermal equilibrium, resulting in expansion and additional mass loss from the accretor. Consequently, during RLOF the fraction of transferred material that is accreted is not taken to be constant, but depends on the thermal timescale of the accretor. For moderately unevolved stars (stars on the MS, HG or helium stars) the accretion efficiency $\beta$ is calculated in our model as follows
\begin{equation}
\label{MaxMaccr}
\beta \equiv \frac{\dot{M}_{\rm a}}{\dot{M}_{\rm d}} = \min \left(\sigma \frac{\dot{M}_{\rm KH,a}}{\dot{M}_{\rm d}},1\right),
\end{equation}
where $\sigma$ is a parameter for which we assume a value of 10 in our standard model \cite[as in][]{Hurley02}. Moreover, in the case of accretion onto a MS or He-MS star, rejuvenation is assumed \cite[]{Hurley02}. The internal structure of the star is changed and new fuel is mixed into the burning region, which results in a star that appears younger. If the accretor is an evolved star on the GB or AGB, mass transfer is assumed to be conservative ($\beta = 1$) because a convective star shrinks as a reaction to mass gain. Accretion onto a WD is a special situation because of its degeneracy and this is be discussed separately below.

When material is lost from the system it removes angular momentum. Angular momentum loss is described with a parameter $\gamma$ that expresses the specific angular momentum of the lost material in terms of the average specific orbital angular momentum, as follows:

\begin{equation}
\label{eq::jorb}
\frac{\dot{J}_{\rm orb}}{J_{\rm orb}} = \gamma (1-\beta) \frac{\dot{M}_{\rm d}}{M_{\rm d}+ M_{\rm a}}.
\end{equation}
In our standard model we assume that during stable RLOF material is lost by isotropic re-emission, removing the specific orbital angular momentum of the accretor ($\gamma$ = $M_{\rm d} / M_{\rm a}$, Table\,\ref{ParTable}).

\subsubsection{Stable RLOF onto a WD\label{sec:RLOFWD}} 
Because a WD is degenerate, it burns accreted material stably only over a small range of mass transfer rates which corresponds to approximately $10^{-7}$\Msun/yr~when hydrogen-rich material is accreted and $10^{-6}$\Msun/yr~when helium-rich material is accreted \cite[][]{Nomoto82}. If the mass transfer rate is too low, the material is not burnt immediately and a layer of material is deposited on the surface. This layer burns unstably, resulting in novae, and, if the mass transfer rate is too high, the WD cannot burn all the accreted material. According to \cite{Nomoto82} the accreted material forms an envelope around the WD and becomes a red giant-like stellar object with a degenerate core and, generally, a CE subsequently forms. \cite{Hachisu96} propose that the burning material on top of the WD drives a wind which blows away the rest of the accreted material. The accreted material burns at the rate of stable burning, but contact is avoided and mass transfer remains stable. We take the latter possibility into account (for a description see Appendix\,\ref{app::efficiencyWD}). The material ejected through the wind from the WD removes specific angular momentum from the WD ($\gamma= M_{\rm d} / M_{\rm a}$). The above also holds for WDs accreting through a wind, but the material transferred to the WD is based on the Bondi-Hoyle accretion efficiency (Eq.\,\ref{eq::BH}).

\subsection{Binary population synthesis\label{Sect::BPS}}
We simulate $N_{M_{\rm 1,i}}\times N_{M_{\rm 2,i}}\times N_{a_{\rm i}}$ binary systems in $\log M_{\rm 1,i}$ - $\log M_{\rm 2,i}$ - $\log a_{\rm i}$ space, with  $M_{\rm 1,i}$ and $M_{\rm 2,i}$ the initial masses of the primary and secondary stars and $a_{\rm i}$ the initial semi-major axis of the binary systems. The volume of each cell in the parameter space is $\delta M_{\rm 1,i} \ \delta M_{\rm 2,i} \ \delta a_{\rm i}$. To compute the SN rate each system is assigned a weight $\Psi$ according to the initial distributions of binary parameters. We normalize the SN rate to the total mass of the stars in our grid,
\begin{equation}
M_{\rm total} = \sum^{M_{1,\rm i,max}}_{M_{1,\rm i,min}} \sum^{M_{2,\rm i,max}}_{M_{2,\rm i,min}} \sum^{a_{\rm i,max}}_{a_{\rm i,min}} (M_{\rm 1,i} + M_{\rm 2,i}) \Psi \, \delta M_{\rm 1,i} \, \delta M_{\rm 2,i} \, \delta a_{\rm i}\,,
\end{equation}
where
\begin{itemize}
\item $M_{1, \rm i,min}= 0.1$\Msun\ and $M_{1, \rm i,max}= 80$\Msun,
\item $M_{2, \rm i,min}=0.01$\Msun~\cite[]{Kouwenhoven07} and $M_{2, \rm i,max}=M_{\rm 1,i}$,
\item $a_{\rm i,min}=5$\Rsun\ and $a_{\rm i,max}=5\cdot 10^6$\Rsun~\cite[]{Kouwenhoven07} and
\item $ \Psi$ is the initial distribution of $M_{\rm 1,i}$, $M_{\rm 2,i}$ and $a_{\rm i}$. 
\end{itemize}
We assume that $\Psi$ is separable, namely
\begin{equation}
\Psi(M_{\rm 1,i}, M_{\rm 2,i}, a_{\rm i}) = \psi(M_{\rm 1,i}) \, \phi(M_{\rm 2,i}) \, \chi(a_{\rm i})\,,
\end{equation}
where 
\begin{itemize}
\item $\psi(M_{\rm 1,i})$ is the initial distribution of primary masses from \citet[][Table\,\ref{ParTable}]{Kroupa93},
\item $\phi(M_{\rm 2,i})$ is the initial secondary masses distribution, which we assume is flat in $M_{\rm 2,i}/M_{\rm 1,i}$ and 
\item $\chi(a_{\rm i})$ is the initial distribution of semi-major axes, which we assume is flat in log $a_{\rm i}$ \cite[]{opik24,Kouwenhoven07}.
\end{itemize}

We calculate the delay time distribution (DTD), which is the SN~rate as a function of time per unit mass of stars formed in a starburst at $t = 0$, as follows:
\begin{equation}
\text{DTD}(t) = \frac{\sum^{M_{\rm 1,i,max}}_{M_{\rm 1,i,min}} \sum^{M_{2,\rm i, max}}_{M_{\rm 2,i,min}} \sum^{a_{\rm i,max}}_{a_{\rm i,min}} \delta\text{(SN~Ia)} \, \Psi \, \delta M_{\rm 1,i} \, \delta M_{\rm 2,i} \, \delta a_{\rm i}} {M_{\rm total}  \, \delta t}\,,
\label{Form::CalcDTD}
\end{equation}
where $\delta\text{(SN~Ia)} = 1$ if the binary system leads to a SN~Ia event during a time interval $t$ to $t+\delta t$, otherwise $\delta\text{(SN~Ia)} = 0$. We assume all stars are formed in binaries, which is an overestimate of the binary fraction, because for low-mass stars the observed fraction of system in binaries is less than 50\% \cite[]{Lada06}. However, in intermediate mass stars, \cite{Kouwenhoven07} find a best fit with 100\% binaries and \cite{DucheneKraus13} conclude that different surveys are consistent with a multiplicity higher than 50\%. Additionally, \cite{Kobul07} and \cite{Sana12} find that more than 70\% of massive stars are in binary systems.

The results of Sect.\,\ref{Sect::BinEvol} are calculated by simulating a grid with $M_{1,\rm i}$ between 2.5 and 9\Msun, $M_{2, \rm i}$ between 1\Msun~and $M_{1}$ and $a_{\rm i}$ between 5 and $5 \cdot 10^3$\Rsun, with $N_{M_{\rm 1,i}}=N_{M_{\rm 2,i}}=N_{a_{\rm i}}=N=150$.

\section{Binary progenitor evolution\label{Sect::BinEvol}}
In this section we discuss the general progenitor evolution of the different SN~Ia channels, SD and DD, and their contribution according to our standard model (Sect.\,\ref{sec:StandAssum}). We describe binary evolution in terms of the number of stable or unstable phases of RLOF and the stellar types at the onset of mass transfer. The latter illustrates the influence of different aspects of binary evolution and initial binary distributions (Sects.\,\ref{sec:inflPar}\,and\,\ref{sec:inflDistr}). In the following sections the mass ratio $q$ is $M_2/M_1$ were the suffixes 1 and 2 denote the initially more and less massive star, respectively.

\subsection{Double degenerate channel\label{sec:DDchann}}

\begin{figure*}
\centering
\includegraphics[width=9.15cm]{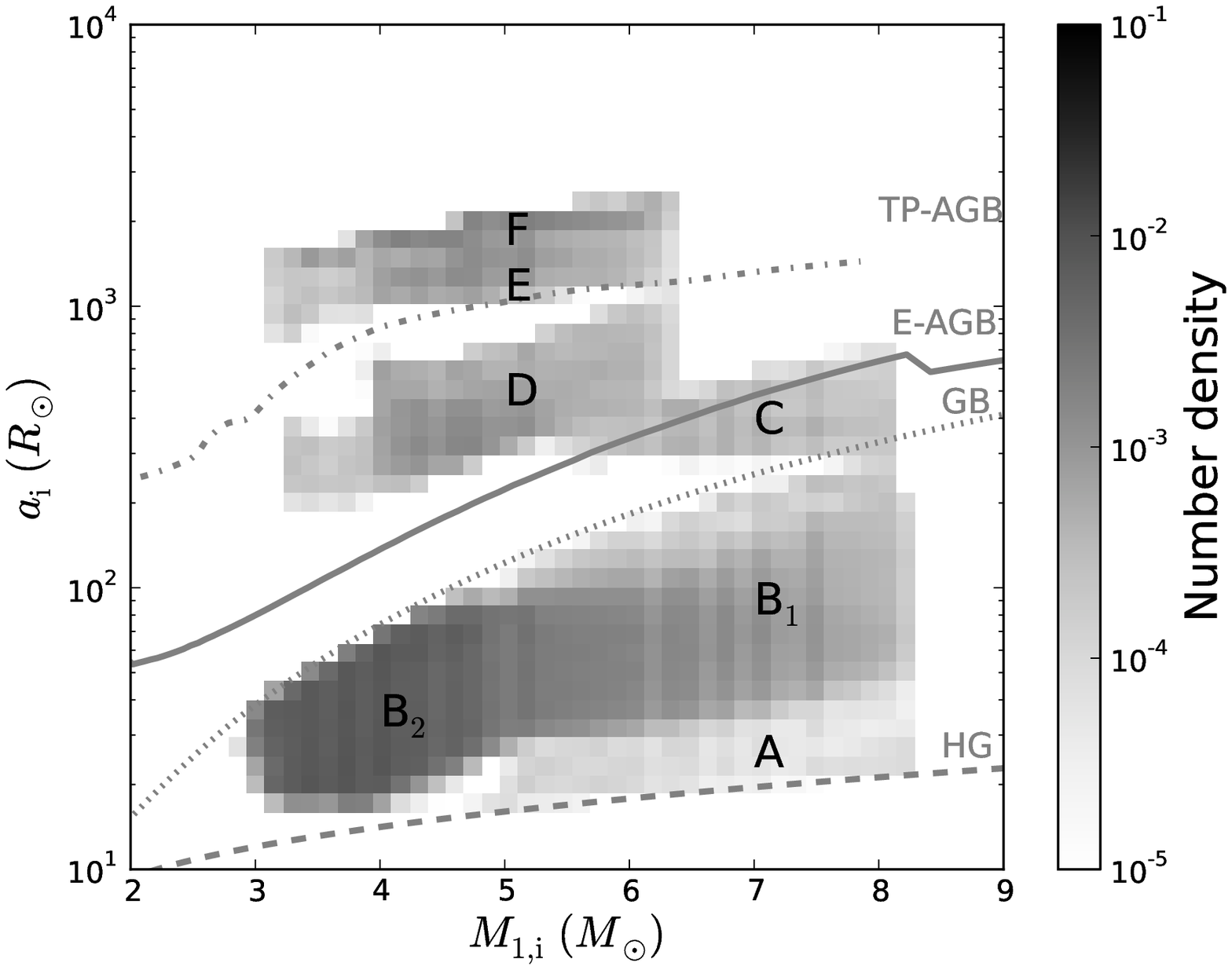}
\includegraphics[width=9.15cm]{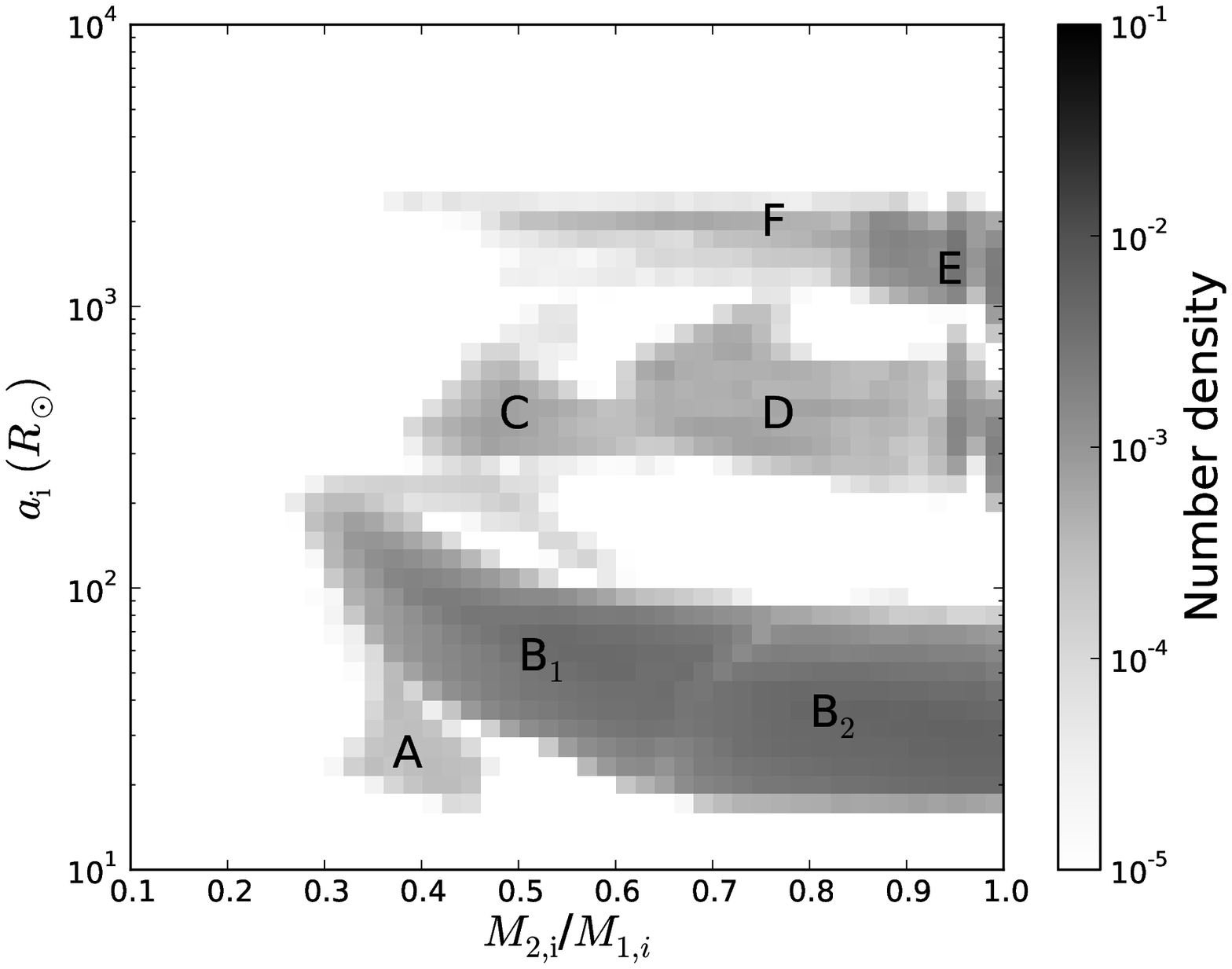}
\caption{Initial separation ($a_{\rm i}$) versus initial mass of the primary star ($M_{\rm i}$, left) and versus initial mass ratio ($M_{2,\rm i}$/$M_{1,\rm i}$, right) of systems that form SNe~Ia through the DD channel in our standard model (Sect.\,\ref{sec:StandAssum}). Number density (greyscale) represents the number of systems normalized to the total number of systems forming a SN~Ia through the DD channel. Lines show the minimum  separation (assuming $q = 1$) for which the primary fills its Roche lobe in a certain evolution stage, as indicated. Symbols A-F indicate differences in the evolution stage when the primary fills its Roche lobe and in the number of CE phases necessary to evolve to a SN~Ia: A = HG, no CE; B = HG, 1 CE (the difference between B$_1$ and B$_2$ is based on which of the two stars first forms a WD); C = GB, 1 CE; D  = E-AGB, 1 CE; E  = TP-AGB, 1 CE; F = TP-AGB, 2 CEs. }
\label{DDCEa}
\end{figure*}

The double degenerate channel (DD) needs two CO WDs, with a combined mass greater than $M_{\rm Ch}$ in a short enough orbit ($a \lesssim 4\Rsun$), to merge within a Hubble time. Consequently, the rate of this channel depends on the number of systems that, after phases of stable RLOF and/or CE evolution, are in such a short orbit. 
The results of \citet{Pakmor10, Pakmor11} indicate that the violent merger of two CO WDs only results in an explosion for a limited range of mass ratios. In combination with possible mass loss during the merger, this restriction can reduce the SN Ia rate from the DD channel by about a factor of five \citep{Chen12}. However, in this work we do not impose a restriction on the mass ratio of the two CO WDs.

Multiple formation channels can form close double WD systems \cite[]{Mennekens10,Toonen12}. \cite{Mennekens10} distinguish between the common envelope channel, which needs two consecutive CE phases, and the Roche-lobe overflow channel, in which a phase of stable Roche-lobe overflow is followed by a CE phase. Additionally, \cite{Toonen12} discuss the formation reversal channel where the secondary forms a WD first. Fig.\,\ref{DDCEa} shows the distribution of systems that evolve towards SNe~Ia according to the DD channel in terms of their initial separation $a_{\rm i}$, initial primary mass $M_{1,\rm i}$ and initial mass ratio $q_{\rm i}$. We distinguish six main regions: two lower regions A and B, with $a_{\rm i}/R_{\odot} \lesssim$ 300, in which the primary fills its Roche lobe during the HG; two intermediate regions C and D, with 300 $\lesssim a_{\rm i}/R_{\odot} \lesssim$ 1000, in which the primary fills its Roche lobe during the GB or E-AGB; and two upper regions E and F, with $a_{\rm i}/R_{\odot} \gtrsim$ 1000, in which one or more CE phases are needed and the primary fills its Roche lobe during the TP-AGB.

\paragraph{\textit{Close systems: regions A and B.}}
If the initial separation is shorter than about 300\Rsun~the primary first fills its Roche lobe on the HG when the star has a radiative envelope. The first phase of mass transfer is thus stable and a CE is avoided. 

The separation and the mass of the secondary determine whether the second phase of mass transfer, from the secondary, is stable. The separation determines both the moment the secondary fills its Roche lobe and the stellar type of the secondary at RLOF. The initially closest systems with the least massive secondaries avoid a CE and two CO WDs form in a short orbit without any CE phase (region A), while the wider systems experience a CE phase when the secondary fills its Roche lobe (region B).

Region A consists of systems that avoid a CE phase during their entire evolution. The range of initial mass ratios that make SNe~Ia is strongly restricted (Fig.\,\ref{DDCEa}), because systems with initial mass ratios larger than 0.46 form double WD systems in an orbit too wide to merge in a Hubble time. Because of this restriction, only 0.7\% of the systems that become a SN~Ia via the DD channel follow this evolutionary channel. However, when RLOF is followed with a detailed stellar evolution code, some of these systems with a mass ratio smaller than 0.46 that contain an accreting MS-star are expected to form a contact system \cite[]{DeMink07} which eventually merges before the formation of two CO WDs.

Region B represents the most common evolutionary channel and corresponds to the Roche-lobe overflow channel \citep{Mennekens10,Toonen12}. In this region the primary starts mass transfer during the HG and a CE phase occurs when the secondary fills its Roche lobe. This channel accounts for 84\% of SNe~Ia formed through the DD channel in our model, a similar fraction to \cite{Mennekens10}. Binary systems with initial masses lower than 2.9\Msun~do not form a merger product with a combined mass greater than 1.4\Msun, while systems with primary masses higher than 8.2\Msun~form an ONe WD. Binary systems with mass ratios lower than 0.25 have an unstable first mass transfer phase and merge.

In region B we distinguish between regions B$_1$ and B$_2$, based on which of the two stars becomes a WD first. In general, the initially more massive star is expected to form the first WD of the binary system (region B$_1$). However, sometimes the evolution of the secondary catches up with the primary because of previous binary interaction and becomes a WD first (region B$_2$), which is the formation reversal channel \cite[][]{Toonen12}. Region B$_2$ contains 49\% of the systems forming region B (41\% of all DD systems).

The systems of region B$_2$ have primary masses between 2.9 and 5\Msun~and mass ratios between 0.67 and 1. When a hydrogen-rich star loses its envelope before the TP-AGB it becomes a helium star before evolving into a WD (Sect.\,\ref{sec:SSE}). In this case the resulting helium stars have a mass between 0.5 and 0.85\Msun. The time it takes for these low-mass helium stars to become a WD is long (between about 40 and 160~Myr) which gives the secondary, which has increased in mass, time to evolve and fill its Roche lobe. It becomes the most massive helium star of the binary system after the subsequent CE phase. The short orbit after the CE phase allows the secondary to fill its Roche lobe again  and evolve into the first WD of the binary system. Afterwards, the primary fills its Roche lobe for the second time and becomes the second WD of the binary system. The range of primary masses that make SNe~Ia is restricted because the lifetime of the corresponding helium star has to be long enough for the secondary to evolve. This restriction originates from the necessity for the secondary to fill its Roche lobe before the primary becomes a WD.

Systems with initial separations in the gap between region B and regions C and D do not form a double WD system. The primary fills its Roche lobe at the end of the HG or on the GB, which leads to a CE phase in which the two stars merge.

\paragraph{\textit{Wide systems that undergo one CE phase: regions C, D and E.}} In systems with separations longer than about 300\Rsun~the primary fills its Roche lobe after the HG while having a convective envelope which results in a CE phase. The primary becomes a WD immediately (region E) or after subsequent evolution as a helium star (regions C and D). Afterwards, the secondary fills its Roche lobe. In regions C, D and E mass transfer from the secondary is stable, permitting accretion onto the initial WD. However, the WD cannot reach $M_{\rm Ch}$ before the secondary loses its entire envelope and the binary becomes a short-period double WD system. This imposes a restriction on the range of initial mass ratios that make SNe~Ia, which determines the separation after the phase of stable RLOF, because the orbit should be short enough when the secondary becomes a WD. This channel produces 12.8\% of DD progenitors (1.9\% in region C, 6.3\% in region D and 4.6\% in region E). The difference between the three regions is the stellar type of the primaries at the onset of mass transfer: in region C the primary is on the GB, in region D on the E-AGB and in region E on the TP-AGB (Fig.\,\ref{DDCEa}).

The ranges of the initial masses of the three evolutionary channels are defined by the necessity for both stars to become a CO WD (upper limit on the masses) and form a massive enough merger (lower limit on the masses). In some systems with a mass ratio close to one ($q_{\rm i} \gtrsim 0.93$) the secondary is already evolved at the moment the primary fills its Roche lobe and during the CE phase both envelopes are lost. The secondary still fills its Roche lobe after the primary becomes a WD, but as a helium star, viz. on the He-HG.

The systems with initial separation in the gap between regions D and E do not form double WD systems. These systems survive the first CE phase and form a helium star with a non-evolved companion. However, because of the longer initial separation, the helium star fills its Roche lobe again as a He-giant. This leads to a second CE phase, during which the two stars merge.

\paragraph{\textit{Wide systems that undergo two CE phases: region F.}} When the initial separation is longer than about 1000\Rsun~the first mass transfer phase starts when the primary is on the TP-AGB. This leads to unstable mass transfer and therefore a CE phase, during which the separation decreases. The primary becomes a CO WD immediately. Subsequently, the secondary fills its Roche lobe which also results in unstable mass transfer. This evolutionary channel is similar to the common envelope channel discussed in \cite{Mennekens10}\ and\ \cite{Toonen12}. This channel produces 2.4\% of the DD systems. These systems have the first phase of mass transfer during the TP-AGB (Fig.\,\ref{DDCEa}), however this evolutionary channel also occurs when mass transfer starts during the E-AGB. Nevertheless, the systems which start RLOF during the E-AGB and have two consecutive CE phases generally merge before the formation of a double WD system and only account for 0.1\% of all DD systems.

In systems with longer initial separations ($a \gtrsim 2500$\Rsun), both stars do not fill their Roche lobe, and therefore do not form double WD systems in a short orbit (Fig.\ref{DDCEa}).

\subsection{Single degenerate channel\label{sec:SDchann}}
The single degenerate channel needs a CO WD and a non-degenerate companion which provides enough mass to the WD at a high enough rate. We distinguish between hydrogen- and helium-rich companions.

\subsubsection{SD with hydrogen-rich donor (SD$_{\rm H}$) \label{Sec::SDH}}

\begin{figure}
\begin{center}
\includegraphics[width=9.15cm]{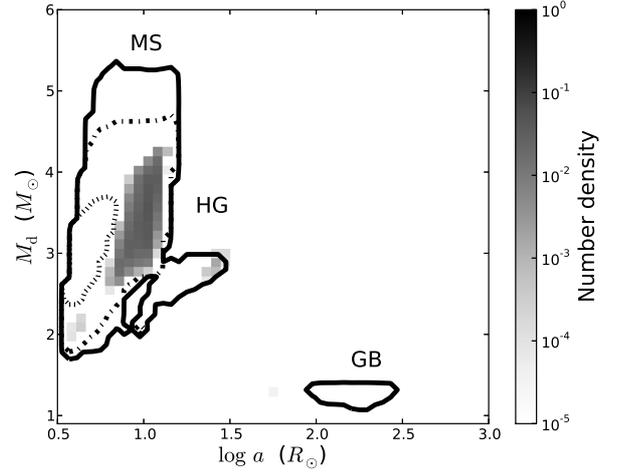}
\caption{Donor mass ($M_{\rm d}$) vs. separation ($a$) at the moment of formation of the CO WD of the systems producing SNe~Ia through the SD$_{\rm H}$ channel, for different CO WD masses: 1.15\Msun~(solid line), 1.0\Msun~(dash-dotted line) and 0.8\Msun~(dotted line). The greyscale shows how these regions are populated in our standard model. Labels indicate the different stellar types of donors stars.}
\label{HachisuSDH}
\end{center}
\end{figure}

\begin{figure*}
\begin{center}
\includegraphics[width=9.15cm]{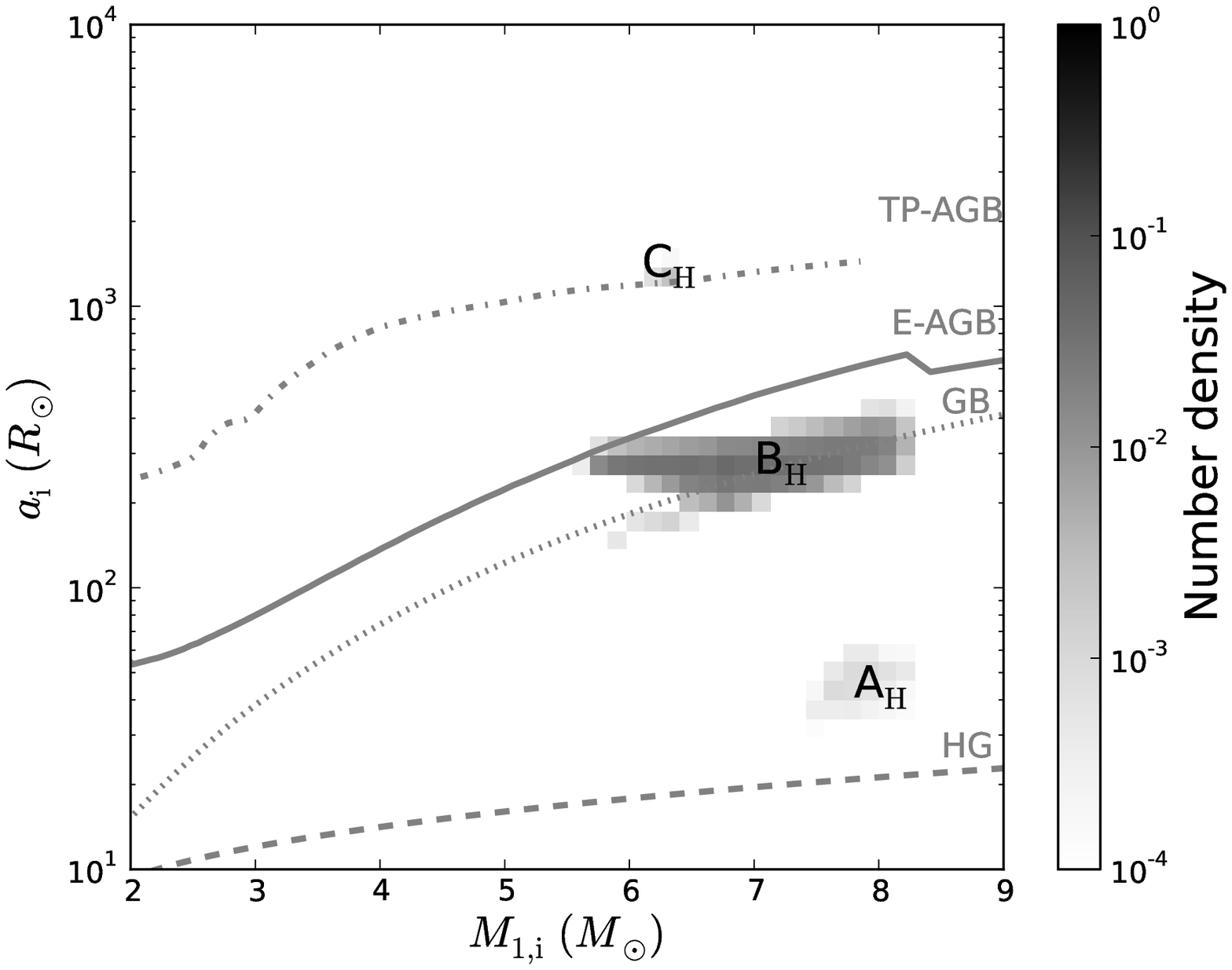}
\includegraphics[width=9.15cm]{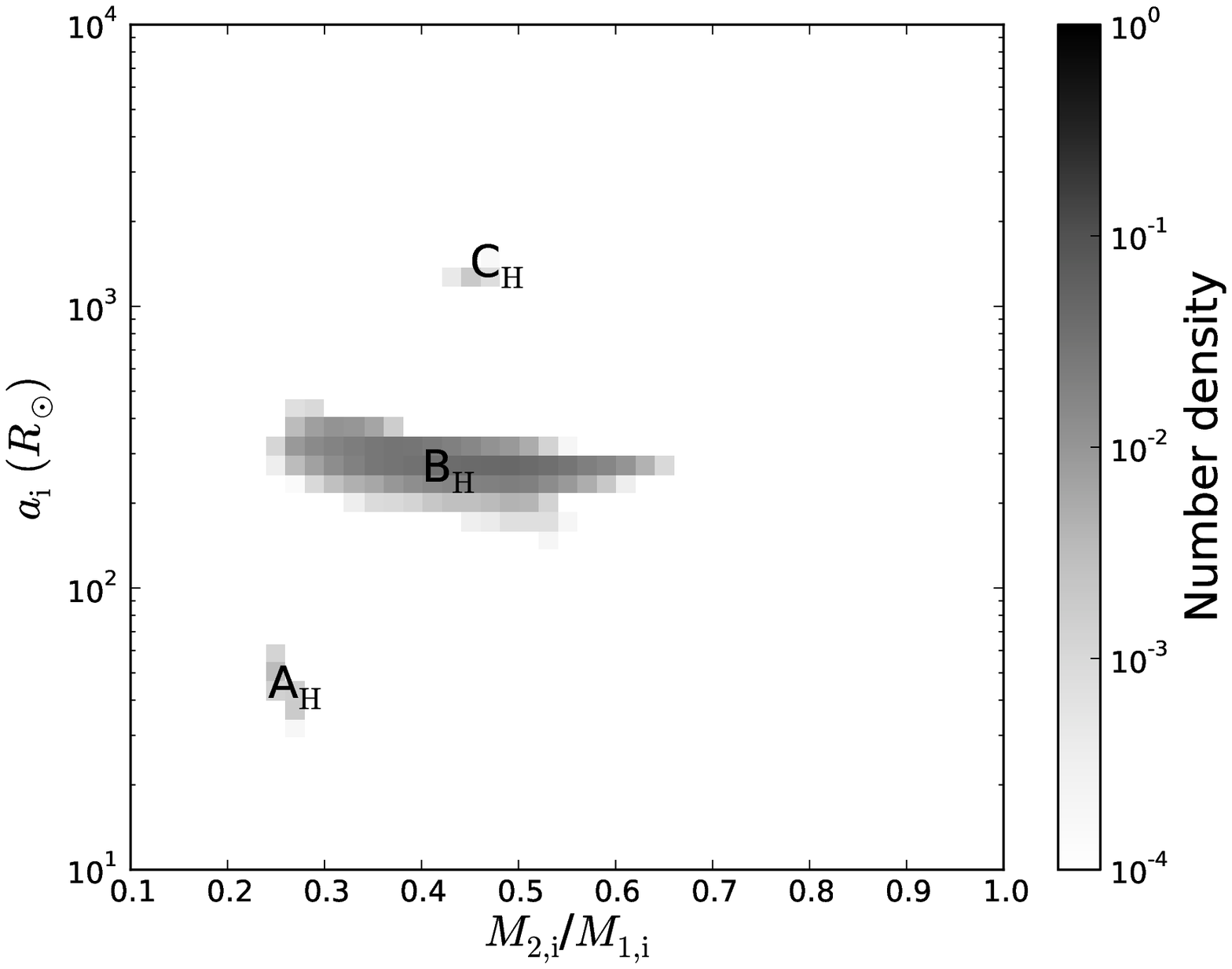}
\caption{As Fig.\,\ref{DDCEa} for systems that form SNe~Ia through the SD$_{\rm H}$ channel. Symbols A$_{\rm H}$, B$_{\rm H}$ and C$_{\rm H}$ indicate groups which are distinguished based on the evolutionary stage when the primary fills its Roche lobe and the number of CE phases necessary to evolve to a SN~Ia: A$_{\rm H}$ = HG, no CE; B$_{\rm H}$ = end of HG/GB, 1 CE; C$_{\rm H}$ = TP-AGB, 1 CE.}
\label{SDHCEa}
\end{center}
\end{figure*}

A hydrogen-rich companion can be in any evolutionary stage between the MS and the AGB. The stability criterion for mass transfer (Table\,\ref{TableQ}) and the rate at which mass is transferred determine which donor stars transfer enough material to the WD to make SNe~Ia. Evolved stars (GB or AGB) have a convective envelope, which results in a smaller critical mass ratio for stable mass transfer compared to non-evolved stars (MS or HG) with a radiative envelope (Table\,\ref{TableQ}). Consequently, donors on the MS can be more massive than evolved donors without the system evolving into a CE. In addition, the mass transfer rate determines the amount of material that is accreted by the WD. Ideally, the mass transfer rate is about $10^{-7}$\Msun\,yr$^{-1}$~\cite[][Sect.\,\ref{sec:StandAssum}]{Nomoto82}. Mass transfer is generally faster from GB donors than from MS donors with the same mass. Consequently, initially less massive stars on the GB than stars on the MS can donate enough mass to a CO WD to grow to\,$M_{\rm Ch}$~(Fig.\,\ref{HachisuSDH}).

The contours in Fig.\,\ref{HachisuSDH} show the results of a simulation with our BPS code in $\log M_{\rm 1,i}$ - $\log M_{\rm 2,i}$ - $\log a_{\rm i}$ space as discussed in Sect.\,\ref{Sect::BPS} with $N=125$, in which one of two stars is a WD at $t=0$ and the other is on the ZAMS. The initial masses range between 0.7 and 1.15\Msun~for the WD and between 0.7 and 6\Msun~for the companion MS star, the separation is varied between the minimum separation for a MS star to fill its Roche lobe and 10$^3$\Rsun~and we assume $e_{\rm i}=0$. The figure shows the possible ranges in donor mass and separation for which WDs with a mass of 0.8, 1.0 or 1.15\Msun~can grow to $M_{\rm Ch}$.

In Fig.\,\ref{HachisuSDH} three main regions are distinguished with $M_{\rm WD} < 1.15\Msun$. The donor stars in the region with $\log a/\Rsun \lesssim 1.2$ and $1.5 \lesssim M_{\rm d}/\Msun \lesssim 5.2$ start transferring mass to the WD during the MS, the donor stars in the region with $1.0 \lesssim \log a/\Rsun \lesssim 1.5$ and $2.4 \lesssim M_{\rm d}/\Msun \lesssim 3.0$ during the HG, and those in the region with $ 2.0 \lesssim \log a/\Rsun \lesssim 2.5$ and $1.15 \lesssim M_{\rm d}/\Msun \lesssim 1.4$ during the GB. For future reference we name these channels the WD+MS, WD+HG, WD+RG channel, respectively, as in \cite{Mennekens10}.

Fig.\,\ref{HachisuSDH} in this article can be compared with Fig.\,3 of \cite{Han04}, which was calculated with a detailed binary stellar evolution code and which discussed the WD+MS channel. Like \cite{Han04} we find that SNe~Ia can originate from accreting WDs with donor masses between 1.5 and 3.5\Msun, but in addition we find systems with donor masses greater than 3.5\Msun, which they find to become dynamically unstable and hence do not become SNe~Ia. This arises because of a different treatment of the stability of RLOF between the two codes. For the WD+RG channel we compare our results with Fig.\,2 of \citet[][]{Wang10b} who find donor masses down to 0.6\Msun. However, these systems do not become a SN~Ia within a Hubble time and therefore are not shown in our Fig.\,\ref{HachisuSDH}.

The greyscale in Fig.\,\ref{HachisuSDH} depicts the number density at WD formation of the systems evolving through the SD$_{\rm H}$ channel with our standard model. To understand why our standard model does not form WD+MS systems over the entire mass and separation range shown in Fig\,\ref{HachisuSDH}, a closer look at the distribution of the initial characteristics of the systems which evolve into a SN~Ia through the SD$_{\rm H}$ channel is necessary (Fig.\,\ref{SDHCEa}). In this figure three regions are distinguished: systems with short (A$_{\rm H}$), intermediate (B$_{\rm H}$) and long initial separations (C$_{\rm H}$).

\paragraph{\textit{Close systems: region A$_{\rm H}$.}} 
Systems with initial separation between between 20 and 70\Rsun~have the first phase of RLOF during the HG, which is stable. Subsequently a WD forms in a short orbit because of the initially low mass ratio. Afterwards the systems evolve through the WD+MS channel (Fig.\,\ref{HachisuSDH}). Only 1\% of the SD$_{\rm H}$ systems evolve through this channel, because of the small initial mass ratios involved and the possibility of the formation of a contact system because of the reaction of the secondary during accretion (cf. region A of the DD channel, Sect.\,\ref{sec:DDchann}).

\paragraph{\textit{Wide systems: regions B$_{\rm H}$ and C$_{\rm H}$.}} 
The most common evolutionary channel (99\%) of the SD$_{\rm H}$ channel is through a CE after which the WD is in a short orbit with an unevolved companion. The systems of region B$_{\rm H}$ follow the WD+MS channel, while the systems of region C$_{\rm H}$ follow the WD+HG channel  or the WD+RG channel.

In region B$_{\rm H}$, systems have initial separations longer than 300\Rsun~and the primary fills its Roche lobe at the end of the HG or the onset of the GB. After the CE phase a helium star forms, which subsequently overflows its Roche lobe stably and evolves into a WD. The secondary fills its Roche lobe during the MS and follows the WD+MS channel. This evolutionary channel is followed by 99\% of the SD$_{\rm H}$ systems, with 21\% starting RLOF at the end of the HG and 78\% during the GB. A star with an initial mass between 5.7 and 8.2\Msun~evolves into a CO WD with a mass between 0.8 and 0.95\Msun~when stripped of its hydrogen envelope at the end of the HG or during the GB. This explains why the greyscale in Fig.\,\ref{HachisuSDH} is limited in the mass range of the donors. Binary systems with initial separations longer than 400\Rsun, after a CE phase, as well as systems with initial separations shorter than 200\Rsun, after a phase of stable mass transfer, form a WD binary system that is too wide to go through the WD+MS channel (Fig.\,\ref{HachisuSDH}).

In region C$_{\rm H}$, systems have initial separations longer than 1000\Rsun~and the primary fills its Roche lobe during the TP-AGB and forms a CO WD immediately. After the CE phase the secondary fills its Roche lobe as an evolved star on the HG or GB. 0.3\% of the SD$_{\rm H}$ systems follow this evolutionary channel. The number of systems going through this evolutionary channel is limited because only systems with CO WDs with initial masses larger than about 1.1\Msun~can grow to $M_{\rm Ch}$ with an evolved donor star. Additionally, the WD+RG channel is almost non-existent as binary systems with a massive WD ($\gtrsim1.0$\Msun) and a low-mass MS star ($\lesssim1.5$\Msun) rarely form with separations between 100 and 300\Rsun~with the assumed CE efficiency and therefore cannot contribute to the WD+RG channel.

Systems with initial separations longer than about 3000\Rsun~do not experience RLOF and therefore do not become a SN~Ia. However, these systems still interact in the form of a stellar wind. Nevertheless, in our standard model wind mass transfer is insufficient for a CO WD to grow to $M_{\rm Ch}$ (Sect.\ref{sec:IntWind}).

\subsubsection{SD with helium-rich donors (SD$_{\rm He}$) \label{Sec::SDHe}}
\begin{figure}
\begin{center}
\includegraphics[width=9.15cm]{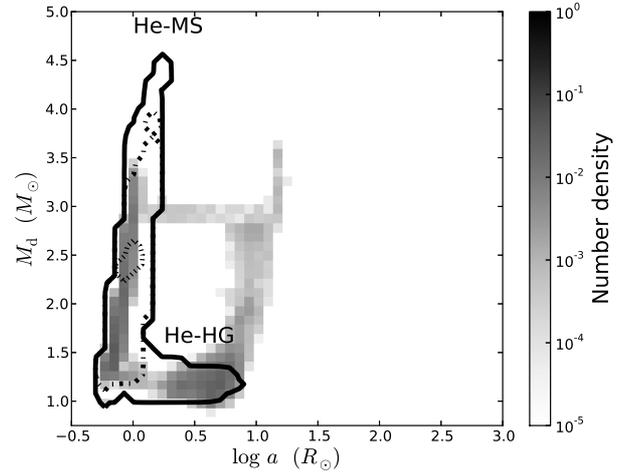}
\caption{As Fig.\ref{HachisuSDH} for systems at formation of the WD and He-MS binary which produce SNe~Ia through the SD$_{\rm He}$ channel.}
\label{HachisuSDHe}
\end{center}
\end{figure}

\begin{figure*}
\begin{center}
\includegraphics[width=9.15cm]{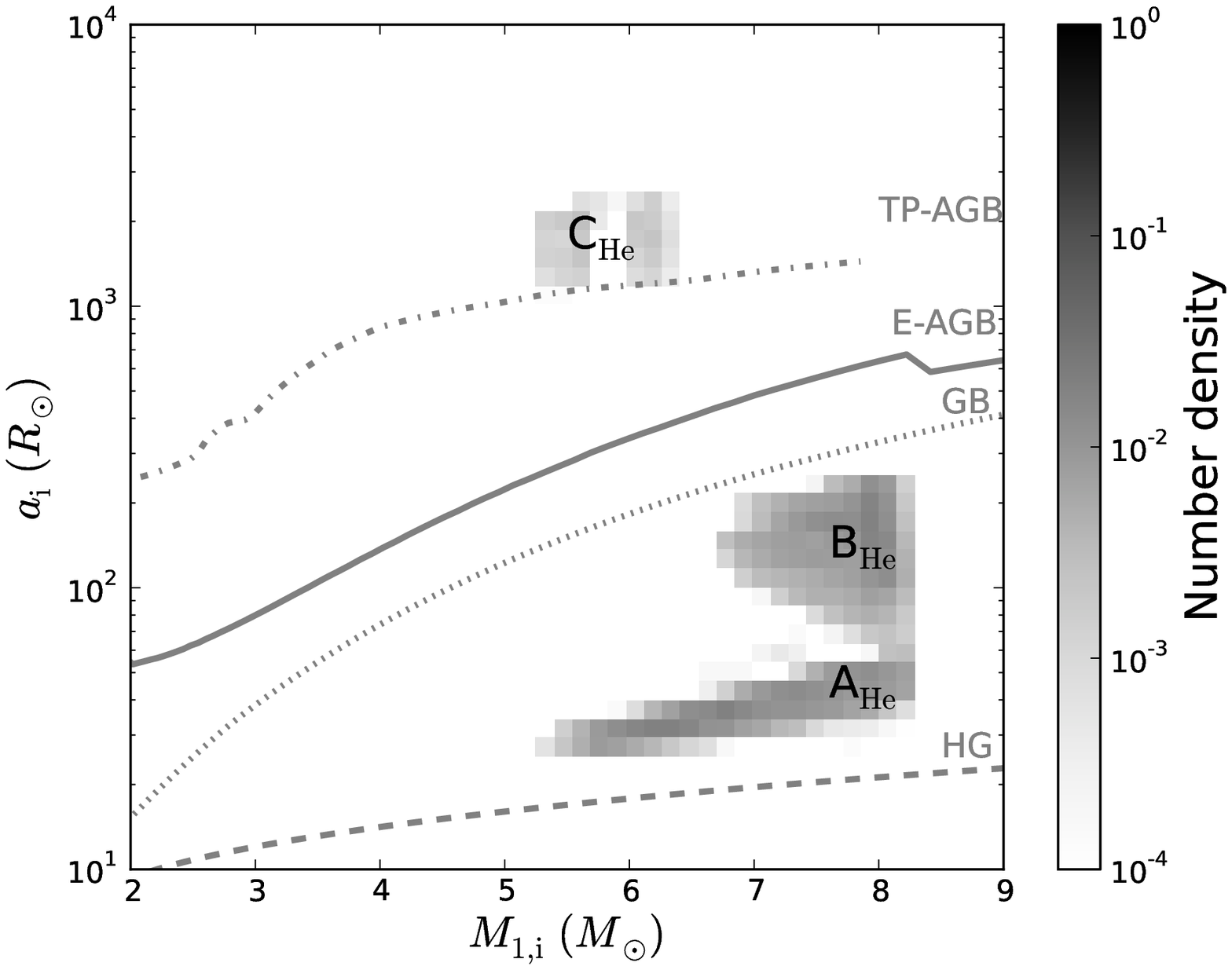}
\includegraphics[width=9.15cm]{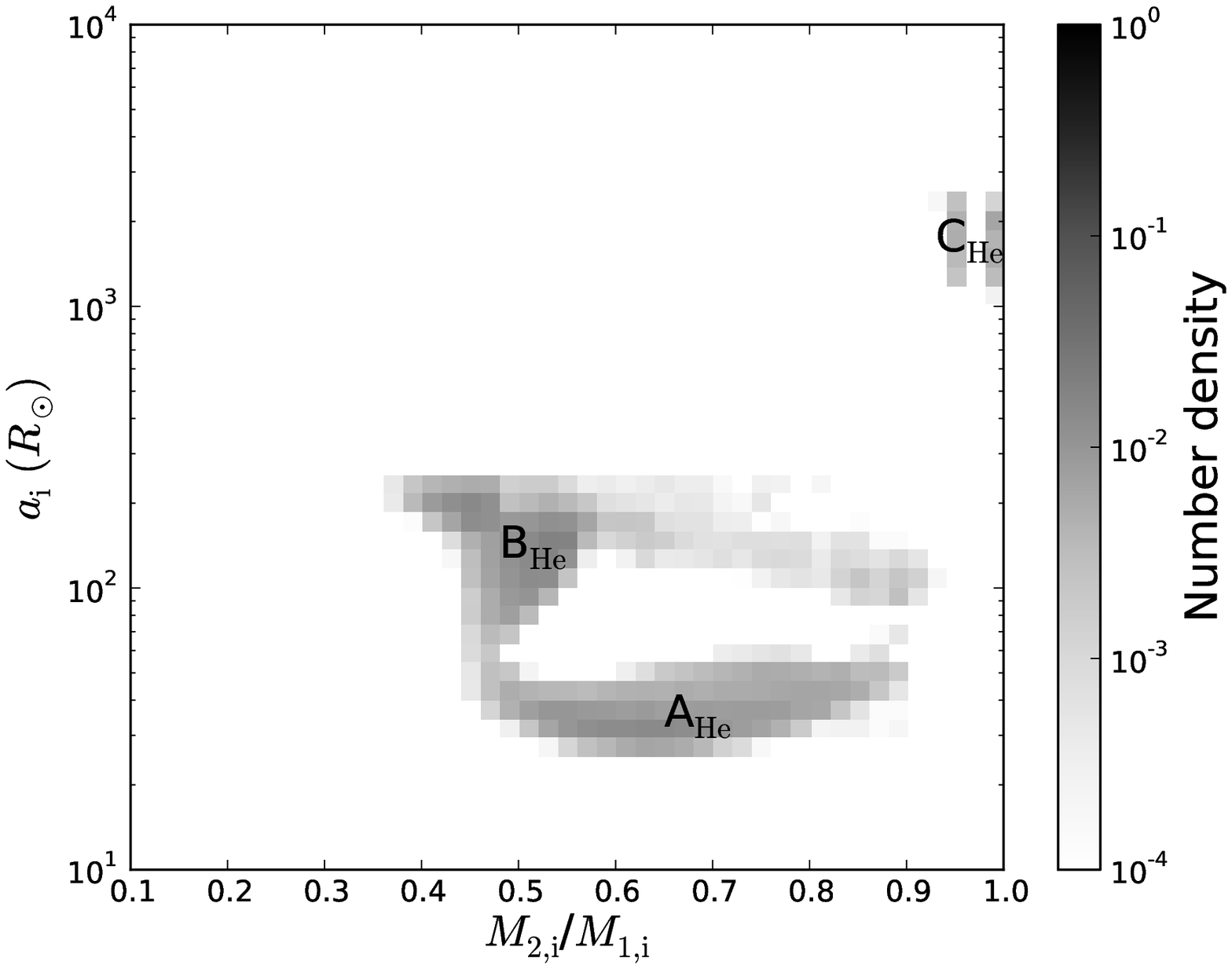}
\caption{As Fig.\,\ref{DDCEa} of systems that form SNe~Ia through the SD$_{\rm He}$ channel. Symbols A$_{\rm He}$, B$_{\rm He}$ and C$_{\rm He}$ indicate groups which are distinguished based on the evolutionary stage when the primary fills its Roche lobe, the number of CE phases necessary to evolve to a SN~Ia and the type of donor star transferring mass to the CO WD: A$_{\rm He}$ = HG, no CE, with He-MS donors; B$_{\rm He}$ = HG, 1 CE, with He-HG donors;  C$_{\rm He}$ = TP-AGB, 1 CE.}
\label{SDHeCEa}
\end{center}
\end{figure*}

Helium rich donors must be massive enough ($M \gtrsim 1$\Msun) to transfer enough material to the WD to make a SN~Ia. Consequently, their H-rich progenitor stars must be massive enough initially (at least 4\Msun). The initial helium star can be non-evolved (He-MS) or evolved (He-HG or He-GB). Fig.\,\ref{HachisuSDHe} shows the results of a simulation similar to that of Sect.\,\ref{Sec::SDH}, with one of the stars at $t=0$ a WD and the other star on the zero-age He-MS. The ranges of both masses, the separation and the resolution are equal to those discussed in Sect.\,\ref{Sec::SDH}.

Fig.\,\ref{HachisuSDHe} depicts the helium donor masses versus the separation of the systems with an initial WD mass $\leq$ 1.15\Msun~which become SNe~Ia. Two regions are distinguished. The left region with $\log a/\Rsun \lesssim 0.2$ and $0.9 \lesssim M_{\rm d}/\Msun \lesssim 4.5$ shows systems that start RLOF to the WD during the He-MS. The middle region with $ 0.0 \lesssim \log a/\Rsun \lesssim 0.85$ and $1.05 \lesssim M_{\rm d}/\Msun \lesssim 1.5$ shows donor stars that start RLOF to the WD during the He-HG. He-GB donors do not form SNe~Ia when the initial WD mass is less than 1.15\Msun, because the evolutionary timescale of these stars is too short to transfer much material to a WD.

He-MS stars more massive than 1.6\Msun~explode as core collapse SNe (CC SNe) if they evolve as single stars \citep{Pols98}. This potentially allows some binary systems to produce \emph{both} a SN~Ia, from the CO WD, and subsequently a core collapse SN from the remaining He star. However, He-MS donors with these masses transfer much material to the WD and have masses lower than 1.6\Msun~when the CO WD reaches $M_{\rm Ch}$. In our model the structure of He-MS adapts during RLOF and the star evolves further as though it started its evolution with this new mass. These initially massive helium stars then become a WD instead of exploding as a CC~SN. However, if the He-MS star does not adapt to its new mass during RLOF, we find that one binary system in our grid produces both a SN~Ia and a CC~SN.

Our Fig.\,\ref{HachisuSDHe} should be compared with Fig.\,8 in \cite{Wang09b}, which depicts the systems evolving through the WD+He-MS and WD+He-HG channel calculated with a detailed binary stellar evolution code. Both figures indicate that helium donors with masses down to about 0.9\Msun~transfer at least 0.25\Msun~to a 1.15\Msun~WD. However, we find that helium stars more massive than 3\Msun~can also transfer this amount stably to a WD, while \cite{Wang09b} find that RLOF is dynamically unstable in these binary systems, because of differences in the stability criteria of RLOF between the two codes. In addition, they find slightly higher donor masses in the WD+He-HG channel, which in our model do not transfer enough material to the WD companion.

The greyscale of Fig.\,\ref{HachisuSDHe} represents the number density of the systems at formation of a WD and helium star binary for the SD$_{\rm He}$ channel. Below we describe the different progenitor evolutionary channels, distinguishing between systems with initially short separations which have He-MS donors (A$_{\rm He}$) or He-HG donors (B$_{\rm He}$) and systems with initially long separations (C$_{\rm He}$). These different groups can be distinguished in Fig.\,\ref{SDHeCEa}.

\paragraph{\textit{Close systems: regions A$_{\rm He}$ and B$_{\rm He}$.}}
Systems with initial separations between 25 and 300\Rsun~have primaries that fill their Roche lobe during the HG when they have radiative envelopes which results in stable mass transfer. After mass transfer a helium star is formed which becomes a WD without interaction or after a phase of stable RLOF. Afterwards a CO WD and a MS star, which has increased in mass, remain. Because of the great difference in mass between the two stars a CE phase follows when the secondary fills its Roche lobe during the HG or beyond, but before the TP-AGB. Subsequently a CO WD and helium star in a short orbit remain. When the secondary fills its Roche lobe for the second time the WD increases in mass and reaches $M_{\rm Ch}$.

The main difference between regions A$_{\rm He}$ and B$_{\rm He}$ is the moment the companion star fills its Roche lobe as a helium star, i.e. during the He-MS at the shortest separations (region A$_{\rm He}$) and during the He-HG at the longest separations (region B$_{\rm He}$).

The initial primary mass and mass ratio of both groups are determined by the need to form a CO WD (upper mass and mass ratio limit) and a massive enough merger product (lower mass and mass ratio limit). The lower primary mass boundary of region A$_{\rm He}$ is lower than of region B$_{\rm He}$, because less massive WDs can grow to $M_{\rm Ch}$ with He-MS donors (Fig.\,\ref{HachisuSDHe}).

Of the SD$ _{\rm He}$ systems, 48\% follow the evolutionary channel of region A$_{\rm He}$. The range of initial separations is limited by the small range of radii of He-MS stars (Fig.\,\ref{HachisuSDHe}). Systems with shorter separations than 25\Rsun~merge during the CE phase before the formation of a helium star, while systems with longer separations than about 55\Rsun~the helium stars fill their Roche lobes after the He-MS. The evolutionary channel of region B$_{\rm He}$ is followed by 48\% of the SD$ _{\rm He}$ systems. Systems with separations shorter than about 55\Rsun~have the helium star filling its Roche lobe during the He-MS, while initially longer separations than 300\Rsun~evolve into a CE phase when the primary fills its Roche lobe.

In our standard model some binary systems form WDs with masses greater than 1.15\Msun\ because the core of the helium star forming the WD grows beyond 1.15\Msun~(see Sect.\,\ref{sec:HeStar}). This results in higher mass donors which can transfer enough material to these massive WDs to reach $M_{\rm Ch}$ than indicated by the solid line of Fig.\,\ref{HachisuSDHe}.  This explains why our model produces He-star donors with masses larger than 1.5\Msun\ at $a \approx 10\Rsun$, outside the solid contour in Fig.\,\ref{HachisuSDHe}.

\paragraph{\textit{Wide systems: region C$_{\rm He}$.}}
SN~Ia progenitor systems with initial separations longer than~1000\Rsun~have an initial mass ratio close to one ($q \gtrsim$ 0.91). The primary fills its Roche lobe during the TP-AGB which results in a CE and a CO WD is formed. Moreover, because the two stars stars have comparable masses, generally the secondary is an evolved star as well at the moment of RLOF. Afterwards, a CO WD and a helium star in a short orbit are formed. However, some systems fill their Roche lobe during the GB shortly after the primary, which results in a second CE phase after which the secondary becomes a helium star. In both cases, the helium star (He-MS or He-HG) fills its Roche lobe afterwards, which increases the mass of the WD to $M_{\rm Ch}$. Only a small range of mass ratios follow this channel because smaller companion masses evolve into too low-mass helium donor stars. This channel accounts for only 4\% of systems in the SD$_{\rm He}$ channel.

\section{Comparison with observations\label{sec:DTD}}
\begin{figure}
\begin{center}
\includegraphics[width=9cm]{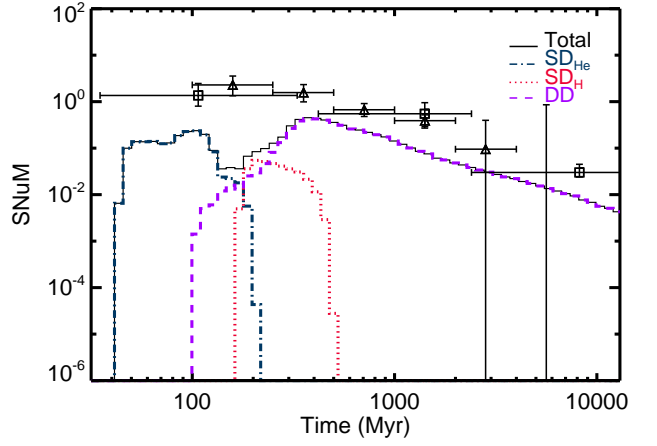}
\caption{Delay time distribution of the different channels with our standard model (Sect.\,\ref{sec:StandAssum}): the SD$_{\rm He}$ channel (blue dash-dotted line), the SD$_{\rm H}$ channel (red dotted line), the DD channel (purple dashed line) and the overall SN~Ia rate (black thin full line), which is the sum of the three channels. The rate is presented in units of SNuM which is the SN~Ia rate per 100 yr per $10^{10}$\Msun~in stars. Data points represent the observed DTD from \citet[triangles]{Totani08} and \citet[squares]{Maoz11}.}
\label{GenDTD}
\end{center}
\end{figure}

\begin{figure}
\begin{center}
\includegraphics[width=9cm]{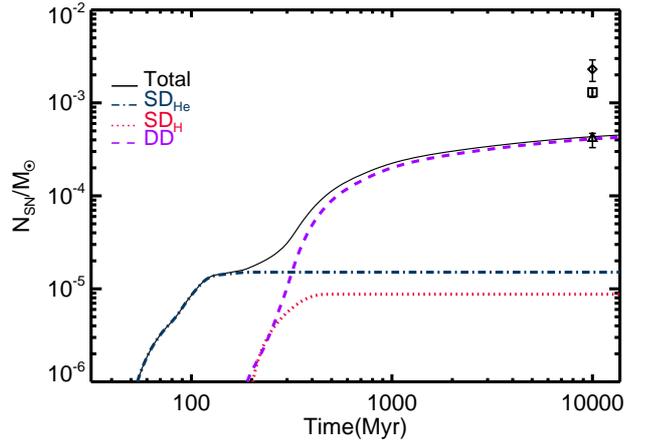}
\caption{Number of SNe~Ia ($N_{\rm SN}$) per unit mass vs. time, with our standard model (Sect.\,\ref{sec:StandAssum}). Line styles have the same meaning as in Fig.\,\ref{GenDTD}. Data points show the integrated rate (over a Hubble time) determined by \citet[][diamond]{Maoz11}, \citet[][square]{Maoz12} and \citet[][triangle]{Graur13}.}
\label{SFDTD}
\end{center}
\end{figure}
In this section we compare the rate of the previously discussed channels and the sum of the three channels, the overall SN~Ia rate, with observations. The delay time distribution (DTD) represents the SN~Ia rate per unit mass of stars formed as a function of time, assuming a starburst at $t=0$. The DTD allows us to investigate the validity of the different progenitor models, by providing a direct comparison with observations.

In early studies the observed DTD has been described by a two-component model \citep{Scannapieco05, Mannucci06}. The first component accounts for the prompt SNe~Ia before 300~Myr, while the second component accounts for the delayed SNe~Ia which have delay times longer than 300~Myr. More recent observations show that the DTD is best described by a continuous power-law function with an index of $-1$ \citep{Totani08,Maoz10,Maoz11,MaozMannucci12,Graur11, Graur13}, more specifically \cite[][]{MaozMannucci12} 
\begin{equation}
\label{eq::DTD}
\text{DTD} (t)  \approx 0.4\,(t/\rm Gyr)^{-1}\, \rm SNuM,
\end{equation}
where SNuM is the supernova rate per 100 yr per 10$^{10}$\Msun. According to \cite{Totani08} this relation supports the DD channel and arises from a combination of the initial separation distribution of the systems ($a_{\rm i}^{-1}$) and the timescale of gravitational radiation ($\tau_{\rm GWR} \propto a^4$). Some groups find a different slope, e.g. \cite{Pritchet08} find a $t^{-0.5 \pm 0.2}$ relation. To compare our models to observed DTDs, we use Eq.\,\ref{Form::CalcDTD}.

We also compare the integrated SN~Ia rate, i.e. the number of SNe~Ia ($N_{\rm SN}$) per unit of stellar mass formed in stars over the history of the Universe. \cite{MaozMannucci12} conclude that the number of SNe~Ia between 35~Myr and the Hubble time, 13.7~Gyr, is consistent with $2 \cdot 10^{-3} \Msun^{-1}$. However, more recent determinations of the observed SN~Ia rate show that the integrated rate may be lower than previously assumed. Different groups find an integrated rate between 0.33$\cdot 10^{-3}$ and 2.9$\cdot 10^{-3} \Msun^{-1}$ \cite[][]{Maoz11, Maoz12, Graur11, Graur13,Perrett12}. \cite{Maoz12} discuss that the divergence may be explained by enhancement of SNe~Ia in galaxy cluster environments at long delay times compared to field environments. 
The lower limit on these observations, however, is found by \cite{Graur13} by applying a $t^{-1}$ relation for the DTD and they did not consider the previously discussed uncertainties of the slope. Because their integrated rate is based on SNe~Ia with long delay times, a steeper power law results in a higher integrated rate. To compare with our models, we use an integrated rate of $2.3 \cdot 10^{-3} \Msun^{-1}$ as found by \cite{Maoz11}. Below we describe the rate of the different channels and the overall SN~Ia rate resulting from our standard model (Figs.\,\ref{GenDTD}\,and\,\ref{SFDTD}).

\subsection{Double degenerate rate \label{sec:DDrate}}

\begin{figure}
\begin{center}
\includegraphics[width=9cm]{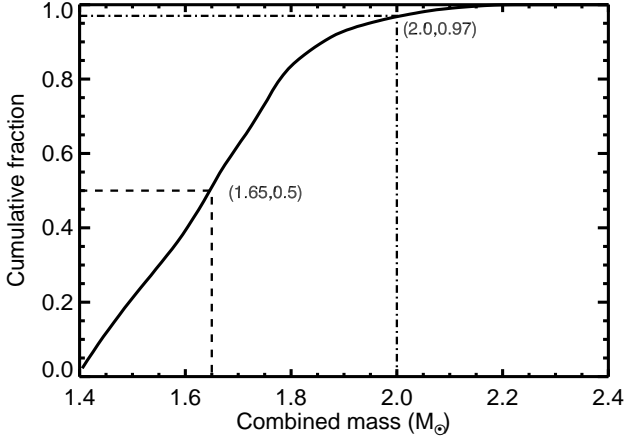}
\caption{Cumulative distribution of the combined mass of the two CO WDs merging within a Hubble time for WD mergers having a combined mass greater than 1.4\Msun~in our standard model. Two lines indicate specific masses and cumulative fractions (Sect.\,\ref{sec:DDrate}).}
\label{DDWDmass}
\end{center}
\end{figure}

In our standard model, the DD channel begins after about 100~Myr and dominates the SN~Ia rate from about 200~Myr up to a Hubble time (Fig.\,\ref{GenDTD}, dashed line). The delay time  of the DD channel can be described by a continuous power-law function from about 400~Myr. However, because the DTD is the combination of different evolutionary channels, it does not exactly follow a $t^{-1}$ relationship, but rather a $t^{-1.3}$ relation. Region B (Fig.\,\ref{DDCEa}), the Roche-lobe overflow channel, contributes from about 100~Myr up to a Hubble time and is always the dominant evolutionary channel. Regions D and E, which form double WD systems after one CE phase followed by a phase of stable RLOF, also contribute with a few percent to the DTD after about 200~Myr up to a Hubble time. The rate following from our standard model for long delay times (averaged between 300~Myr and a Hubble time) is 0.031~SNuM and peaks at about 400~Myr. The integrated rate is $4.3\cdot 10^{-4} \Msun^{-1}$.

Another observable prediction of the DD channel is the mass of the merger product, which can be up to 2.4\Msun. A merger product of 2.4\Msun~can lead to an overluminous SN~Ia if all its mass is burned \cite[e.g.][]{YoonLanger05}, or --if only 1.4\Msun~is burned-- in about 1\Msun~of carbon and oxygen which remains unburned, which can be observed in the spectrum. Fig.\,\ref{DDWDmass} shows that of all double WD systems with a combined mass higher than~$M_{\rm Ch}$~50\% have a combined mass larger than 1.65\Msun. However, only about 3\%, which have a combined mass greater than 2\Msun, would be classified as super-Chandrasekhar because of the current limitations in defining the total WD mass of the progenitor systems leading to observed SNe~Ia \cite[e.g.][]{Howell06}.

\subsection{Single degenerate rate}
The delay time of the SD channel depends on the evolutionary timescale of the secondary, the donor star to the exploding CO WD. The range of initial masses of the different types of donor stars (Sect.\,\ref{sec:SDchann}) is apparent in their respective DTDs. The helium-rich donors are initially the most massive and therefore the resulting SNe~Ia occur earlier than the SNe~Ia formed through the SD$_{\rm H}$ channel (Fig.\,\ref{GenDTD}, dash-dotted line and dotted line).

The SD channel with helium rich donors contributes between about 45~Myr and about 200~Myr. This short time frame arises because only initially massive secondary stars, which have a helium core greater than 1\Msun, can transfer enough material to the CO WD as a helium star. Assuming a starburst the average rate between 40 and 200~Myr is 0.092~SNuM and the integrated rate is $1.5\cdot 10^{-5} \Msun^{-1}$.

In the SD channel with hydrogen rich donors we distinguish between non-evolved and evolved donors. The shortest delay times occur for MS and HG donors, while the more delayed type Ia~SNe originate from GB donors. SNe~Ia formed through the WD+MS channel occur from about 170~Myr until 500~Myr, the WD+HG channel contributes at about 450~Myr. The WD+RG channel contributes from about 4000~Myr, but is not significant and cannot be distinguished in the DTD.

\subsection{Overall SN~Ia rate}
The sum of the three channels results in a DTD best described by a broken power-law, slightly increasing before 100~Myr, a dip between 200 and 400~Myr and $t^{-1.3}$ relation afterwards (Fig.\,\ref{GenDTD}). The dominant formation channel of prompt SNe~Ia is the SD$_{\rm He}$ channel, while the dominating channel of longer delay times is the DD channel. Assuming a starburst, the rate between 40 and 100~Myr is on average 0.14~SNuM and is dominated by the SD$_{\rm He}$ channel. The average rate between 100 and 400~Myr is 0.22~SNuM with approximately equal contributions from the SD and DD channel. At longer delay times, the DD channel dominates (Table\,\ref{rateStarburst} and Fig.\,\ref{GenDTD}). The integrated rate is $4.8\cdot 10^{-4} \Msun^{-1}$, with about 95\% of the SNe~Ia formed through the DD channel, and is approximately a factor five lower than the \cite{Maoz11} rate but compatible with the lowest estimates for the SN~Ia rate \citep[][Fig.\,\ref{SFDTD}]{Graur13}.

Additionally, we find that 2.4\% of intermediate-mass stars with a primary mass between 3 and 8\Msun\ explode as a SN~Ia. This is compatible with the lower limit, expressed as $\eta$, given by \cite{Maoz08} based on several observational estimates, more specifically they find that $\eta \approx 2-40\%$. However, \cite{Maoz08} discusses that $\eta=15\%$ is consistent with all the different observational estimates discussed in the paper, which is about a factor six higher than our results.

The SN~Ia rate versus the rate of core collapse SNe (CC SNe) is another observable prediction. We find that $N_{\rm SNIa}/N_{\rm SNCC} = 0.07 - 0.14$, where the upper limit is determined by assuming that all primaries with a mass between 8 and 25\Msun\ explode as a CC~SN and the lower limit is determined by assuming that both primaries and secondaries with a primary mass between 8 and 25\Msun\ explode as a CC~SN. \cite{Mannucci05} determine this ratio based on star forming galaxies and find $N_{\rm SNIa}/N_{\rm SNCC} = 0.35 \pm 0.08$. \cite{Cappellaro99} estimate a lower ratio of $0.28 \pm 0.07$, as well as \cite{Li11III}, who find that $N_{\rm SNIa}/N_{\rm SNCC} = 0.220 \pm 0.067$ and $0.248 \pm 0.071$, for the prompt and the delayed SN~Ia component, respectively. On the other hand \cite{Plaa07} find a higher ratio of $0.79 \pm 0.15$ based on X-ray observation of the hot gas in clusters of galaxies. Our ratio is between two and four times smaller than the ratio found by \cite{Cappellaro99}, and more than a factor five lower than the ratio estimated by \cite{Plaa07}.

\section{Uncertainties in binary evolution\label{sec:inflPar}}
The theoretical rate and delay time distribution of type Ia~SNe depends on many aspects of binary evolution, such as CE evolution, angular momentum loss and the stability criterion of Roche-lobe overflowing stars, some of which are not well constrained. In order to test the dependence of the progenitor evolution and the rate on the assumptions we performed 35 additional simulations with our BPS code, labelled \#A to \#R, with each letter indicating a variation in a parameter, and numerical subscripts indicating different values for important parameters. We only discuss the most important parameters, while in Table\,\ref{rateSF}\,and\,\ref{rateStarburst} we give an overview of our results. Variations of the initial distributions of binary parameters are discussed in Sect.\,\ref{sec:inflDistr}. The results discussed in this section and the next are calculated with a resolution of $N=100$ (Sect.\ref{Sect::BPS}), except for the simulation of conservative RLOF ($N=150$) and the simulation with $\gamma_{\rm wind}=2$ ($N=125$), in which the ranges of both initial masses and separation are longer than in our standard model. For comparison we list in Tables\,\ref{rateSF}~and~\ref{rateStarburst} the results of our standard model with $N=150$ to show that choosing a higher resolution only has a small effect.

\subsection{Common envelope evolution}
In both the SD and DD channels the prescription of the CE phase is crucial because almost all progenitor systems go through at least one CE phase. This phase is modelled by comparing the binding energy of the envelope and the orbital energy, the $\alpha_{\rm ce}$-prescription, which is parametrized by the parameters $\alpha_{\rm ce}$ and $\lambda_{\rm ce}$ (Sect.\,\ref{sec:CE}). Below we describe the effect of varying both parameters separately.

\subsubsection{Common envelope efficiency\label{sec::varalphace}}
The CE efficiency $\alpha_{\rm ce}$ plays a crucial role in the progenitor evolution. We vary $\alpha_{\rm ce}$ between 0.2 and 10 (models \#A$_1$ to \#A$_4$). Different groups determine a CE efficiency smaller than 1 after fitting the $\alpha_{\rm ce}$-prescription to a population of observed post-CE binaries \cite[][]{Zorotovic10,DeMarco11,Davis12}. \cite{Zorotovic10} found that only a CE efficiency between 0.2 and 0.3 reproduces their entire observed sample. A CE efficiency of 10 is an extreme assumption to demonstrate the effect of an efficient extra energy source during the CE phase.

\paragraph{\textit{SD-channel}.}

\begin{figure}
\begin{center}
\includegraphics[width=9cm]{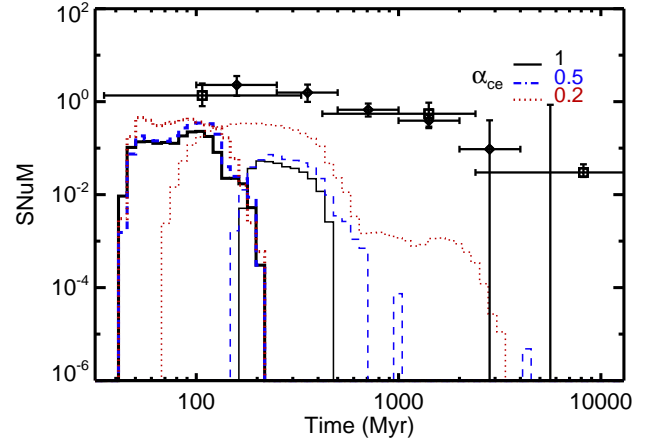}
\caption{DTD of the SD-channel for the assumptions discussed in Sect.\,\ref{sec:StandAssum} and with variable CE efficiency, $\alpha_{\rm ce}$, which is 1 (black full line), 0.5 (blue dashed line) and 0.2 (red dotted line). The thick lines show the SD$_{\rm He}$ channel, while the thin lines show SD$_{\rm H}$ channel. Data points have the same meaning as in Fig.\,\ref{GenDTD}.}
\label{CompSD}
\end{center}
\end{figure}

In our models with lower common-envelope efficiencies than our standard model (models \#A$_1$ and \#A$_2$, with $\alpha_{\rm ce}=0.2$ and 0.5, respectively), systems with initially longer separations than in our standard model survive a CE phase and contribute to the SD$_{\rm H}$ channel, more specifically to the WD+MS channel and the WD+HG channel. Stars which are stripped of their envelopes at a later stage of their evolution have more massive cores and form more massive WDs. Consequently, generally the mass range of accreting WDs increases with decreasing CE efficiency. More specifically, in the WD+MS channel for $\alpha_{\rm ce}=1$ the initial mass of the CO WD ranges between 0.8 and 0.95\Msun, while for $\alpha_{\rm ce}=0.2$ (model \#A$_1$) the initial mass of the CO WD ranges between 0.8 and 1.15\Msun, with an associated increase of the integrated rate (Fig.\,\ref{HachisuSDH} and Table\,\ref{rateSF}), from $0.9\cdot 10^{-5}\Msun^{-1}$ when $\alpha_{\rm ce}$ equals 1, to $9.0\cdot 10^{-5}\Msun^{-1}$ when $\alpha_{\rm ce}$ equals 0.2.

Additionally, the DTD changes significantly (Fig.\,\ref{CompSD} and Table\,\ref{rateStarburst}). As the mass range of the donor stars of the WD+MS and WD+HG channels is enlarged, the rate of the SD channel increases and contributes for a longer time. The rate of the WD+RG channel does not increase in models \#A$_1$ and \#A$_2$, because binary systems formed after RLOF with massive CO WDs ($M_{\rm WD}>1.0$\Msun) and low mass MS stars ($M<2.0$\Msun) have separations shorter than 30\Rsun~and therefore do not contribute to the WD+RG channel (Fig.\,\ref{HachisuSDH}).

In the case of higher CE efficiencies we find the opposite (models \#A$_3$ and \#A$_4$): the WD+MS and WD+HG channels decrease and the WD+RG channel increases (Table\,\ref{rateSF}). In model \#A$_4$ ($\alpha_{\rm ce}=10$) the DTD of the SD$_{\rm H}$ channel mainly consists of a delayed component from SNe~Ia formed through the WD+RG channel. Likewise the integrated rate of the SD$_{\rm He}$ channel increases in models \#A$_1$ and \#A$_2$, but only by a factor two to three, because of the already large mass range of CO WDs formed when $\alpha_{\rm ce}$ equals 1.

\paragraph{\textit{DD-channel.}}

\begin{figure}
\begin{center}
\includegraphics[width=9cm]{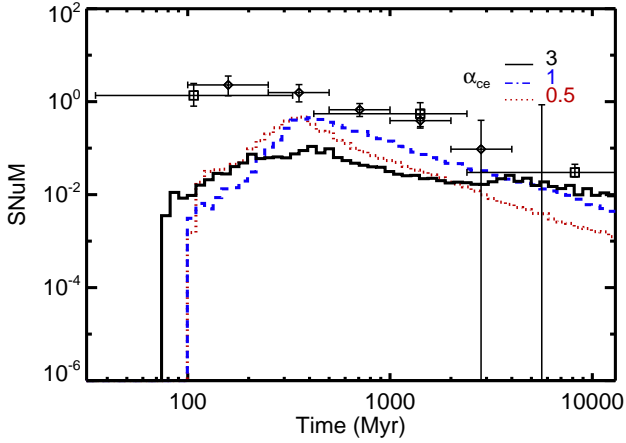}
\caption{DTD of the DD-channel for our standard model (Sect.\,\ref{sec:StandAssum}) and with variable CE efficiency, $\alpha_{\rm ce}$, which is 3 (black full line), 1 (blue dashed line) and 0.5 (red dotted line). Data points have the same meaning as in Fig.\,\ref{GenDTD}.}
\label{CompDD}
\end{center}
\end{figure}

\begin{figure}
\begin{center}
\includegraphics[width=9cm]{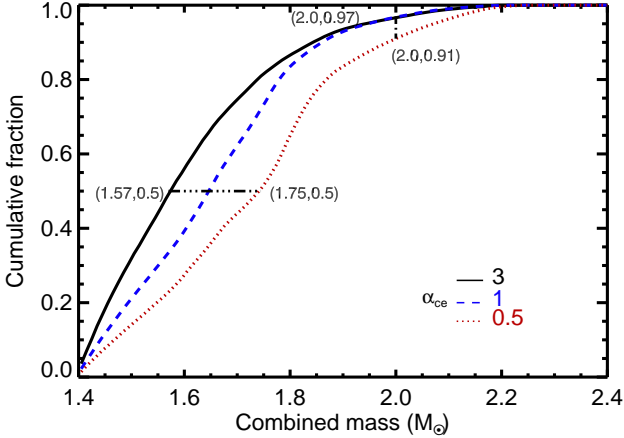}
\caption{Cumulative distribution of the combined mass of the two CO WDs merging within a Hubble time, for WD mergers having a combined mass higher than 1.4\Msun, assuming different CE efficiencies. Different lines have same meaning as Fig.\,\ref{CompDD}. Vertical and horizontal lines indicate specific masses and cumulative fractions (Sect.\,\ref{sec::varalphace}).}
\label{CompDDWDmass}
\end{center}
\end{figure}

Two effects play a role when changing the CE efficiency (Fig.\,\ref{CompDD} and Tables\,\ref{rateSF}~and~\ref{rateStarburst}). The first effect is the selective evolution towards a formation reversal (Table\,\ref{rateSF}, region B$_2$). In the formation reversal channel, the CE phase preceding the formation of the second helium star brings two helium stars together in a short orbit. If $\alpha_{\rm ce}=1$ the more massive helium star fills its Roche lobe when it is an evolved helium star. If the separation is too long after the CE phase and the helium star does not fill its Roche lobe, the subsequently formed WDs do not merge within a Hubble time. At small $\alpha_{\rm ce}$ the more massive helium star fills its Roche lobe at an earlier evolutionary stage, during the He-MS, and the double WD system does not form a massive enough merger product. This evolutionary channel completely disappears at low CE efficiency (e.g. $\alpha_{\rm ce}=0.2$, model \#A$_1$) and almost completely for high CE efficiencies (e.g. $\alpha_{\rm ce}=10$, model \#A$_4$). It appears that $\alpha_{\rm ce}=1$ is an optimal value for this formation path (Table\,\ref{rateSF}).

The second effect is that the lower and upper separation boundaries for the regions B to F would change to longer separations in the models with lower CE efficiencies compared to our standard model. However, systems with longer separations are less common and in some regions the movement of the upper boundary is limited, such as in region F. In region F, the initially widest binary systems in which both stars fill their Roche lobe already form SNe~Ia  when $\alpha_{\rm ce}=1$. These two implications decrease the DD rate assuming lower CE efficiencies (Table\,\ref{rateSF}).

On the other hand, if the CE efficiency is very high (e.g. $\alpha_{\rm ce}=10$, Table\,\ref{rateSF}), the orbital energy does not decrease much during a CE phase and fewer double WD systems evolve into a short enough orbit to merge within a Hubble time. Consequently, the rate of the DD channel decreases for very high CE efficiencies. Regions C to E almost disappear when $\alpha_{\rm ce}=10$. The rate of region F increases when $\alpha_{\rm ce}=3$ and 10, but these SNe~Ia originate from binary systems which have the first CE phase when the primary is on the E-AGB or GB, with initially shorter separations compared to our standard model.

The DTD of the DD channel is a combination of different evolutionary paths. Fig.\,\ref{CompDD} shows that the slope of the DTD changes with variation of the CE efficiency. When $\alpha_{\rm ce}=1$ the DTD can be approximated by a $t^{-1.3}$ power-law, however the slope flattens when $\alpha_{\rm ce}$ increases ($t^{-0.5}$ when $\alpha_{\rm ce}=3$) or steepens when $\alpha_{\rm ce}$ decreases ($t^{-1.5}$ when $\alpha_{\rm ce}=0.5$).

Moreover, when the CE efficiency decreases the double WD systems that merge originate from initially wider systems. Consequently, the resulting WDs are more massive and therefore also the merger product (Fig.\,\ref{CompDDWDmass}). In our standard model, 50\% of the merging DD systems have a mass lower than 1.65\Msun, while this is 1.57\Msun~in model \#A$_3$ with $\alpha_{\rm ce}=3$. Additionally, in our standard model only 3\% of the merger products have a total mass higher than 2\Msun, which is 9\% when $\alpha_{\rm ce}=0.5$ (model \#A$_2$). The dominance of region B is less strong in the most extreme models \#A$_1$ and \#A$_4$ ($\alpha_{\rm ce} = 0.2$ and 10), therefore these behave differently than the models discussed above, however the 50\% and 2\Msun~boundaries for both models are within the ranges discussed above.

\paragraph{\textit{Overall SN~Ia rate.}} 
Decreasing the CE efficiency increases the rate of both SD channels and decreases the rate of the DD channel. The DD channel peaks when $\alpha_{\rm ce}=1$. Generally, the DD channel is the dominant formation channel, but the importance of the SD channel increases at lower CE efficiencies. The theoretical integrated SN~Ia rate varies between $9\cdot 10^{-5} \Msun^{-1}$ ($\alpha_{\rm ce}=10$) and $45\cdot 10^{-5} \Msun^{-1}$ ($\alpha_{\rm ce}=1$), which is a factor of 5 to 26 times lower than the \cite{Maoz11} rate.

\subsubsection{Mass distribution of the envelope}
We investigate the influence of $\lambda_{\rm ce}$, which describes the mass distribution of the envelope, on the SN~Ia rate. We also consider a specific extra energy source which is expressed by $\lambda_{\rm ion}$. In our standard model $\lambda_{\rm ce}$ is a function of the type of star and its evolutionary state (Appendix\,\ref{app::lambda}). As this prescription is not available in all BPS codes, $\lambda_{\rm ce}$ is often taken to be 1 in other BPS studies.

Compared to the results of our standard model, the results of model \#B$_1$ with $\lambda_{\rm ce}=1$ show an increase of the integrated rate by only 3\% and small changes in the different channels (Table\,\ref{rateSF}). However, Table\,\ref{rateStarburst} shows that more SNe~Ia occur at shorter delay times compared to the DTD from our standard model. If we assume $\lambda_{\rm ce}$ equals 1 for all stars, the envelopes of HG and helium stars are less strongly bound than in our standard model, while those of stars on the GB and AGB are more strongly bound. 
This assumption leaves the systems interacting during the GB or AGB in a shorter orbit after the CE phase compared to our standard model. In double WD systems formed through the dominant evolutionary channel, corresponding to region B, the secondary generally fills its Roche lobe for the first time during the GB or beyond. Consequently, the double WD systems formed through this evolutionary channel merge at relatively shorter delay times compared to our standard model.

During the CE phase, extra energy sources may facilitate the loss of the envelope. One example is the ionization energy which is modelled with $\lambda_{\rm ion}$. In model \#B$_2$ we assume that 50\% of the ionization energy is used to eject the envelope during the CE phase.  This has the largest effect on the most massive progenitor systems and the systems which interact during the AGB. In general, the rate of the progenitor channels in which one star interacts during the AGB increases because AGB stars lose their envelope more easily during the CE phase. Consequently, SNe~Ia formed through these progenitor channels originate from systems with shorter separations than in our standard model (e.g. region F of the DD channel and regions C$_{\rm H}$ and C$_{\rm He}$ of both SD channels, Table\,\ref{rateSF}). Even though a variation of $\lambda_{\rm ion}$ has a large effect on individual systems, the SN~Ia rate increases by only 5\% compared to our standard model.

\subsection{Stability criterion of Roche-lobe overflowing stars}

\subsubsection{Helium stars\label{sec::varqhe}}

\begin{figure}
\begin{center}
\includegraphics[width=9cm]{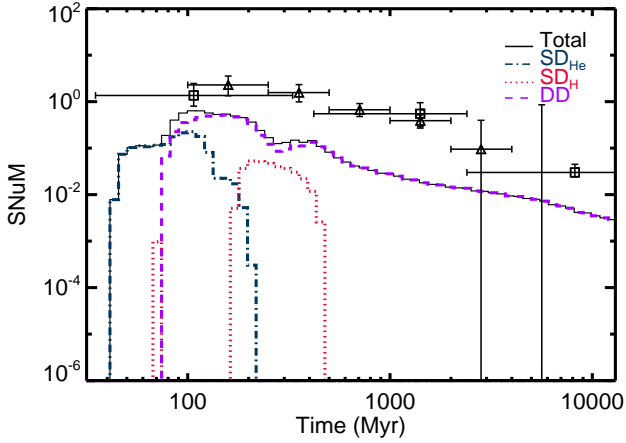}
\caption{Delay time distribution of the different channels with model \#C which assumes $Q_{\rm crit,He-HG}$ =  $Q_{\rm crit,He-GB}$. 
Line styles and data points have the same meaning as in Fig.\,\ref{GenDTD}.}
\label{DTD_OS}
\end{center}
\end{figure}

The evolution of helium stars is ill-constrained and therefore there remain uncertainties in this evolutionary phase. We updated the stability criterion of Roche-lobe overflowing helium stars in the HG. In our standard model the stability criterion of He-HG stars is the same as the criterion of hydrogen-rich stars in the HG, which is expected as both types of stars have a similar structure of their outer layers and therefore react similarly to mass loss. In older results published with our code the stability criterion of He-HG stars is equal to the criterion of  He-GB stars (model \#C, Table\,\ref{TableQ}). With this assumption stable RLOF from helium stars in the HG is less likely. As many results have been published with the latter assumption, we compare the effect of a variation of the stability criterion of Roche-lobe overflowing helium stars.

One might expect that a different prescription for Roche-lobe overflowing helium stars mainly affects the SD$_{\rm He}$ channel, more specifically the progenitors with He-HG donors, i.e.\,region B$_{\rm He}$. Table\,\ref{rateSF} shows that in model \#C region B$_{\rm He}$ is altered and the integrated rate of the SD$_{\rm He}$ channel decreases from 1.50 $\cdot 10^{-5} \, M_{\odot}^{-1}$ in our standard model to 1.40 $\cdot 10^{-5} \, M_{\odot}^{-1}$ in model \#C. In our standard model only a small number of WD and helium star systems with a mass ratio smaller than 0.87 form a SN~Ia through the He-HG + WD channel because of the high mass transfer rates involved, which explains the small difference in the rate of the SD$_{\rm He}$ channel (Fig.\,\ref{HachisuSDHe}).

The helium star stability criterion mainly affects the DD channel, more specifically region B$_2$ (the formation reversal channel, Table\,\ref{rateSF}). This is related to the crucial role of the stability criterion of helium stars in the formation reversal channel because these systems go through a phase where both stars are helium stars simultaneously. In this evolutionary channel, the difference between the stability criteria determines if a double WD system is formed in a short orbit or the binary system merges during a CE phase. No SNe~Ia are formed through the formation reversal channel in model \#C and the integrated rate of region B decreases to $1.4\cdot 10^{-4} \, M_{\odot}^{-1}$, compared to  $3.6\cdot 10^{-4} \, M_{\odot}^{-1}$ with our standard model.

Another difference between the results of our standard model and model \#C is the DTD (Figs.\,\ref{GenDTD}\,and\,\ref{DTD_OS}~and Table\,\ref{rateStarburst}). The DTD of systems at short delay times changes substantially. The average rate at delay times between 100 and 300~Myr is 0.052~SNuM with our standard model, while it is 0.31 SNuM with model \#C (Table\,\ref{rateStarburst}). Our results suggest a $t^{-1.3}$ relationship from about 400~Myr with the former model, and a $t^{-1.1}$ relationship from about 115~Myr with the latter. These differences originate from double WD systems that are formed after an evolved helium star has interacted with the first formed WD. In our standard model, when the helium star fills its Roche lobe mass transfer is stable. In model \#C, RLOF is unstable for some of these systems. In both cases a double WD system forms but systems that result from a CE phase are in a shorter orbit afterwards than those that result from a phase of stable RLOF and therefore merge within a shorter time.

In conclusion, a different stability criterion of Roche-lobe overflowing helium stars affects: 1) the DTD, more specifically the shorter delay times and 2) the integrated rate by a factor two. Model \#C has an integrated SN~Ia rate of $2.3\cdot 10^{-4} \Msun^{-1}$ (Table\,\ref{rateStarburst}), which is a factor ten lower than the observed \cite{Maoz11} rate.

\subsubsection{Stars in the Hertzsprung gap}

A non-degenerate accreting star can undergo thermal expansion when it is brought out of thermal equilibrium. The binary can then evolve into a contact system, which possibly leads to a CE and/or merger. This process is more likely to occur in unequal mass binary systems \cite[]{DeMink07}. However, the details of this process depend on the accretion efficiency and how energy is transported in the accreting star, both of which are uncertain. Because in our model the formulae used to determine the evolution of a star are based on stars in thermal equilibrium, we do not take thermal expansion into account. It is expected that the effect is equivalent to increasing the critical mass ratio $Q_{\rm crit}$ for donor stars in the HG (Table\,\ref{TableQ}).

To estimate the effect of this uncertainty on the rate and the DTD, we consider model \#D with a different stability criterion of donor stars in the HG and He-HG with a non-degenerate accretor, $Q_{\rm crit,HG}=0.5$, compared to our standard model ($Q_{\rm crit,HG}=0.25$). This implies that stable RLOF of HG donors is less likely to occur, and therefore influences regions A, B, A$_{\rm H}$, B$_{\rm H}$, A$_{\rm He}$ and B$_{\rm He}$  (Table\,\ref{rateSF}). Because the systems that start RLOF during the HG have short initial separations and more binary systems evolve into a CE in model \#D compared to our standard model, fewer systems survive the first RLOF phase and form a SN~Ia. In model \#D the integrated rate decreases by 23\% compared to our standard model (Table\,\ref{rateStarburst}).

\subsection{Accretion efficiency}

Apart from the stability criterion for RLOF, one of the main uncertainties in binary evolution is the RLOF accretion efficiency $\beta$. The RLOF accretion efficiency depends on the reaction of the accretor to mass accretion and on the angular momentum of the accreted material, which determines how fast the accretor spins up in reaction to accretion. It is uncertain how efficiently the star can lose the gained angular momentum. In our standard model the accretion efficiency is a function of the thermal timescale of the accretor (Eq.\,\ref{MaxMaccr}) in which the uncertainty is expressed by a parameter $\sigma$ which equals 10 in our standard model.

In models \#E$_1$ and \#E$_2$ we assume $\sigma$ is 1 and 1000, respectively. A variation of the accretion efficiency of non-degenerate accretors has an impact on regions A to B of the DD-channel and A$_{\rm H}$, B$_{\rm H}$, A$_{\rm He}$ and B$_{\rm He}$ of the SD channels. In our standard model, mass transfer during RLOF is approximately conservative for high mass ratios because there is only a small difference between the thermal timescales of the two stars. In model \#E$_2$ RLOF is approximately conservative for all mass ratios, while in model \#E$_1$ RLOF is non-conservative over the entire mass ratio range. Consequently, a variation of $\sigma$ mainly affects binary systems with unequal masses. A higher accretion efficiency increases the mass of the initially lowest mass companions after RLOF, which results in a higher SN~Ia rate, and vice-versa. However, region A of the DD channel disappears in model \#E$_2$ because the two WDs are in too long an orbit after conservative mass transfer to merge within a Hubble time. Compared to our standard model, the total integrated rate decreases by 12\% assuming that the accretor can only accept accreted material with mass transfer rates lower than its thermal timescale (model \#E$_1$, Table\,\ref{rateSF}) and it increases by 14\% assuming RLOF is approximately conservative (model \#E$_2$, Table\,\ref{rateSF}).

\subsubsection{Unlimited accretion onto WDs}

\begin{figure}
\begin{center}
\includegraphics[width=9cm]{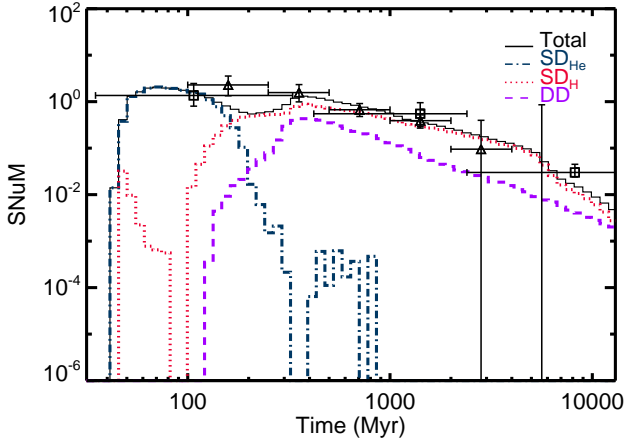}
\caption{Delay time distribution of the SN~Ia channels of model \#F$_1$ with conservative mass transfer only to the WD. The line styles and data points have the same meaning as in Fig.\,\ref{GenDTD}.}
\label{ConWDDTD}
\end{center}
\end{figure}

The models that are used to determine the WD retention efficiency assume a WD that has cooled for $10^8$~yrs \cite[Appendix\,\ref{app::efficiencyWD} and][]{Nomoto82}. These models do not consider WDs with different temperatures, e.g. because of previous nova outbursts, which can alter the accretion efficiency. For this and other reasons, the retention efficiency of WDs is uncertain and several models exist which describe this efficiency \cite[][and reference therein]{Bours13}. Rather than investigating the effect of different retention efficiencies, as was done by \cite{Bours13}, we consider an extreme possibility where all mass transferred to the WD by RLOF remains on the WD and is burnt into carbon and oxygen (model \#F$_1$). We do not consider such unlimited accretion onto a WD as a realistic model, but it represents an upper limit to the SD channel. In this model the critical mass ratio $Q_{\rm crit}$ for stable RLOF onto WDs is the same as onto non-degenerate stars (Table\,\ref{TableQ}). During CE evolution we assume no mass is accreted. During phases of wind accretion, the amount accreted is calculated according to the Bondi-Hoyle prescription (Eq.\,\ref{eq::BH}) for both degenerate and non-degenerate stars.

\paragraph{\textit{DD channel.}}
Unlimited accretion onto a WD does not significantly affect the rate of this channel, which is not determined by the amount of material which is accreted but by how close together two WDs are formed. However, conservative RLOF results in changes in the initial parameter space compared to our non-conservative model (Table\,\ref{rateSF}). The systems which undergo two CE phases (region F) are unaffected. The systems forming regions C to E in our standard model --which have stable mass transfer when the secondary fills its Roche lobe-- disappear in model \#F$_1$. In the latter model, after the primary fills its Roche lobe during the GB or E-AGB, either RLOF is unstable when the secondary fills its Roche lobe and the binary evolves into a CE and merges, or RLOF is stable and everything is accreted by the WD which explodes as a SN~Ia through the SD$_{\rm H}$ channel. The DTD of the DD channel does not alter significantly, however the overall rate decreases compared to our standard model by about 10\% (Table\,\ref{rateSF}~and~Fig.\,\ref{GenDTD}~and~\ref{ConWDDTD}).

\paragraph{\textit{SD channel.}}
For both SD channels large changes are expected because of the prominent role of the accretion efficiency in these channels. Figs.\,\ref{HachisuSDH}~and~\ref{HachisuSDHe} change significantly, because lower masses can provide enough mass to the WD through conservative mass transfer. This is partly counteracted by the fact that stable RLOF is only possible for somewhat lower donor masses.

Assuming conservative mass transfer towards the WD decreases the lower limit of donor masses in the SD$_{\rm H}$ channel to 0.45\Msun~(Fig.\,\ref{HachisuSDH}) and the minimum mass of WDs at formation to 0.3\Msun.  Consequently, the rate increases significantly in both models with conservative mass transfer. The SD$_{\rm H}$ channel is dominant for delay times longer than 200~Myr (Fig.\,\ref{ConWDDTD} and Table\,\ref{rateStarburst}). The integrated rate is about three times larger than the rate of the DD channel with our standard model (Table\,\ref{rateSF}).

In the SD$_{\rm He}$ channel the minimum He-star donor mass decreases to 0.7\Msun, significantly lower than without conservative mass transfer (Fig.\,\ref{HachisuSDHe}). The SD$_{\rm He}$ channel still starts contributing to the DTD from about 45~Myr and continues until about 300~Myr in model \#F$_1$ (Fig.\,\ref{ConWDDTD} and Table\,\ref{rateStarburst}). It remains the dominant channel for short delay times and increases by a factor 10 compared to our standard model (Table\,\ref{rateSF}).

\paragraph{\textit{Overall SN~Ia rate.}}
The DTD is completely dominated by the SD channel. Prompt SNe~Ia are mainly produced through the SD$_{\rm He}$ channel, while delayed SNe~Ia are mainly produced through the SD$_{\rm H}$ channel. The delayed component approximately follows a $t^{-1}$ relation from 300~Myr, however the rate drops slightly from about 5~Gyr. In the models with unlimited accretion onto the WD the observed DTD is well reproduced. The theoretical DTD has a rate averaged between 0 and 13.7~Gyr of 0.14~SNuM in model \#F$_1$ (Figs.\,\ref{ConWDDTD}). The resulting integrated rate is compatible with the \cite{Maoz11} rate (Table\,\ref{rateSF}).

\subsection{Angular momentum loss}
When material is lost from a binary system it also removes angular momentum. Several prescriptions for angular momentum loss exist and influence the evolution of a binary system in different ways, depending on whether mass is lost through a stellar wind or RLOF. Below we discuss the effect of the uncertainties in angular momentum loss on both types of mass loss from a binary system (models \#G, \#H$_1$ and \#H$_2$).

\subsubsection{Stellar wind}
When mass is lost in a stellar wind, it is often assumed that this material does not interact with the binary system and that it is lost in a spherically symmetric way. However, this assumption only holds when winds are very fast compared to the orbital velocity of the system. AGB stars have slow winds \cite[speed\,$\approx$\,10-15 km$\rm s^{-1}$, see e.g.][]{VassiliadisWood93} and when they are in relatively close orbits the wind can interact with the orbit and remove specific orbital angular momentum from the binary system \citep[][]{Jahanara05, Izzard10}.

In model \#G we assume that the material lost through a stellar wind carries twice the specific orbital angular momentum of the binary system. Some systems which do not interact in our standard model do interact in model \#G, which increases the rate of regions F, C$_{\rm H}$ and C$_{\rm He}$ (Table\,\ref{rateSF}). Additionally, even though the DD channel remains dominant, the SD$_{\rm H}$ channel, more specifically the WD+RG path, is more important at longer delay times (Table\,\ref{rateStarburst}). Because angular momentum loss through a stellar wind does not affect the most common regions, the integrated rate of the two models differs by only 15\%.

\subsubsection{Roche-lobe overflow}
Variation of the prescription of angular momentum loss when material is removed during a phase of stable RLOF affects the channels which have stable RLOF that is far from conservative. Our model of accreting WDs that drive an optically thick wind requires that the material lost removes the specific orbital angular momentum of the accreting star (Sect.\,\ref{sec:RLOFWD}) and therefore we only vary the prescription of angular momentum loss during RLOF for other types of accretor.

A different assumption for angular momentum loss during RLOF affects regions A and B of the DD channel and A$_{\rm H}$, B$_{\rm H}$, A$_{\rm He}$ and B$_{\rm He}$ of the SD channels. In general, when more angular momentum is lost compared to our standard model initially wider systems become a SN~Ia after a phase of stable RLOF, and vice versa. In model \#H$_1$ mass lost takes the specific orbital angular momentum of the donor star, while in model \#H$_2$ it is twice the specific orbital angular momentum of the binary. Although angular momentum loss during RLOF influences the most common evolutionary channels, it does not alter the rate drastically (Table\,\ref{rateSF}).

\subsection{Wind prescription}
In our standard model, wind mass loss from stars up to the E-AGB is described by \citet[][Eq.\,\ref{Reimers}, with $\eta=0.5$]{Reimers75} and by \cite{VassiliadisWood93} for the TP-AGB. Mass loss from helium stars is described by \cite{Reimers75} or \cite{Hamann98} depending on which of the two is stronger (Eq.\ref{eq::Mhestar}). Although the general trend of the evolution of the wind is known, the rate is not well constrained \cite[]{Wachter02}. Below we discuss the effect of this uncertainty on the SN~Ia rate.

\subsubsection{Stars on the E-AGB and helium stars}
In model \#I the strength of the wind of stars up to the E-AGB and helium stars is increased ($\eta=5$, Eq.\,\ref{Reimers}) compared to our standard model ($\eta=0.5$). This affects all the evolutionary channels, even those which show interaction on the TP-AGB because the prior evolution is altered. The DD channel is mainly affected in regions D to F because these binary systems have the strongest winds. Region F increases significantly, because more systems survive two CE phases as more material is lost before RLOF sets in. The SD$_{\rm H}$ channel is mainly affected in region C$_{\rm H}$ and the SD$_{\rm He}$ channel in regions B$_{\rm He}$ and C$_{\rm He}$. The rate of both channels decreases because less material is accreted than through stable RLOF. In conclusion, the rate of the SD channel decreases in model \#I compared to our standard model. The opposite is true for the DD channel. The integrated SN~Ia rate changes by only 6\%.

\subsubsection{TP-AGB}
Several alternative prescriptions for wind mass loss during the TP-AGB phase are used to describe this evolution phase. Changing the wind prescription of stars on the TP-AGB only affects regions E and F of the DD channel and regions C$_{\rm H}$ and C$_{\rm He}$ of the SD channels (Table\,\ref{rateStarburst}, models \#J$_1$ to \#J$_4$). The high mass loss rate of \citet[][model \#J$_3$]{Bloecker95} results in the shortest TP-AGB phase and forms the lowest mass WDs,  which accordingly results in the lowest SN~Ia rate of all the prescriptions for the TP-AGB under considerations. The prescription of \citet[][with $\eta=1$ during the TP-AGB, model \#J$_2$]{Reimers75} describes the longest wind phase, which results in the highest SN~Ia rate. Because a change in the wind prescription of stars on the TP-AGB does not affect the most common evolution paths, a variation of it only changes the integrated rate up to 5\%.

\subsection{The combined effect of different binary parameters\label{sec::varextr1}}
In the above sections we tested the influence on the SN~Ia rate of different binary evolution aspects separately. However, we do not necessarily expect that the effect of varying two parameters simultaneously is the same as the sum of the effects of varying each parameter separately. Therefore we change some parameters under study at the same time and investigate their combined effect. We combine those parameters that have the largest influence on the SN~Ia rate, to determine the range of variation caused by uncertainties in binary evolution.

\subsubsection{Common envelope efficiency and stability criterion of RLOF}
Model \#P  combines  a low CE efficiency ($\alpha_{\rm ce} = 0.2$) and a different stability criterion of stars on the He-HG ($Q_{\rm crit,He-HG} = Q_{\rm crit,He-GB}$). Both assumptions separately decrease the rate (models \#A$_1$ and \#C). When combined, the rate also decreases, but not by as much as the sum of the two effects separately. This is because model \#A$_1$ and \#C both affect region B$_2$ of the DD channel, moreover in both models this region disappears. In model \#P, as in model \#A$_1$, the SD channel dominates. The DTD is similar to that of model \#A$_1$, except for an additional decrease of the DD channel because of the extra change of the stability criterion of Roche-lobe overflowing helium stars. The integrated rate decreases by about 65\% compared to our standard model.

\subsubsection{Accretion efficiency and angular momentum loss during wind mass transfer}
Model \#Q combines a high Bondi-Hoyle accretion efficiency ($\alpha_{\rm BH}=5$, Eq.\,\ref{eq::BH}) and a high angular momentum loss during wind mass loss ($\gamma_{\rm wind}= 2$, Eq.\,\ref{eq::jorb}). It increases the SN~Ia rate by 13\%.

However, the rate of model \#Q is lower than that of model \#G. When more material is accreted during wind mass transfer and less material is lost, less angular momentum is lost from the system. Therefore fewer double WD systems form in a short enough orbit to merge within a Hubble time in model \#Q compared to model \#G. The same holds for regions C$_{\rm H}$ and C$_{\rm He}$ of both SD channels, where at WD formation the binary system is in a longer orbit in model \#Q than in model \#G. One might expect that the same reasoning holds for model \#L, in which less material is lost from the binary system compared to our standard model. However, in both model \#L and our standard model, when material is removed from the binary system the specific angular momentum of the donor star is lost, which is smaller for our progenitor systems than in models \#G and \#Q. Therefore, the rate of model \#L increases compared to our standard model because more material is accreted by the companion star.

\section{Influence of initial binary distributions\label{sec:inflDistr}}

The distribution functions of initial binary parameters are uncertain because of limitations in the different techniques used to determine them, such as difficulties in resolving binary companions or the incorrect determination of masses because of rotation \cite[][]{Scalo98,Kroupa01,DucheneKraus13}. In addition, it is not clear if the distribution functions are universal. Therefore we compare the theoretical SN~Ia rates calculated assuming different initial binary distribution functions.

\subsection{Initial mass ratio distribution}
The initial mass ratio distribution of intermediate-mass stars is highly uncertain because of difficulties in the observations of the companion star \cite[]{DucheneKraus13}. A flat distribution function of the initial mass ratio is widely used for BPS studies. However, some observations suggest a different distribution function for intermediate-mass stars, e.g. $\phi(q)$ $\propto$ $q^{-0.4}$ \cite[][]{Kouwenhoven07}. In models \#N$_1$ to \#N$_5$ we investigate initial mass ratio distributions of this form with slopes between $-1$ and 1 (Table\,\ref{ParTable}).

Figs.\,\ref{DDCEa},\,\ref{SDHCEa}~and~\ref{SDHeCEa} show that most evolutionary channels require binary systems with a high initial mass ratio. Therefore the rate strongly depends on the distribution of the initial mass ratio and peaks when equal masses are favoured. Table\,\ref{rateSF}~shows that the rate of every channel increases if the initial mass-ratio distribution is skewed towards equal masses, except in the most extreme case of model \#N$_5$. In model \#N$_5$ the rate corresponding to the regions which do not favour equal masses decreases compared to the rate of model \#N$_4$, such as regions B$_{\rm H}$ and B$_{\rm He}$ of the SD channels, although the overall rate of model \#N$_5$ is higher compared to models \#N$_1$ to \#N$_4$. The integrated rate differs by a factor 4.5 between the two extremes and is a factor 4 to 17 times lower than the \cite{Maoz11} rate. The most extreme model which strongly disfavours equal masses (model \#N$_1$) is probably not realistic, as in this model about 50\% of the systems with an initial primary mass between 2.5 and 10\Msun~have a secondary mass lower than 0.2\Msun.

\subsection{Initial mass function}

\begin{table} 
\caption{The fraction $\eta$ of the binary systems with a primary mass between 3 and 8\Msun\ that result in a SN~Ia and the expected SN~Ia rate versus CC~SN~rate of different models with varying IMFs compared with the observations.}
\label{TableCCSNe}
\begin{center}
%\begin{tabular}{l|c|c|c}
\begin{tabular}{lccc}
\hline \noalign{\smallskip}   
Model & IMF & $\eta$ (\%) & $N_{\rm SNe Ia}/N_{\rm CC SNe}$\tablefootmark{a} \\
\hline \noalign{\smallskip} %\hline 
Standard &\cite{Kroupa93} & 2.4  & 0.07 - 0.14 \\
\#O$_1$ &\cite{Scalo98} & 2.4 & 0.05 - 0.09 \\
\#O$_2$ &\cite{Kroupa01} & 2.3 & 0.04 - 0.09 \\
\#O$_3$ &\cite{Chabrier03} & 2.5 & 0.05 - 0.09 \\
\#O$_4$ &\cite{Bell03} & 2.3 & 0.04 - 0.09 \\
\hline \noalign{\smallskip}
Observed & - & 15 (2-40)\tablefootmark{b} & $0.28 \pm 0.07$\tablefootmark{c} \\
\hline 
\end{tabular}
\tablefoot{\tablefoottext{a}{The upper and lower limit are determined by assuming that only the primary or both stars explode as a CC SNe in a binary system with a primary mass between 8 and 25\Msun.} \tablefoottext{b}{\cite{Maoz08}.} \tablefoottext{c}{\cite{Cappellaro99}.}}
\end{center}
\end{table}

The initial mass function (IMF) of \cite{Kroupa93} is widely used, especially in SN~Ia population synthesis studies. However, other prescriptions for the IMF exist. The IMF chosen defines the normalization of the population under study, therefore it is important to know which IMF is assumed when comparing results from different BPS codes and to realize how it affects the SN~Ia rate (model \#O$_1$ to \#O$_4$). In general the IMF is described as a broken power-law function $M^{-\Gamma}$, with $\Gamma$ the slope of the power law. \cite{Kroupa93} determine the IMF based on the low-mass stellar population in the Galactic disc.  A break in the slope is observed around 0.5\Msun, with $\Gamma=1.3$ for systems with masses lower than 0.5\Msun~and $\Gamma=2.2$ for higher masses. An additional break arises around 1.0\Msun, with a slope $\Gamma=2.7$ above this mass.

\citet[][model \#O$_1$]{Scalo98} defines an IMF mainly based on different galaxies and stellar associations, and his IMF is commonly used in older BPS studies. He finds that a stellar population contains a similar number of intermediate-mass stars with a mass between 1 and 10\Msun~compared to the IMF of \cite{Kroupa93}, but it contains fewer low-mass stars. As the number of intermediate mass stars remains constant, while it decreases for the low-mass stars, the overall rate of model \#O$_1$ increases compared to our standard model (Table\,\ref{rateSF}).

\citet[][model \#O$_2$]{Kroupa01} and \citet[][model \#O$_4$]{Bell03} find similar IMFs, based on the Galactic field and a large galaxy sample from the local universe, respectively. Both groups find a Salpeter-like IMF with $\Gamma \approx 2.35$ for intermediate and massive stars, and a flatter slope below 0.5 or 0.6\Msun, respectively. A name commonly used for the IMF determined by \citet{Bell03} is the `diet Salpeter' IMF and this is used for example by \cite{Maoz12} and \cite{Graur13} to extract the SN~Ia rate. Models \#O$_2$ and \#O$_4$ show an increase of the SN~Ia rate of approximately 40\% compared to our standard model. Because the slope in the mass range between 2.5 and 10\Msun~is different from that in our standard model, the rate corresponding to the different evolution paths does not increase by the same amount, e.g.\ the rates from regions C and D of the DD channel change differently because the former favours less massive primaries than the latter.

\citet[][model \#O$_3$]{Chabrier03} compares different present-day IMFs and shows that low-mass primaries ($<1$\Msun) are distributed according to a log-normal function, while the best fit to the more massive stars is a power-law function in between the Salpeter-function and the function determined by \cite{Scalo98} for the mass range between 1 and 10\Msun. This results in a stellar population that contains fewer low-mass primaries than our standard model. In model \#O$_3$ our integrated rate increases by about 45\% compared to our standard model and is about a factor 3.5 lower than the \cite{Maoz11} rate.

While other prescriptions of the IMF result in an increase of the SN~Ia rate with respect to our standard model, the rate of CC SNe increases as well. In models 
\#O$_1$ to \#O$_4$ the ratio of the SN~Ia rate and the CC~SN~rate decreases compared to our standard model and is about a factor three to seven lower than the observationally estimated ratio by \citet[][Table\,\ref{TableCCSNe}]{Cappellaro99}. The fraction $\eta$ of intermediate mass systems leading to a SN~Ia varies between 2.3\% and 2.5\% between the different models (Table\,\ref{TableCCSNe}) and is about about a factor six smaller than the observational estimate \cite[][Table\,\ref{TableCCSNe}]{Maoz08}, although the latter is quite uncertain. Both comparisons in Table\,\ref{TableCCSNe} indicate that although a variation of the IMF increases the integrated SN~Ia rate, it does not reproduce other observational predictions regarding SN~Ia.

\subsection{Combining two distribution functions}

\begin{figure}
\begin{center}
\includegraphics[width=9cm]{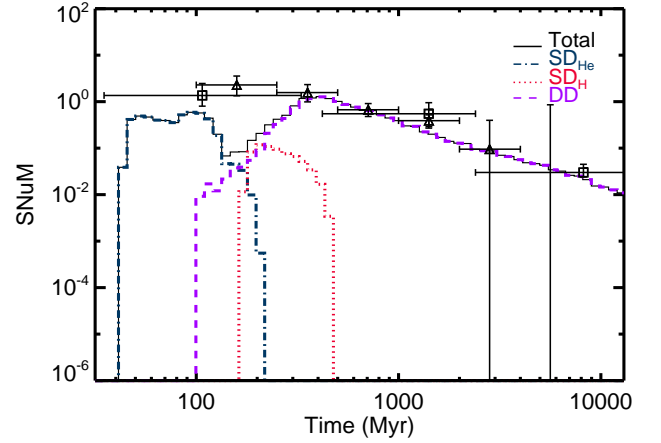}
\caption{Delay time distribution of the SN~Ia channels of model \#R, our most optimistic model, with an IMF according to \cite{Chabrier03} and and initial mass ratio distribution $\phi(q) \propto q^{1}$ . The line styles and data points have the same meaning as in Fig.\,\ref{GenDTD}.}
\label{OptimisticDTD}
\end{center}
\end{figure}

In a population of stars, both the IMF and the initial mass ratio distribution can differ from our standard model. To test the most favourable situation we change both initial binary distribution functions simultaneously. We assume in model \#R an IMF according to \cite{Chabrier03} and an initial mass ratio distribution function with $\phi(q) \propto q$ (see Table\,\ref{rateSF}~and~\ref{rateStarburst}).

The rate corresponding to all the regions increases in model \#R, similar to the amount expected from the combination of the two distributions separately. The overall rate is about a factor three lower than the \cite{Maoz11} rate and is compatible with the \cite{Maoz12} rate (Fig.\,\ref{OptimisticDTD}).

\section{Discussion \& conclusion\label{sec::disc}}

\subsection{Progenitor evolution}

We find that the dominant progenitor evolutionary path for the DD channel is the Roche-lobe overflow path, in which the primary WD forms after a phase of stable RLOF and the secondary WD forms after a CE phase. In our standard model, this path accounts for 84\% of the systems evolving though the DD channel, which is comparable to the fraction determined by \cite{Mennekens10}. In addition, 41\% of the DD channel evolves through the formation reversal path, in which the initially less massive star forms the first WD \cite[]{Toonen12}. This path is less common according to the results of \cite{Toonen12}, mainly because stable RLOF of helium stars in the HG is less likely to occur in the model of \citet[][]{Toonen12}. The common envelope channel, which forms a double WD system after two CE phases, is uncommon and only accounts for about 2.4\% of the DD channel in our standard model. This is lower than the fraction of about 17\% found by \citet{Mennekens10}, which is partly explained  by the fact that they only consider two formation scenarios for the DD channel. We also find another formation channel, in which the first WD forms after a CE phase and the secondary WD forms after a phase of stable RLOF, which accounts for about 13\% of the DD systems in our standard model.

The dominant progenitor evolutionary path for the SD channel with hydrogen-rich donors (SD$_{\rm H}$) involves the formation of a WD after a CE phase and subsequent evolution through the WD+MS path. The least common path in our standard model is the WD+RG path, in contrast to the results of \cite{Ruiter09} who find that this path is more common than the WD+MS path. A possible origin of the differences between these results is that in the models of \cite{Ruiter09} stable RLOF of stars in the HG is less likely to occur, while stable RLOF of early GB stars is more likely to occur than in our model \citep{popcorn}. \cite{Mennekens10} also find a DTD from the SD$_{\rm H}$ channel dominated by the WD+MS path.

The dominant progenitor evolutionary path for the SD channel with helium-rich donors (SD$_{\rm He}$) is through a phase of stable RLOF, which forms the WD, followed by a CE phase which forms the helium star donor. In our standard model and in most other models, the WD+He-MS path and WD+He-HG path contribute equally to the rate.

\subsection{The theoretical SN Ia rate and comparison to other work}

The DD channel contributes to the SN Ia rate from about 100~Myr up to a Hubble time. The different models agree that the respective DTD follows a $t^{-x}$ power law, with $x=1.3$ in our standard model. The DD channel does not contribute at the shortest delay times. The DTD of SNe~Ia at short delay times, $\lesssim 100$~Myr, is formed by the SD$_{\rm He}$ channel. Our models show that the SD$_{\rm H}$ channel may contribute through the WD+MS path from about 70 to 3500~Myr, depending on the mass range of the initially formed CO WDs, and until about 8000~Myr through the WD+RG path. The SD channel does not contribute to the longest delay times, $\gtrsim 8000$~Myr. In our standard model, however, the SD$_{\rm H}$ channel mainly contributes between 100 and 500~Myr. Generally, the DD channel is the dominant formation channel, comprising 95\% of the SN Ia rate in our standard model, but it cannot reproduce the prompt channel. Only the SD$_{\rm He}$ channel can account for SNe~Ia at short delay times.

Additionally, our models show that the $t^{-1}$ relation is not a standard characteristic of the DD channel. Because the DD channel is a combination of different progenitor channels, the DTD depends on the contribution of each of them (Sect.\,\ref{sec::varalphace}).

Although we produce SNe~Ia both at short and long delay times, we do not reproduce the observed number of SNe~Ia. We find an integrated rate which is a factor 3 to 26 times lower than \cite{Maoz11} rate. However, the highest integrated rate we find, in model \#R, is compatible with the \cite{Maoz12} rate.

The results of other groups are similar. \cite{Ruiter09} find that the DD channel reproduces the observed $t^{-1}$ relation, but is not able to reproduce the observed height of the DTD. \cite{Wang09} also discuss the SD$_{\rm He}$ channel and find that the SD$_{\rm He}$ channel is the progenitor channel responsible for the prompt SNe~Ia. \cite{Greggio10} determines the DTD with an analytical approach. Her models which assume contributions of both the SD and the DD channel indicate that SNe~Ia  with delay times shorter than 0.1~Gyr originate from a combination of both channels, while at longer delay times the DD channel dominates. \cite{Greggio10} finds that the SD channel cannot reproduce the observed SN~Ia rate at delay times longer than 10~Gyr.

\cite{Hachisu08} describe the possibility of a radiation driven wind from the WD that strips extra material from the donor star and stabilizes RLOF more than the model adopted in this research. This results in more massive donor stars which can steadily transfer material to the WD. \cite{Mennekens10} consider this model to derive the DTD with a BPS code, but they cannot reproduce the observed rate. Their rate is at least a factor three too low. In addition, the delayed component, produced by the SD WD+RG channel, drops at about 10~Gyr.

\subsection{Uncertainties in binary evolution}

The influence of uncertain aspects of single and binary star evolution on the SN~Ia rate and DTD are studied in this paper. We find that most uncertainties only have a small effect on the SN~Ia rate ($<15\%$). This is because in our standard model the DD channel dominates, which is less sensitive to the masses of the WDs. However, the DD channel mainly depends on the spiral-in time after the formation of the two WDs and most uncertainties  only marginally affect the spiral-in time. The uncertainties with the largest effect are the CE efficiency and helium star evolution, more specifically the stability of Roche-lobe-overflowing helium stars. Both result in a variation of at least a factor two in the integrated rate (Table\,\ref{rateStarburst}). Moreover, the ratio of the SD channel and the DD channel changes significantly when the CE efficiency is varied. The DD channel peaks when $\alpha_{\rm ce}=1$ and the SD channel increases with decreasing CE efficiency. In the model with a low CE efficiency ($\alpha_{\rm ce}=0.2$) the SD channel even dominates.

Other uncertainties are the initial binary distributions, where the two most extreme models give integrated rates which differ by a factor of five. The integrated rate of the model which combines the most optimistic distribution function of the initial primary mass and initial mass ratio is only a factor three lower than the \cite{Maoz11} rate. In addition, the number of SN~Ia versus the number of CC~SN remains at least a factor 2.5 lower than estimated by \cite{Cappellaro99}.

We also consider the extreme situation of unlimited accretion onto a WD (model \#F$_1$), which we do not believe to be realistic, but is adopted to determine an upper limit to the contribution of the SD channel. We find that rate corresponding to model~\#F$_1$ reproduces the \cite{Maoz11} rate. More realistic assumptions yield a rate from the SD channel that is a factor of 10--100 lower. This does not indicate that the SD channel can be excluded, only that with normal assumptions it is hard to reproduce the observed rate with only the SD channel.

We mainly vary one parameter at a time to show its effect on the SN~Ia progenitor evolution and the theoretical rate separately. We demonstrate that the result of changing two parameters at the same time is not always equal to the sum of the effects of each individual change (Sect.\,\ref{sec::varextr1}).

In addition, as mentioned in Sect.\,\ref{sec:DTD}, the observed rate is also uncertain. Most recent observations show a rate which is about a factor two and maybe a factor four lower than the previously determined rate \citep{Maoz12,Graur13,Perrett12}. \cite{Maoz12} conclude that their rate at long delay times based on a sample dominated by field galaxies is more than 2$\sigma$ lower than the rate at long delay times of \cite{Maoz11} based on galaxies in cluster environments. In addition, the metallicities of clusters of galaxies \cite[]{Plaa07} indicate a higher fraction of SNe~Ia than in our own Galaxy. \cite{Maoz12} suggest a possible enhancement of SNe~Ia in cluster environments, while \cite{Sarazin86} suggests a difference in the IMF between the two types of environments. Even though it is not clear exactly where the differences come from, we point out that the theoretical rate found in our standard model is compatible with the lower limit of the observed rate.

\subsection{Other uncertainties in the results}

We show that variation of the binary physics assumptions has a great influence on the rate. However, our research does not give a complete overview of the uncertainties that dominate the SN~Ia rate. \cite{Nelemans13} show that different BPS codes show conflicting results, mainly for the SD channel. Recently four different BPS codes have been compared \cite[]{popcorn}, including ours. \cite{popcorn} show that the results found with the four codes are similar, when the same approximate assumptions are made. The differences that they find are caused by differences in the inherent assumptions in the codes, such as the initial-final mass relation, helium star evolution, the stability criterion of Roche-lobe overflowing stars and the mass transfer rate. In addition, \cite{Bours13} show that the lack of understanding of the retention efficiency of WDs gives an integrated rate of the SD channel which varies between $ < 10^{-7}$~and~$1.5 \cdot 10^{-4}$\Msun$^{-1}$.

Varying the binary fraction affects the rate as well. Our models assume a binary fraction of 100\%, which is a reasonable assumptions for O and B stars \cite[e.g.][]{Kouwenhoven07, Sana12}, but an overestimation for lower-mass stars \cite[e.g.][]{Raghavan10}. As we do not expect that SNe~Ia originate from single stars, the rate decreases when we adopt more realistic binary fractions. In addition, differences in the metallicity of the progenitor systems can alter the observed SN~Ia rate. \cite{Toonen12} discuss that a lower metallicity than solar does not affect the DTD from the DD channel, however, the integrated SN~Ia rate increases with 30 to 60\%. \cite{Meng11} find that the DTD of the SD$_{\rm H}$ channel is more delayed in models with lower metallicities than Solar.

There are also newly proposed channels possibly leading to SNe~Ia which we did not consider for the work presented here, such as the core degenerate model \citep{Kashi11,Ilkov12,Ilkov13}, in which a WD merges with the core of an AGB star during the CE phase; the double-detonation sub-Chandrasekhar explosion \citep{Fink10,Kromer10}, in which a WD explodes after the detonation of a thin helium layer accreted onto the WD with a low mass-transfer rate; the violent merger model \citep{Pakmor10,Pakmor11}, in which a massive enough WD explodes because of the accretion of material from another WD with an almost equal mass as the exploding WD; and the spin-up/spin-down model \citep{Distefano11,Hachisu12}, in which a WD gains angular momentum from the accreted material which spins up the WD, which can possible lead to super-Chandrasekhar WD and a delay between the explosion of the WD and the accretion of the material. Finally, we do not consider triple star evolution, which produces double WD mergers in eccentric orbits \cite[]{Hamers13} and, according to \cite{Rosswog09}, is an alternative scenario to produce SNe~Ia. Even though we do not investigate the rate of these channels, it is expected that similar uncertainties as discussed in this work influence the rate.

\subsection{Outlook}
Our models indicate that the progenitor evolution of SNe~Ia does not consist of one evolutionary channel, but has many different branches with the relevance of each depending on different aspects of binary evolution. With upcoming supernova surveys we will get more detailed information on the differences between individual SNe~Ia. Therefore it is possible to gain insight in the sub-populations of SNe~Ia. As a next step the characteristics of newly proposed progenitor channels should be investigated in more detail, including the properties of the merger products or the remaining companion stars. The outcome of such a study, combined with the results of this paper, can be linked with the different sub-populations of SNe~Ia. However, when studying the rates of the other progenitor channels the uncertainties discussed in this work should always be kept in mind.

\acknowledgements
JSWC would like to thank Alex Chiotellis for the discussions about wind mass loss and for his help to determine the critical mass ratios of accreting WDs. We also would like to thank the referee, Zhanwen Han, for the useful feedback. The work of JSWC was supported by the Dutch Research School for Astronomy (NOVA). RGI thanks Alexander von Humboldt Foundation for funding his current position.

\bibliography{lit}

\begin{thebibliography}{130}
\expandafter\ifx\csname natexlab\endcsname\relax\def\natexlab#1{#1}\fi

\bibitem[{{Bell} {et~al.}(2003){Bell}, {McIntosh}, {Katz}, \&
  {Weinberg}}]{Bell03}
{Bell}, E.~F., {McIntosh}, D.~H., {Katz}, N., \& {Weinberg}, M.~D. 2003, \apjs,
  149, 289

\bibitem[{{Bloecker}(1995)}]{Bloecker95}
{Bloecker}, T. 1995, \aap, 297, 727

\bibitem[{{Bloom} {et~al.}(2012){Bloom}, {Kasen}, {Shen}, {Nugent}, {Butler},
  {Graham}, {Howell}, {Kolb}, {Holmes}, {Haswell}, {Burwitz}, {Rodriguez}, \&
  {Sullivan}}]{Bloom12}
{Bloom}, J.~S., {Kasen}, D., {Shen}, K.~J., {et~al.} 2012, \apjl, 744, L17

\bibitem[{{Bondi} \& {Hoyle}(1944)}]{BondiHoyle44}
{Bondi}, H. \& {Hoyle}, F. 1944, \mnras, 104, 273

\bibitem[{{Bours} {et~al.}(2013){Bours}, {Toonen}, \& {Nelemans}}]{Bours13}
{Bours}, M.~C.~P., {Toonen}, S., \& {Nelemans}, G. 2013, ArXiv e-prints

\bibitem[{{Cappellaro} {et~al.}(1999){Cappellaro}, {Evans}, \&
  {Turatto}}]{Cappellaro99}
{Cappellaro}, E., {Evans}, R., \& {Turatto}, M. 1999, \aap, 351, 459

\bibitem[{{Chabrier}(2003)}]{Chabrier03}
{Chabrier}, G. 2003, \pasp, 115, 763

\bibitem[{{Chen} {et~al.}(2012){Chen}, {Jeffery}, {Zhang}, \& {Han}}]{Chen12}
{Chen}, X., {Jeffery}, C.~S., {Zhang}, X., \& {Han}, Z. 2012, \apjl, 755, L9

\bibitem[{{Chevalier}(2012)}]{Chevalier12}
{Chevalier}, R.~A. 2012, \apjl, 752, L2

\bibitem[{{Chiotellis} {et~al.}(2012){Chiotellis}, {Schure}, \&
  {Vink}}]{Chiotellis12}
{Chiotellis}, A., {Schure}, K.~M., \& {Vink}, J. 2012, \aap, 537, A139

\bibitem[{{Chomiuk} {et~al.}(2012){Chomiuk}, {Soderberg}, {Moe}, {Chevalier},
  {Rupen}, {Badenes}, {Margutti}, {Fransson}, {Fong}, \&
  {Dittmann}}]{Chomiuk12}
{Chomiuk}, L., {Soderberg}, A.~M., {Moe}, M., {et~al.} 2012, \apj, 750, 164

\bibitem[{{Dan} {et~al.}(2013){Dan}, {Rosswog}, {Brueggen}, \&
  {Podsiadlowski}}]{Dan13}
{Dan}, M., {Rosswog}, S., {Brueggen}, M., \& {Podsiadlowski}, P. 2013, ArXiv
  e-prints

\bibitem[{{Davis} {et~al.}(2012){Davis}, {Kolb}, \& {Knigge}}]{Davis12}
{Davis}, P.~J., {Kolb}, U., \& {Knigge}, C. 2012, \mnras, 419, 287

\bibitem[{{De Marco} {et~al.}(2011){De Marco}, {Passy}, {Moe}, {Herwig}, {Mac
  Low}, \& {Paxton}}]{DeMarco11}
{De Marco}, O., {Passy}, J.-C., {Moe}, M., {et~al.} 2011, \mnras, 411, 2277

\bibitem[{{de Mink} {et~al.}(2007){de Mink}, {Pols}, \& {Hilditch}}]{DeMink07}
{de Mink}, S.~E., {Pols}, O.~R., \& {Hilditch}, R.~W. 2007, \aap, 467, 1181

\bibitem[{{de Plaa} {et~al.}(2007){de Plaa}, {Werner}, {Bleeker}, {Vink},
  {Kaastra}, \& {M{\'e}ndez}}]{Plaa07}
{de Plaa}, J., {Werner}, N., {Bleeker}, J.~A.~M., {et~al.} 2007, \aap, 465, 345

\bibitem[{{Dewi} \& {Tauris}(2000)}]{DewiTauris00}
{Dewi}, J.~D.~M. \& {Tauris}, T.~M. 2000, \aap, 360, 1043

\bibitem[{{Di Stefano}(2010)}]{DiStefano10}
{Di Stefano}, R. 2010, \apj, 712, 728

\bibitem[{{Di Stefano} {et~al.}(2011){Di Stefano}, {Voss}, \&
  {Claeys}}]{Distefano11}
{Di Stefano}, R., {Voss}, R., \& {Claeys}, J.~S.~W. 2011, \apjl, 738, L1

\bibitem[{{Dilday} {et~al.}(2012){Dilday}, {Howell}, {Cenko}, {Silverman},
  {Nugent}, {Sullivan}, {Ben-Ami}, {Bildsten}, {Bolte}, {Endl}, {Filippenko},
  {Gnat}, {Horesh}, {Hsiao}, {Kasliwal}, {Kirkman}, {Maguire}, {Marcy},
  {Moore}, {Pan}, {Parrent}, {Podsiadlowski}, {Quimby}, {Sternberg}, {Suzuki},
  {Tytler}, {Xu}, {Bloom}, {Gal-Yam}, {Hook}, {Kulkarni}, {Law}, {Ofek},
  {Polishook}, \& {Poznanski}}]{Dilday12}
{Dilday}, B., {Howell}, D.~A., {Cenko}, S.~B., {et~al.} 2012, Science, 337, 942

\bibitem[{{Duch{\^e}ne} \& {Kraus}(2013)}]{DucheneKraus13}
{Duch{\^e}ne}, G. \& {Kraus}, A. 2013, ArXiv e-prints

\bibitem[{{Eggleton}(2006)}]{Eggleton06}
{Eggleton}, P. 2006, {Evolutionary Processes in Binary and Multiple Stars}

\bibitem[{{Eggleton}(1971)}]{Eggleton71}
{Eggleton}, P.~P. 1971, \mnras, 151, 351

\bibitem[{{Eggleton}(1983)}]{Eggleton83}
{Eggleton}, P.~P. 1983, \apj, 268, 368

\bibitem[{{Fink} {et~al.}(2010){Fink}, {R{\"o}pke}, {Hillebrandt},
  {Seitenzahl}, {Sim}, \& {Kromer}}]{Fink10}
{Fink}, M., {R{\"o}pke}, F.~K., {Hillebrandt}, W., {et~al.} 2010, \aap, 514,
  A53

\bibitem[{{Garc{\'{\i}}a-Senz} {et~al.}(2012){Garc{\'{\i}}a-Senz}, {Badenes},
  \& {Serichol}}]{GarciaBadenes11}
{Garc{\'{\i}}a-Senz}, D., {Badenes}, C., \& {Serichol}, N. 2012, \apj, 745, 75

\bibitem[{{Gilfanov} \& {Bogd{\'a}n}(2010)}]{Gilfanov10}
{Gilfanov}, M. \& {Bogd{\'a}n}, {\'A}. 2010, \nat, 463, 924

\bibitem[{{Glebbeek} {et~al.}(2008){Glebbeek}, {Pols}, \&
  {Hurley}}]{Glebbeek08}
{Glebbeek}, E., {Pols}, O.~R., \& {Hurley}, J.~R. 2008, \aap, 488, 1007

\bibitem[{{Graur} \& {Maoz}(2013)}]{Graur13}
{Graur}, O. \& {Maoz}, D. 2013, \mnras, 430, 1746

\bibitem[{{Graur} {et~al.}(2011){Graur}, {Poznanski}, {Maoz}, {Yasuda},
  {Totani}, {Fukugita}, {Filippenko}, {Foley}, {Silverman}, {Gal-Yam},
  {Horesh}, \& {Jannuzi}}]{Graur11}
{Graur}, O., {Poznanski}, D., {Maoz}, D., {et~al.} 2011, \mnras, 417, 916

\bibitem[{{Greggio}(2010)}]{Greggio10}
{Greggio}, L. 2010, \mnras, 406, 22

\bibitem[{{Hachisu} {et~al.}(1996){Hachisu}, {Kato}, \& {Nomoto}}]{Hachisu96}
{Hachisu}, I., {Kato}, M., \& {Nomoto}, K. 1996, \apjl, 470, L97+

\bibitem[{{Hachisu} {et~al.}(1999){Hachisu}, {Kato}, \& {Nomoto}}]{Hachisu99}
{Hachisu}, I., {Kato}, M., \& {Nomoto}, K. 1999, \apj, 522, 487

\bibitem[{{Hachisu} {et~al.}(2008){Hachisu}, {Kato}, \& {Nomoto}}]{Hachisu08}
{Hachisu}, I., {Kato}, M., \& {Nomoto}, K. 2008, \apjl, 683, L127

\bibitem[{{Hachisu} {et~al.}(2012){Hachisu}, {Kato}, {Saio}, \&
  {Nomoto}}]{Hachisu12}
{Hachisu}, I., {Kato}, M., {Saio}, H., \& {Nomoto}, K. 2012, \apj, 744, 69

\bibitem[{{Hamann} \& {Koesterke}(1998)}]{Hamann98}
{Hamann}, W.-R. \& {Koesterke}, L. 1998, \aap, 335, 1003

\bibitem[{{Hamann} {et~al.}(1995){Hamann}, {Koesterke}, \&
  {Wessolowski}}]{Hamann95}
{Hamann}, W.-R., {Koesterke}, L., \& {Wessolowski}, U. 1995, \aap, 299, 151

\bibitem[{{Hamers} {et~al.}(2013){Hamers}, {Pols}, {Claeys}, \&
  {Nelemans}}]{Hamers13}
{Hamers}, A.~S., {Pols}, O.~R., {Claeys}, J.~S.~W., \& {Nelemans}, G. 2013,
  \mnras, 704

\bibitem[{{Hamuy} {et~al.}(2003){Hamuy}, {Phillips}, {Suntzeff}, {Maza},
  {Gonz{\'a}lez}, {Roth}, {Krisciunas}, {Morrell}, {Green}, {Persson}, \&
  {McCarthy}}]{Hamuy03}
{Hamuy}, M., {Phillips}, M.~M., {Suntzeff}, N.~B., {et~al.} 2003, \nat, 424,
  651

\bibitem[{{Han} \& {Podsiadlowski}(2004)}]{Han04}
{Han}, Z. \& {Podsiadlowski}, P. 2004, \mnras, 350, 1301

\bibitem[{{Howell}(2011)}]{Howell11}
{Howell}, D.~A. 2011, Nature Communications, 2

\bibitem[{{Howell} {et~al.}(2006){Howell}, {Sullivan}, {Nugent}, {Ellis},
  {Conley}, {Le Borgne}, {Carlberg}, {Guy}, {Balam}, {Basa}, {Fouchez}, {Hook},
  {Hsiao}, {Neill}, {Pain}, {Perrett}, \& {Pritchet}}]{Howell06}
{Howell}, D.~A., {Sullivan}, M., {Nugent}, P.~E., {et~al.} 2006, \nat, 443, 308

\bibitem[{{Hurley} {et~al.}(2000){Hurley}, {Pols}, \& {Tout}}]{Hurley00}
{Hurley}, J.~R., {Pols}, O.~R., \& {Tout}, C.~A. 2000, \mnras, 315, 543

\bibitem[{{Hurley} {et~al.}(2002){Hurley}, {Tout}, \& {Pols}}]{Hurley02}
{Hurley}, J.~R., {Tout}, C.~A., \& {Pols}, O.~R. 2002, \mnras, 329, 897

\bibitem[{{Iben} \& {Tutukov}(1984)}]{IbenTutukov84}
{Iben}, Jr., I. \& {Tutukov}, A.~V. 1984, \apjs, 54, 335

\bibitem[{{Ilkov} \& {Soker}(2012)}]{Ilkov12}
{Ilkov}, M. \& {Soker}, N. 2012, \mnras, 419, 1695

\bibitem[{{Ilkov} \& {Soker}(2013)}]{Ilkov13}
{Ilkov}, M. \& {Soker}, N. 2013, \mnras, 428, 579

\bibitem[{{Ivanova}(2011)}]{Ivanova11}
{Ivanova}, N. 2011, in Astronomical Society of the Pacific Conference Series,
  Vol. 447, Evolution of Compact Binaries, ed. L.~{Schmidtobreick}, M.~R.
  {Schreiber}, \& C.~{Tappert}, 91

\bibitem[{{Izzard}(2004)}]{IzzardThesis04}
{Izzard}, R.~G. 2004, PhD thesis, University of Cambridge

\bibitem[{{Izzard} {et~al.}(2010){Izzard}, {Dermine}, \& {Church}}]{Izzard10}
{Izzard}, R.~G., {Dermine}, T., \& {Church}, R.~P. 2010, \aap, 523, A10

\bibitem[{{Izzard} {et~al.}(2006){Izzard}, {Dray}, {Karakas}, {Lugaro}, \&
  {Tout}}]{Izzard06}
{Izzard}, R.~G., {Dray}, L.~M., {Karakas}, A.~I., {Lugaro}, M., \& {Tout},
  C.~A. 2006, \aap, 460, 565

\bibitem[{{Izzard} {et~al.}(2009){Izzard}, {Glebbeek}, {Stancliffe}, \&
  {Pols}}]{Izzard09}
{Izzard}, R.~G., {Glebbeek}, E., {Stancliffe}, R.~J., \& {Pols}, O.~R. 2009,
  \aap, 508, 1359

\bibitem[{{Izzard} {et~al.}(2004){Izzard}, {Tout}, {Karakas}, \&
  {Pols}}]{Izzard04}
{Izzard}, R.~G., {Tout}, C.~A., {Karakas}, A.~I., \& {Pols}, O.~R. 2004,
  \mnras, 350, 407

\bibitem[{{Jahanara} {et~al.}(2005){Jahanara}, {Mitsumoto}, {Oka}, {Matsuda},
  {Hachisu}, \& {Boffin}}]{Jahanara05}
{Jahanara}, B., {Mitsumoto}, M., {Oka}, K., {et~al.} 2005, \aap, 441, 589

\bibitem[{{Karakas} {et~al.}(2002){Karakas}, {Lattanzio}, \&
  {Pols}}]{Karakas02}
{Karakas}, A.~I., {Lattanzio}, J.~C., \& {Pols}, O.~R. 2002, \pasa, 19, 515

\bibitem[{{Kashi} \& {Soker}(2011)}]{Kashi11}
{Kashi}, A. \& {Soker}, N. 2011, \mnras, 417, 1466

\bibitem[{{Kato} \& {Hachisu}(2004)}]{KatoHachisu04}
{Kato}, M. \& {Hachisu}, I. 2004, \apjl, 613, L129

\bibitem[{{Kerzendorf} {et~al.}(2009){Kerzendorf}, {Schmidt}, {Asplund},
  {Nomoto}, {Podsiadlowski}, {Frebel}, {Fesen}, \& {Yong}}]{Kerzendorf09}
{Kerzendorf}, W.~E., {Schmidt}, B.~P., {Asplund}, M., {et~al.} 2009, \apj, 701,
  1665

\bibitem[{{Kobulnicky} \& {Fryer}(2007)}]{Kobul07}
{Kobulnicky}, H.~A. \& {Fryer}, C.~L. 2007, \apj, 670, 747

\bibitem[{{Kouwenhoven} {et~al.}(2007){Kouwenhoven}, {Brown}, {Portegies
  Zwart}, \& {Kaper}}]{Kouwenhoven07}
{Kouwenhoven}, M.~B.~N., {Brown}, A.~G.~A., {Portegies Zwart}, S.~F., \&
  {Kaper}, L. 2007, Astronomy and Astrophysics, 474, 77

\bibitem[{{Kromer} {et~al.}(2010){Kromer}, {Sim}, {Fink}, {R{\"o}pke},
  {Seitenzahl}, \& {Hillebrandt}}]{Kromer10}
{Kromer}, M., {Sim}, S.~A., {Fink}, M., {et~al.} 2010, \apj, 719, 1067

\bibitem[{{Kroupa}(2001)}]{Kroupa01}
{Kroupa}, P. 2001, \mnras, 322, 231

\bibitem[{{Kroupa} {et~al.}(1993){Kroupa}, {Tout}, \& {Gilmore}}]{Kroupa93}
{Kroupa}, P., {Tout}, C.~A., \& {Gilmore}, G. 1993, \mnras, 262, 545

\bibitem[{{Lada}(2006)}]{Lada06}
{Lada}, C.~J. 2006, \apjl, 640, L63

\bibitem[{{Leonard}(2007)}]{Leonard07}
{Leonard}, D.~C. 2007, \apj, 670, 1275

\bibitem[{{Li} {et~al.}(2011{\natexlab{a}}){Li}, {Bloom}, {Podsiadlowski},
  {Miller}, {Cenko}, {Jha}, {Sullivan}, {Howell}, {Nugent}, {Butler}, {Ofek},
  {Kasliwal}, {Richards}, {Stockton}, {Shih}, {Bildsten}, {Shara}, {Bibby},
  {Filippenko}, {Ganeshalingam}, {Silverman}, {Kulkarni}, {Law}, {Poznanski},
  {Quimby}, {McCully}, {Patel}, {Maguire}, \& {Shen}}]{Li11}
{Li}, W., {Bloom}, J.~S., {Podsiadlowski}, P., {et~al.} 2011{\natexlab{a}},
  \nat, 480, 348

\bibitem[{{Li} {et~al.}(2011{\natexlab{b}}){Li}, {Chornock}, {Leaman},
  {Filippenko}, {Poznanski}, {Wang}, {Ganeshalingam}, \& {Mannucci}}]{Li11III}
{Li}, W., {Chornock}, R., {Leaman}, J., {et~al.} 2011{\natexlab{b}}, \mnras,
  412, 1473

\bibitem[{{Mannucci} {et~al.}(2006){Mannucci}, {Della Valle}, \&
  {Panagia}}]{Mannucci06}
{Mannucci}, F., {Della Valle}, M., \& {Panagia}, N. 2006, \mnras, 370, 773

\bibitem[{{Mannucci} {et~al.}(2005){Mannucci}, {Della Valle}, {Panagia},
  {Cappellaro}, {Cresci}, {Maiolino}, {Petrosian}, \& {Turatto}}]{Mannucci05}
{Mannucci}, F., {Della Valle}, M., {Panagia}, N., {et~al.} 2005, \aap, 433, 807

\bibitem[{{Maoz}(2008)}]{Maoz08}
{Maoz}, D. 2008, \mnras, 384, 267

\bibitem[{{Maoz} \& {Mannucci}(2012)}]{MaozMannucci12}
{Maoz}, D. \& {Mannucci}, F. 2012, \pasa, 29, 447

\bibitem[{{Maoz} {et~al.}(2012){Maoz}, {Mannucci}, \& {Brandt}}]{Maoz12}
{Maoz}, D., {Mannucci}, F., \& {Brandt}, T.~D. 2012, \mnras, 426, 3282

\bibitem[{{Maoz} {et~al.}(2011){Maoz}, {Mannucci}, {Li}, {Filippenko}, {Della
  Valle}, \& {Panagia}}]{Maoz11}
{Maoz}, D., {Mannucci}, F., {Li}, W., {et~al.} 2011, \mnras, 412, 1508

\bibitem[{{Maoz} {et~al.}(2010){Maoz}, {Sharon}, \& {Gal-Yam}}]{Maoz10}
{Maoz}, D., {Sharon}, K., \& {Gal-Yam}, A. 2010, \apj, 722, 1879

\bibitem[{{Margutti} {et~al.}(2012){Margutti}, {Soderberg}, {Chomiuk},
  {Chevalier}, {Hurley}, {Milisavljevic}, {Foley}, {Hughes}, {Slane},
  {Fransson}, {Moe}, {Barthelmy}, {Boynton}, {Briggs}, {Connaughton}, {Costa},
  {Cummings}, {Del Monte}, {Enos}, {Fellows}, {Feroci}, {Fukazawa}, {Gehrels},
  {Goldsten}, {Golovin}, {Hanabata}, {Harshman}, {Krimm}, {Litvak},
  {Makishima}, {Marisaldi}, {Mitrofanov}, {Murakami}, {Ohno}, {Palmer},
  {Sanin}, {Starr}, {Svinkin}, {Takahashi}, {Tashiro}, {Terada}, \&
  {Yamaoka}}]{Margutti12}
{Margutti}, R., {Soderberg}, A.~M., {Chomiuk}, L., {et~al.} 2012, \apj, 751,
  134

\bibitem[{{Meng} {et~al.}(2009){Meng}, {Chen}, \& {Han}}]{Meng09}
{Meng}, X., {Chen}, X., \& {Han}, Z. 2009, \mnras, 395, 2103

\bibitem[{{Meng} {et~al.}(2011){Meng}, {Li}, \& {Yang}}]{Meng11}
{Meng}, X.~C., {Li}, Z.~M., \& {Yang}, W.~M. 2011, \pasj, 63, L31

\bibitem[{{Mennekens} {et~al.}(2010){Mennekens}, {Vanbeveren}, {De Greve}, \&
  {De Donder}}]{Mennekens10}
{Mennekens}, N., {Vanbeveren}, D., {De Greve}, J.~P., \& {De Donder}, E. 2010,
  \aap, 515, A89+

\bibitem[{{Nelemans} {et~al.}(2013){Nelemans}, {Toonen}, \&
  {Bours}}]{Nelemans13}
{Nelemans}, G., {Toonen}, S., \& {Bours}, M. 2013, in IAU Symposium, Vol. 281,
  IAU Symposium, 225--231

\bibitem[{{Nielsen} {et~al.}(2013){Nielsen}, {Dominik}, {Nelemans}, \&
  {Voss}}]{Nielsen13}
{Nielsen}, M.~T.~B., {Dominik}, C., {Nelemans}, G., \& {Voss}, R. 2013, \aap,
  549, A32

\bibitem[{{Nomoto}(1982)}]{Nomoto82}
{Nomoto}, K. 1982, \apj, 253, 798

\bibitem[{{Nomoto} \& {Kondo}(1991)}]{NomotoKondo91}
{Nomoto}, K. \& {Kondo}, Y. 1991, \apjl, 367, L19

\bibitem[{{{\"O}pik}(1924)}]{opik24}
{{\"O}pik}, E. 1924, Publications of the Tartu Astrofizica Observatory, 25, 1

\bibitem[{{Pakmor} {et~al.}(2011){Pakmor}, {Hachinger}, {R{\"o}pke}, \&
  {Hillebrandt}}]{Pakmor11}
{Pakmor}, R., {Hachinger}, S., {R{\"o}pke}, F.~K., \& {Hillebrandt}, W. 2011,
  \aap, 528, A117

\bibitem[{{Pakmor} {et~al.}(2010){Pakmor}, {Kromer}, {R{\"o}pke}, {Sim},
  {Ruiter}, \& {Hillebrandt}}]{Pakmor10}
{Pakmor}, R., {Kromer}, M., {R{\"o}pke}, F.~K., {et~al.} 2010, \nat, 463, 61

\bibitem[{{Pakmor} {et~al.}(2013){Pakmor}, {Kromer}, {Taubenberger}, \&
  {Springel}}]{Pakmor13}
{Pakmor}, R., {Kromer}, M., {Taubenberger}, S., \& {Springel}, V. 2013, \apjl,
  770, L8

\bibitem[{{Passy} {et~al.}(2012){Passy}, {De Marco}, {Fryer}, {Herwig},
  {Diehl}, {Oishi}, {Mac Low}, {Bryan}, \& {Rockefeller}}]{Passy12}
{Passy}, J.-C., {De Marco}, O., {Fryer}, C.~L., {et~al.} 2012, \apj, 744, 52

\bibitem[{{Patat} {et~al.}(2007){Patat}, {Benetti}, {Justham}, {Mazzali},
  {Pasquini}, {Cappellaro}, {Della Valle}, {-Podsiadlowski}, {Turatto},
  {Gal-Yam}, \& {Simon}}]{Patat07}
{Patat}, F., {Benetti}, S., {Justham}, S., {et~al.} 2007, \aap, 474, 931

\bibitem[{{Patat} {et~al.}(2013){Patat}, {Cordiner}, {Cox}, {Anderson},
  {Harutyunyan}, {Kotak}, {Palaversa}, {Stanishev}, {Tomasella}, {Benetti},
  {Goobar}, {Pastorello}, \& {Sollerman}}]{Patat13}
{Patat}, F., {Cordiner}, M.~A., {Cox}, N.~L.~J., {et~al.} 2013, \aap, 549, A62

\bibitem[{{Perlmutter} {et~al.}(1999){Perlmutter}, {Aldering}, {Goldhaber},
  {Knop}, {Nugent}, {Castro}, {Deustua}, {Fabbro}, {Goobar}, {Groom}, {Hook},
  {Kim}, {Kim}, {Lee}, {Nunes}, {Pain}, {Pennypacker}, {Quimby}, {Lidman},
  {Ellis}, {Irwin}, {McMahon}, {Ruiz-Lapuente}, {Walton}, {Schaefer}, {Boyle},
  {Filippenko}, {Matheson}, {Fruchter}, {Panagia}, {Newberg}, {Couch}, \& {The
  Supernova Cosmology Project}}]{Perlmutter99}
{Perlmutter}, S., {Aldering}, G., {Goldhaber}, G., {et~al.} 1999, \apj, 517,
  565

\bibitem[{{Perrett} {et~al.}(2012){Perrett}, {Sullivan}, {Conley},
  {Gonz{\'a}lez-Gait{\'a}n}, {Carlberg}, {Fouchez}, {Ripoche}, {Neill},
  {Astier}, {Balam}, {Balland}, {Basa}, {Guy}, {Hardin}, {Hook}, {Howell},
  {Pain}, {Palanque-Delabrouille}, {Pritchet}, {Regnault}, {Rich},
  {Ruhlmann-Kleider}, {Baumont}, {Lidman}, {Perlmutter}, \&
  {Walker}}]{Perrett12}
{Perrett}, K., {Sullivan}, M., {Conley}, A., {et~al.} 2012, \aj, 144, 59

\bibitem[{{Phillips}(1993)}]{Phillips93}
{Phillips}, M.~M. 1993, \apjl, 413, L105

\bibitem[{{Pols} {et~al.}(1998){Pols}, {Schroder}, {Hurley}, {Tout}, \&
  {Eggleton}}]{Pols98}
{Pols}, O.~R., {Schroder}, K.-P., {Hurley}, J.~R., {Tout}, C.~A., \&
  {Eggleton}, P.~P. 1998, \mnras, 298, 525

\bibitem[{{Pols} {et~al.}(1995){Pols}, {Tout}, {Eggleton}, \& {Han}}]{Pols95}
{Pols}, O.~R., {Tout}, C.~A., {Eggleton}, P.~P., \& {Han}, Z. 1995, \mnras,
  274, 964

\bibitem[{{Pritchet} {et~al.}(2008){Pritchet}, {Howell}, \&
  {Sullivan}}]{Pritchet08}
{Pritchet}, C.~J., {Howell}, D.~A., \& {Sullivan}, M. 2008, \apjl, 683, L25

\bibitem[{{Raghavan} {et~al.}(2010){Raghavan}, {McAlister}, {Henry}, {Latham},
  {Marcy}, {Mason}, {Gies}, {White}, \& {ten Brummelaar}}]{Raghavan10}
{Raghavan}, D., {McAlister}, H.~A., {Henry}, T.~J., {et~al.} 2010, \apjs, 190,
  1

\bibitem[{{Reimers}(1975)}]{Reimers75}
{Reimers}, D. 1975, Memoires of the Societe Royale des Sciences de Liege, 8,
  369

\bibitem[{{Riess} {et~al.}(1998){Riess}, {Filippenko}, {Challis},
  {Clocchiatti}, {Diercks}, {Garnavich}, {Gilliland}, {Hogan}, {Jha},
  {Kirshner}, {Leibundgut}, {Phillips}, {Reiss}, {Schmidt}, {Schommer},
  {Smith}, {Spyromilio}, {Stubbs}, {Suntzeff}, \& {Tonry}}]{Riess98}
{Riess}, A.~G., {Filippenko}, A.~V., {Challis}, P., {et~al.} 1998, \aj, 116,
  1009

\bibitem[{{Riess} {et~al.}(1996){Riess}, {Press}, \& {Kirshner}}]{Riess96}
{Riess}, A.~G., {Press}, W.~H., \& {Kirshner}, R.~P. 1996, \apj, 473, 88

\bibitem[{{Rosswog} {et~al.}(2009){Rosswog}, {Kasen}, {Guillochon}, \&
  {Ramirez-Ruiz}}]{Rosswog09}
{Rosswog}, S., {Kasen}, D., {Guillochon}, J., \& {Ramirez-Ruiz}, E. 2009,
  \apjl, 705, L128

\bibitem[{{Ruiter} {et~al.}(2009){Ruiter}, {Belczynski}, \& {Fryer}}]{Ruiter09}
{Ruiter}, A.~J., {Belczynski}, K., \& {Fryer}, C. 2009, \apj, 699, 2026

\bibitem[{{Ruiz-Lapuente} {et~al.}(2004){Ruiz-Lapuente}, {Comeron},
  {M{\'e}ndez}, {Canal}, {Smartt}, {Filippenko}, {Kurucz}, {Chornock}, {Foley},
  {Stanishev}, \& {Ibata}}]{RuizLapuente04}
{Ruiz-Lapuente}, P., {Comeron}, F., {M{\'e}ndez}, J., {et~al.} 2004, \nat, 431,
  1069

\bibitem[{{Sana} {et~al.}(2012){Sana}, {de Mink}, {de Koter}, {Langer},
  {Evans}, {Gieles}, {Gosset}, {Izzard}, {Le Bouquin}, \& {Schneider}}]{Sana12}
{Sana}, H., {de Mink}, S.~E., {de Koter}, A., {et~al.} 2012, Science, 337, 444

\bibitem[{{Sarazin}(1986)}]{Sarazin86}
{Sarazin}, C.~L. 1986, Reviews of Modern Physics, 58, 1

\bibitem[{{Scalo}(1998)}]{Scalo98}
{Scalo}, J. 1998, in Astronomical Society of the Pacific Conference Series,
  Vol. 142, The Stellar Initial Mass Function (38th Herstmonceux Conference),
  ed. G.~{Gilmore} \& D.~{Howell}, 201

\bibitem[{{Scannapieco} \& {Bildsten}(2005)}]{Scannapieco05}
{Scannapieco}, E. \& {Bildsten}, L. 2005, \apjl, 629, L85

\bibitem[{{Schaefer} \& {Pagnotta}(2012)}]{Schaefer12}
{Schaefer}, B.~E. \& {Pagnotta}, A. 2012, \nat, 481, 164

\bibitem[{{Shen} {et~al.}(2013){Shen}, {Guillochon}, \& {Foley}}]{Shen13}
{Shen}, K.~J., {Guillochon}, J., \& {Foley}, R.~J. 2013, \apjl, 770, L35

\bibitem[{{Soberman} {et~al.}(1997){Soberman}, {Phinney}, \& {van den
  Heuvel}}]{Soberman97}
{Soberman}, G.~E., {Phinney}, E.~S., \& {van den Heuvel}, E.~P.~J. 1997, \aap,
  327, 620

\bibitem[{{Soker} {et~al.}(2013){Soker}, {Kashi}, {Garc{\'{\i}}a-Berro},
  {Torres}, \& {Camacho}}]{Soker13}
{Soker}, N., {Kashi}, A., {Garc{\'{\i}}a-Berro}, E., {Torres}, S., \&
  {Camacho}, J. 2013, \mnras, 431, 1541

\bibitem[{{Sternberg} {et~al.}(2011){Sternberg}, {Gal-Yam}, {Simon}, {Leonard},
  {Quimby}, {Phillips}, {Morrell}, {Thompson}, {Ivans}, {Marshall},
  {Filippenko}, {Marcy}, {Bloom}, {Patat}, {Foley}, {Yong}, {Penprase},
  {Beeler}, {Allende Prieto}, \& {Stringfellow}}]{Sternberg11}
{Sternberg}, A., {Gal-Yam}, A., {Simon}, J.~D., {et~al.} 2011, Science, 333,
  856

\bibitem[{{Toonen} {et~al.}(2013){Toonen}, {Claeys}, {Mennekens}, \&
  {Ruiter}}]{popcorn}
{Toonen}, S., {Claeys}, J.~S.~W., {Mennekens}, N., \& {Ruiter}, A.~J. 2013,
  ArXiv e-prints

\bibitem[{{Toonen} {et~al.}(2012){Toonen}, {Nelemans}, \& {Portegies
  Zwart}}]{Toonen12}
{Toonen}, S., {Nelemans}, G., \& {Portegies Zwart}, S. 2012, \aap, 546, A70

\bibitem[{{Totani} {et~al.}(2008){Totani}, {Morokuma}, {Oda}, {Doi}, \&
  {Yasuda}}]{Totani08}
{Totani}, T., {Morokuma}, T., {Oda}, T., {Doi}, M., \& {Yasuda}, N. 2008,
  \pasj, 60, 1327

\bibitem[{{Tout} \& {Eggleton}(1988)}]{ToutEggleton88}
{Tout}, C.~A. \& {Eggleton}, P.~P. 1988, \mnras, 231, 823

\bibitem[{{van den Heuvel} {et~al.}(1992){van den Heuvel}, {Bhattacharya},
  {Nomoto}, \& {Rappaport}}]{Vandenheuvel92}
{van den Heuvel}, E.~P.~J., {Bhattacharya}, D., {Nomoto}, K., \& {Rappaport},
  S.~A. 1992, \aap, 262, 97

\bibitem[{{van Loon} {et~al.}(2005){van Loon}, {Cioni}, {Zijlstra}, \&
  {Loup}}]{VanLoon05}
{van Loon}, J.~T., {Cioni}, M.-R.~L., {Zijlstra}, A.~A., \& {Loup}, C. 2005,
  \aap, 438, 273

\bibitem[{{Vassiliadis} \& {Wood}(1993)}]{VassiliadisWood93}
{Vassiliadis}, E. \& {Wood}, P.~R. 1993, \apj, 413, 641

\bibitem[{{Wachter} {et~al.}(2002){Wachter}, {Schr{\"o}der}, {Winters},
  {Arndt}, \& {Sedlmayr}}]{Wachter02}
{Wachter}, A., {Schr{\"o}der}, K.-P., {Winters}, J.~M., {Arndt}, T.~U., \&
  {Sedlmayr}, E. 2002, \aap, 384, 452

\bibitem[{{Wang} {et~al.}(2009{\natexlab{a}}){Wang}, {Chen}, {Meng}, \&
  {Han}}]{Wang09}
{Wang}, B., {Chen}, X., {Meng}, X., \& {Han}, Z. 2009{\natexlab{a}}, \apj, 701,
  1540

\bibitem[{{Wang} \& {Han}(2010)}]{Wang10b}
{Wang}, B. \& {Han}, Z.-W. 2010, Research in Astronomy and Astrophysics, 10,
  235

\bibitem[{{Wang} {et~al.}(2010){Wang}, {Li}, \& {Han}}]{Wang10}
{Wang}, B., {Li}, X.-D., \& {Han}, Z.-W. 2010, \mnras, 401, 2729

\bibitem[{{Wang} {et~al.}(2009{\natexlab{b}}){Wang}, {Meng}, {Chen}, \&
  {Han}}]{Wang09b}
{Wang}, B., {Meng}, X., {Chen}, X., \& {Han}, Z. 2009{\natexlab{b}}, \mnras,
  395, 847

\bibitem[{{Webbink}(1984)}]{Webbink84}
{Webbink}, R.~F. 1984, \apj, 277, 355

\bibitem[{{Weidemann}(2000)}]{Weidemann00}
{Weidemann}, V. 2000, \aap, 363, 647

\bibitem[{{Whelan} \& {Iben}(1973)}]{WhelanIben73}
{Whelan}, J. \& {Iben}, Jr., I. 1973, \apj, 186, 1007

\bibitem[{{Whyte} \& {Eggleton}(1980)}]{WhyteEggleton80}
{Whyte}, C.~A. \& {Eggleton}, P.~P. 1980, \mnras, 190, 801

\bibitem[{{Yoon} \& {Langer}(2005)}]{YoonLanger05}
{Yoon}, S.-C. \& {Langer}, N. 2005, \aap, 435, 967

\bibitem[{{Yungelson} \& {Livio}(2000)}]{YungelsonLivio00}
{Yungelson}, L.~R. \& {Livio}, M. 2000, \apj, 528, 108

\bibitem[{{Zorotovic} {et~al.}(2010){Zorotovic}, {Schreiber}, {G{\"a}nsicke},
  \& {Nebot G{\'o}mez-Mor{\'a}n}}]{Zorotovic10}
{Zorotovic}, M., {Schreiber}, M.~R., {G{\"a}nsicke}, B.~T., \& {Nebot
  G{\'o}mez-Mor{\'a}n}, A. 2010, \aap, 520, A86+

\end{thebibliography}

% Large tables in landscape format:

\begin{landscape}

\begin{table}[tbh!]
\caption{Number of SNe~Ia within a Hubble time per $10^{5}$\Msun of stars formed, corresponding to the different regions discussed in Sect.\,\ref{Sect::BinEvol} and the different channels, calculated with our standard model (Sect.\,\ref{sec:StandAssum}) and the models from our parameter study (Table\,\ref{ParTable}).}
\label{rateSF}
\begin{center}
%\begin{tabular}{ll | cccccccc | cccc | cccc | c}  
\begin{tabular}{ll @{~\qquad} cccccccc @{~\qquad} cccc @{~\qquad} cccc @{~\qquad} c}  
\hline \noalign{\smallskip}
& & & & & & & & & DD & & & & {SD$_{\rm H}$} & & & & {SD$_{\rm He}$} & SNe~Ia\\
N$^{o}$ & Model & A & B$_1$ & B$_2$ & C & D & E & F & Total & A$_{\rm H}$ & B$_{\rm H}$ & C$_{\rm H}$ & Total & A$_{\rm He}$ & B$_{\rm He}$ & C$_{\rm He}$ & Total & Total \\
\hline \noalign{\smallskip} %\hline 

& Standard ($N=150$) & 0.29 & 18.4 & 17.5 & 0.82 & 2.70 &  1.94 &  1.01 &  42.6 & 0.01 & 0.86 & 0.002 &  0.87 & 0.71 &  0.72 &  0.07 &  1.51 & 45.0 \\
& Standard ($N=100$) & 0.29 &  18.4 &  17.5 &  0.82 &  2.53 &  2.08 &  1.03 &  42.7 &  0.01 &  0.86 & 0.001 &  0.87 & 0.72 &  0.72 &  0.03 &  1.47 & 45.0 \\
\hline \noalign{\smallskip}

\#A$_1$ & $\alpha_{\rm ce}$=0.2 &  0.29 &   2.6 &   0.0 &  0.00 &  0.01 &  1.57 &  0.04 & 4.5 & 0.01 &  8.96 & 0.011 &  8.97 & 2.98 &  0.40 &  0.00 &  3.38 & 16.8\\
\#A$_2$ & $\alpha_{\rm ce}$=0.5 &   0.29 &  13.5 &   3.8 &  0.12 &  1.96 &  1.83 &  0.11 &  21.6 & 0.01 &  1.25 & 0.070 &  1.33 & 1.06 &  1.12 &  0.04 &  2.22 & 25.1\\
\#A$_3$ & $\alpha_{\rm ce}$=3 &   0.29 &   6.4 &   7.6 &  0.91 &  2.33 &  1.59 &  4.94 &  24.1 & 0.01 &  0.00 & 0.001 &  0.01 & 1.47 &  0.38 &  0.42 &  2.28 & 26.4\\
\#A$_4$ &  $\alpha_{\rm ce}$=10 &   0.29 &   4.0 &   0.1 &  0.22 &  0.11 &  0.08 &  2.73 &   7.6 & 0.01 &  0.00 & 0.073 &  0.08 &  0.00 &  0.26 &  0.93 &  1.19 &  8.9\\
\#B$_1$ & $\lambda_{\rm CE}$ = 1 & 0.29 &  21.7 &  15.8 &  0.65 &  2.80 &  2.57 &  0.57 &  44.4 &  0.01 &  1.08 & 0.004 & 1.09 &  0.04 &  0.79 &  0.14 &  0.96 & 46.4\\
\#B$_2$ & $\lambda_{\rm ion}$ = 0.5 & 0.29 &  17.1 &  17.6 &  0.82 &  2.34 &  1.33 &  5.54 &  45.0 &  0.01 &  0.77 & 0.004 &  0.78 &  0.68 &  0.67 &  0.08 &  1.42 & 47.3\\
\hline \noalign{\smallskip}

\#C & $Q_{\rm crit,He-HG} = Q_{\rm crit,He-GB}$ & 0.29 &  13.9 &   0.0 &  0.82 &  1.89 &  2.06 &  1.07 &  20.0 & 0.01 &  0.86 & 0.001 &  0.87 & 0.72 &  0.62 &  0.03 &  1.38 & 22.3\\
\#D & $Q_{\rm crit,HG}$ = 0.5 &  0.00 &  13.8 &  12.6 &  0.82 &  2.53 &  2.08 &  1.03 &  32.9 & 0.00 &  0.74 & 0.001 &  0.74 &  0.70 &  0.38 &  0.03 &  1.12 &  34.7\\
\hline \noalign{\smallskip}

\#E$_1$ & $\sigma$ = 1 & 0.17 &  19.7 &  11.1 &  0.75 &  2.42 &  2.08 &  1.07 &  37.3 &  0.00 &  0.64 & 0.001 &  0.64 &  0.61 &  0.97 &  0.03 &  1.62 & 39.6\\
\#E$_2$ & $\sigma$ = 1000 & 0.00 &  22.6 &  19.7 &  0.82 &  2.48 &  2.08 &  1.04 &  48.7 & 0.00 &  0.86 & 0.001 &  0.87 & 1.42 & 0.10 &  0.03 &  1.55 & 51.1\\
\#F$_1$ & $\eta_{\rm H}=1$, $\eta_{\rm He}=1$ &  0.00 &  16.5 &  17.4 &  0.00 &  0.56 &  0.80 &  0.76 &  36.0 & 3.91 &  48.6 &  80.8 & 133.3 & 1.51 & 14.18 &  0.66 & 16.35 & 185.6\\
\#F$_2$ & $\eta_{\rm H}=1$, $\eta_{\rm He}=1$, $\sigma \rightarrow \infty$ &  0.00 &  18.5 &  18.1 &  0.00 &  0.41 &  0.80 &  0.75 &  38.6 &  0.03 &  53.5 &  77.7 & 131.3 &  2.62 & 23.93 &  0.84 & 27.39 & 197.3\\
\hline \noalign{\smallskip}

\#G & $\gamma$ ($J_{\rm orb}$, wind) = 2 & 0.28 &  18.8 &  17.0 &  0.84 &  2.77 &  2.41 &  7.01 &  49.2 & 0.01 &  0.87 & 0.103 &  0.98 & 0.77 &  0.79 &  0.12 &  1.68 & 51.9\\
\#H$_1$ & $\gamma$ ($J_{\rm orb}$, RLOF) = $M_{\rm a}/M_{\rm d}$ & 0.00 &  21.9 &  19.5 &  0.82 &  2.53 &  2.08 &  1.04 &  47.9 & 0.00 &  0.94 & 0.001 &  0.94 & 0.84 &  1.01 &  0.03 &  1.88 & 50.7\\
\#H$_2$ & $\gamma$ ($J_{\rm orb}$,RLOF) = 2 & 0.84 &  15.1 &  17.0 &  0.82 &  2.48 &  2.08 &  1.04 &  39.4 & 0.11 &  0.87 & 0.001 &  0.98 & 0.71 &  0.36 &  0.03 &  1.11 & 41.5\\
\hline \noalign{\smallskip}

\#I & $\eta$ \cite[]{Reimers75} = 5 & 0.29 &  16.3 &  17.6 &  0.82 &  2.54 &  0.61 &  7.60 &  45.7 &  0.01 &  0.86 & 0.000 &  0.87 &  0.65 & 0.10 &  0.10 &  0.85 & 47.5\\
\#J$_1$ & TP-AGB \cite[]{Karakas02} & 0.29 &  18.5 &  17.5 &  0.82 &  2.54 &  1.74 & 1.12 &  42.5 & 0.01 &  0.88 & 0.001 &  0.89 & 0.72 &  0.72 &  0.03 &  1.46 & 44.9\\
\#J$_2$ & TP-AGB \cite[][ $\eta$ = 1]{Reimers75}& 0.29 &  18.5 &  17.5 &  0.82 &  2.54 &  2.39 &  2.82 &  44.9 & 0.01 &  0.88 & 0.004 &  0.89 & 0.72 &  0.72 &  0.03 &  1.47 & 47.2\\
\#J$_3$ & TP-AGB \cite[]{Bloecker95} & 0.29 &  18.5 &  17.5 &  0.82 &  2.54 &  0.72 &  0.80 &  41.2 & 0.01 &  0.88 & 0.001 &  0.89 & 0.72 &  0.72 &  0.02 &  1.45 & 43.5\\
\#J$_4$ & TP-AGB \cite[]{VanLoon05} & 0.29 &  18.5 &  17.5 &  0.82 &  2.54 &  1.59 &  2.18 &  43.4 & 0.01 &  0.88 & 0.001 &  0.89 & 0.72 &  0.72 &  0.03 &  1.46 & 45.8\\
\hline \noalign{\smallskip}

\#K & CE accretion = 0.05\Msun & 0.29 &  17.8 &  20.0 &  0.85 &  2.60 &  2.12 &  1.28 &  45.0 &  0.01 &  0.88 & 0.001 &  0.89 &  0.74 &  0.72 &  0.03 &  1.49 & 47.4\\
\#L & Bondi-Hoyle efficiency $\alpha_{\rm BH}$ = 5 &   0.29 &  18.5 &  17.6 &  0.82 &  2.48 &  2.07 &  2.80 &  44.5 &  0.01 &  0.87 & 0.002 &  0.88 &  0.74 &  0.77 &  0.04 &  1.55 & 46.9\\
\#M & B (CRAP) = 1e3 & 0.29 &  16.2 &  16.7 &  0.81 &  2.82 &  1.12 &  1.61 &  39.5 & 0.01 &  0.86 & 0.000 &  0.87 & 0.60 &  0.04 &  0.04 &  0.67 & 41.0\\
\hline \noalign{\smallskip}

\#N$_1$ & $\phi(q_{\rm i})\propto$ $q_{\rm i}^{-1}$ & 0.14 &   6.4 &   4.2 &  0.31 &  0.63 &  0.44 &  0.28 &  12.4 & 0.01 &  0.39 & 0.001 &  0.39 & 
 0.21 &  0.26 &  0.01 &  0.47 & 13.3\\
\#N$_2$ & $\phi(q_{\rm i})\propto$ $q_{\rm i}^{-0.5}$ & 0.26 &  14.0 &  10.9 &  0.66 &  1.62 &  1.23 &  0.69 &  29.3 & 0.01 &  0.76 & 0.001 &  0.77 & 0.50 &  0.57 &  0.02 &  1.10 & 31.2\\
\#N$_3$ & $\phi(q_{\rm i}) \propto$ $q_{\rm i}^{-0.4}$ & 0.28 &  15.2 &  12.3 &  0.71 &  1.82 &  1.41 &  0.77 &  32.5 & 0.01 &  0.80 & 0.001 &  0.81 & 0.56 &  0.62 &  0.02 &  1.20 & 34.5\\
\#N$_4$ & $\phi(q_{\rm i}) \propto$ $q_{\rm i}^{0.5}$ & 0.26 &  19.9 &  23.1 &  0.82 &  3.22 &  2.85 &  1.27 &  51.4 & 0.01 &  0.80 & 0.001 &  0.81 & 0.83 &  0.73 &  0.04 &  1.61 & 53.8\\
\#N$_5$ & $\phi(q_{\rm i}) \propto$ $q_{\rm i}^{1}$ & 0.21 &  19.7 &  27.6 &  0.75 &  3.74 &  3.54 &  1.41 &  56.9 & 0.00 &  0.68 & 0.001 &  0.68 & 
 0.88 &  0.67 &  0.05 &  1.60 & 59.2\\
\hline \noalign{\smallskip}

\#O$_1$ & $\psi(M_{\rm 1,i})$ \citep{Scalo98}& 0.37 &  23.8 &  22.7 &  1.06 &  3.27 &  2.68 &  1.34 &  55.2 & 0.01 &  1.11 & 0.002 &  1.13 & 0.93 &  0.93 &  0.04 &  1.90 & 58.2\\
\#O$_2$ & $\psi(M_{\rm 1,i})$ \citep{Kroupa01} & 0.46 & 26.5 & 22.2 & 1.32 & 3.52 & 2.86 & 1.44 & 58.3 & 0.01 & 1.39 & 0.002 & 1.41 & 1.16 & 1.21 & 0.05 & 2.41 & 62.2\\
\#O$_3$ & $\psi(M_{\rm 1,i})$ \citep{Chabrier03}  & 0.48 &  27.9 &  23.4 &  1.39 &  3.71 &  3.01 &  1.52 &  61.4 & 0.02 &  1.47 & 0.002 &  1.48 & 
 1.22 &  1.27 &  0.05 &  2.54 & 65.4\\
\#O$_4$ & $\psi(M_{\rm 1,i})$ \citep{Bell03} & 0.45 &  26.7 &  22.8 &  1.31 &  3.56 &  2.90 &  1.46 &  59.1 & 0.01 &  1.38 & 0.002 &  1.40 & 1.15 &  1.19 &  0.05 &  2.39 & 62.9\\
\hline \noalign{\smallskip}

\#P & $Q_{\rm crit,He-HG} = Q_{\rm crit,He-GB}$, $\alpha_{\rm ce}=0.2$ & 0.29 &   1.8 &   0.0 &  0.00 &  0.01 &  1.57 &  0.04 &   3.7 & 0.01 &  8.96 & 0.011 &  8.97 & 2.97 &  0.12 &  0.00 &  3.09 & 15.8\\
\#Q & $\gamma$ ($J_{\rm orb}$, wind) = 2, $\alpha_{\rm BH}$ = 5 & 0.28 &  18.8 &  17.2 &  0.84 &  2.72 &  2.14 &  5.95 &  47.9 &  0.01 &  0.87 & 0.098 &  0.98 &  0.78 &  0.86 &  0.11 &  1.75 & 50.7\\
\#R & $\psi(M_{1})$ \citep{Chabrier03}, $\phi(q) \propto$ $q^{1}$ & 0.34 & 29.4 & 36.9 & 1.27 & 5.46 & 5.12 & 2.06 & 80.5 & 0.01 & 1.14 & 0.002 & 1.15 & 1.48 & 1.19 & 0.09 & 2.76 & 84.4\\

\hline
\end{tabular}
\end{center}
\end{table}

\end{landscape}

\begin{landscape}

\begin{table}[tbh!]
\caption{The averaged SN~Ia rate for the different channels separately and combined (the overall SN~Ia rate) for different time intervals, assuming a starburst, calculated with our standard model (Sect.\,\ref{sec:StandAssum}) and the models from our parameter study (Table\,\ref{ParTable}). The rates are given in SNuM = rate per 100 yr per 10$^{10}$\Msun.}
\label{rateStarburst}
\begin{center}
\begin{tabular}{ll | cccc | cccc | ccc | cccc}
%\begin{tabular}{ll @{\qquad} cccc @{\qquad} cccc @{\qquad} ccc @{\qquad} cccc}
\hline \noalign{\smallskip}
& & \multicolumn{4}{c|}{DD} & \multicolumn{4}{c|}{SD$_{\rm H}$} & \multicolumn{3}{c|}{SD$_{\rm He}$\tablefootmark{a}}  & \multicolumn{4}{c}{SNe~Ia} \\
 & \hfill Time-interval (Gyr): & 0-0.1 & 0.1-0.3 & 0.3-1  & 1-10  & 0-0.1 & 0.1-0.3 & 0.3-1  & 1-10 & 0-0.1 & 0.1-0.3 & 0.3-1  & 0-0.1 & 0.1-0.3 & 0.3-1  & 1-10  \\
N$^{o}$ & Model &  &  &  &  &  &  &  &  &  &  &  &  &  &  &  \\
\hline \noalign{\smallskip} %\hline 

 & Standard ($N=150$) & 2.2e-5 & 0.052 & 0.27 & 0.023 & 0.0 & 0.028 & 0.0045 & 6.1e-8 & 0.085 & 0.033 & 0.0 & 0.085 & 0.11 & 0.28 & 0.023\\
 & Standard ($N=100$) & 0.0 & 0.056 & 0.27 & 0.023 & 0.0 & 0.028 & 0.0046 & 0.0 & 0.084 & 0.031 & 0.0 & 0.084 & 0.11 & 0.28 & 0.023\\

\hline \noalign{\smallskip}
\#A$_1$ & $\alpha_{\rm ce}$=0.2 & 0.0 & 0.026 & 0.034 & 0.0015 & 0.018 & 0.30 & 0.038 & 1.4e-4 & 0.20 & 0.069 & 0.0 & 0.22 & 0.39 & 0.073 & 0.0016\\
\#A$_2$ & $\alpha_{\rm ce}$=0.5 & 0.0 & 0.12 & 0.17 & 0.0078 & 0.0 & 0.037 & 0.0085 & 6.2e-7 & 0.11 & 0.055 & 0.0 & 0.11 & 0.21 & 0.17 & 0.0078\\
\#A$_3$ & $\alpha_{\rm ce}$=3 & 0.0020 & 0.047 & 0.059 & 0.017 & 0.0 & 0.0 & 1.2e-4 & 7.8e-7 & 0.20 & 0.021 & 0.0 & 0.20 & 0.068 & 0.059 & 0.017\\
\#A$_4$ & $\alpha_{\rm ce}$=10 & 6.9e-5 & 0.0026 & 0.021 & 0.0057 & 0.0 & 0.0 & 1.3e-4 & 8.2e-5 & 0.021 & 0.049 & 0.0 & 0.021 & 0.052 & 0.021 & 0.0058\\
\#B$_1$ & $\lambda_{\rm ce}$ = 1 & 0.0 & 0.16 & 0.34 & 0.018 & 0.0 & 0.035 & 0.0056 & 0.0 & 0.045 & 0.026 & 0.0 & 0.045 & 0.22 & 0.35 & 0.018\\
\#B$_2$ & $\lambda_{\rm ion}$ = 0.5 & 0.0 & 0.067 & 0.27 & 0.025 & 0.0 & 0.024 & 0.0041 & 5.0e-6 & 0.080 & 0.031 & 0.0 & 0.080 & 0.12 & 0.27 & 0.025\\

\hline \noalign{\smallskip}
\#C & $Q_{\rm crit,He-HG} = Q_{\rm crit,He-GB}$ & 0.059 & 0.31 & 0.061 & 0.0087 & 6.7e-5 & 0.028 & 0.0046 & 0.0 & 0.075 & 0.031 & 0.0 & 0.13 & 0.37 & 0.066 & 0.0087\\
\#D & $Q_{\rm crit,HG}=0.5$ & 0.0 & 0.037 & 0.19 & 0.019 & 0.0 & 0.023 & 0.0039 & 0.0 & 0.079 & 0.016 & 0.0 & 0.079 & 0.077 & 0.20 & 0.019\\

\hline \noalign{\smallskip}
\#E$_1$ & $\sigma$ = 1 & 0.0 & 0.053 & 0.20 & 0.022 & 0.0 & 0.023 & 0.0027 & 0.0 & 0.087 & 0.037 & 0.0 & 0.087 & 0.11 & 0.21 & 0.022\\
\#E$_2$ & $\sigma$ = 1000 & 7.5e-5 & 0.070 & 0.34 & 0.024 & 0.0 & 0.028 & 0.0044 &  0.0 & 0.11 & 0.024 & 0.0 & 0.11 & 0.12 & 0.35 & 0.024\\
\#F$_1$ & $\eta_{\rm H}=1$, $\eta_{\rm He}=1$ & 0.0 & 0.055 & 0.26 & 0.017 & 0.0023 & 0.39 & 0.62 & 0.089 & 0.93 & 0.35 & 2.0e-4 & 0.93 & 0.80 & 0.89 & 0.11\\
\#F$_2$ & $\eta_{\rm H}=1$, $\eta_{\rm He}=1$ , $\sigma \rightarrow \infty$ & 0.0 & 0.060 & 0.34 & 0.015 & 0.0026 & 0.38 & 0.60 & 0.090 & 1.5 & 0.72 & 1.9e-4 & 1.5 & 1.2 & 0.94 & 0.10\\

\hline \noalign{\smallskip}
\#G & $\gamma$ ($J_{\rm orb}$, wind) = 2 & 0.0 & 0.074 & 0.31 & 0.026 & 0.0 & 0.028 & 0.0046 & 1.1e-4 & 0.099 & 0.034 & 0.0 & 0.099 & 0.14 & 0.32 & 0.026\\
\#H$_1$ & $\gamma$ ($J_{\rm orb}$, RLOF) = $M_{\rm a}/M_{\rm d}$ & 0.0 & 0.066 & 0.30 & 0.026 & 0.0 & 0.029 & 0.0051 & 0.0 & 0.092 & 0.048 & 0.0 & 0.092 & 0.14 & 0.30 & 0.026\\
\#H$_2$ & $\gamma$ ($J_{\rm orb}$, RLOF) = 2 & 7.6e-5 & 0.075 & 0.28 & 0.019 & 0.0 & 0.028 & 0.0060 & 0.0 & 0.076 & 0.018 & 0.0 & 0.076 & 0.12 & 0.29 & 0.019\\

\hline \noalign{\smallskip}
\#I & $\eta$ \cite[]{Reimers75} = 5 & 1.4e-4 & 0.081 & 0.36 & 0.020 & 0.0 & 0.027 & 0.0045 & 2.6e-7 & 0.063 & 0.011 & 0.0 & 0.064 & 0.12 & 0.37 & 0.020\\
\#J$_1$ & TP-AGB \cite[]{Karakas02} & 0.0 & 0.058 & 0.27 & 0.023 & 0.0 & 0.028 & 0.0045 & 0.0 & 0.083 & 0.032 & 0.0 & 0.083 & 0.12 & 0.27 & 0.023\\
\#J$_2$ & TP-AGB \cite[][$\eta=1$]{Reimers75}& 0.0 & 0.067 & 0.29 & 0.024 & 0.0 & 0.028 & 0.0046 & 1.9e-6 & 0.084 & 0.032 & 0.0 & 0.084 & 0.13 & 0.30 & 0.024\\
\#J$_3$ & TP-AGB \cite[]{Bloecker95} & 0.0 & 0.057 & 0.26 & 0.022 & 0.0 & 0.028 & 0.0045 & 0.0 & 0.082 & 0.032 & 0.0 & 0.082 & 0.12 & 0.26 & 0.022\\
\#J$_4$ & TP-AGB \cite[]{VanLoon05} &0.0 & 0.063 & 0.27 & 0.024 & 0.0 & 0.028 & 0.0045 & 0.0 & 0.083 & 0.032 & 0.0 & 0.083 & 0.12 & 0.28 & 0.024\\

\hline \noalign{\smallskip}
\#K & CE accretion = 0.05\Msun & 0.0 & 0.058 & 0.27 & 0.026 & 0.0 & 0.027 & 0.0049 & 0.0 & 0.084 & 0.032 & 0.0 & 0.084 & 0.12 & 0.27 & 0.026\\
\#L & Bondi-Hoyle efficiency $\alpha_{\rm BH}$ = 5 & 0.0 & 0.061 & 0.29 & 0.023 & 0.0 & 0.028 & 0.0046 &  6.2e-7 & 0.088 & 0.034 & 0.0 & 0.088 & 0.12 & 0.29 & 0.023\\
\#M & B (CRAP) = 1e3 & 2.8e-4 & 0.061 & 0.31 & 0.017 & 0.0 & 0.028 & 0.0045 & 0.0 & 0.056 & 0.0058 & 0.0 & 0.056 & 0.095 & 0.32 & 0.017\\

\hline \noalign{\smallskip}
\#N$_1$ & $\phi(q_{\rm i}) \propto$ $q_{\rm i}^{-1}$ & 0.0 & 0.015 & 0.075 & 0.0070 & 0.0 & 0.011 & 0.0024 & 0.0 & 0.024 & 0.011 & 0.0 & 0.024 & 0.038 & 0.077 & 0.0070\\
\#N$_2$ & $\phi(q_{\rm i}) \propto$ $q_{\rm i}^{-0.5}$ & 0.0 & 0.038 & 0.18 & 0.016 & 0.0 & 0.023 & 0.0043 & 0.0 & 0.060 & 0.025 & 0.0 & 0.060 & 0.086 & 0.19 & 0.016\\
\#N$_3$ & $\phi(q_{\rm i}) \propto$ $q_{\rm i}^{-0.4}$ & 0.0 & 0.042 & 0.20 & 0.018 & 0.0 & 0.025 & 0.0045 & 0.0 & 0.066 & 0.027 & 0.0 & 0.066 & 0.093 & 0.21 & 0.018\\
\#N$_4$ & $\phi(q_{\rm i}) \propto$ $q_{\rm i}^{0.5}$ & 0.0 & 0.068 & 0.34 & 0.027 & 0.0 & 0.027 & 0.0039 & 0.0 & 0.096 & 0.032 & 0.0 & 0.096 & 0.13 & 0.34 & 0.027\\
\#N$_5$ & $\phi(q_{\rm i}) \propto$ $q_{\rm i}^{1}$ & 0.0 & 0.077 & 0.39 & 0.029 & 0.0 & 0.024 & 0.0030 & 0.0 & 0.10 & 0.030 & 0.0 & 0.10 & 0.13 & 0.39 & 0.029\\

\hline \noalign{\smallskip}
\#O$_1$ & $\psi(M_{\rm 1,i})$ \citep{Scalo98} & 0.0 & 0.072 & 0.35 & 0.030 & 0.0 & 0.036 & 0.0059 & 0.0 & 0.11 & 0.041 & 0.0 & 0.11 & 0.15 & 0.36 & 0.030\\
\#O$_2$ & $\psi(M_{\rm 1,i})$ \citep{Kroupa01} & 0.0 & 0.079 & 0.36 & 0.032 & 0.0 & 0.044 & 0.0074 & 0.0 & 0.14 & 0.051 & 0.0 & 0.14 & 0.17 & 0.37 & 0.032\\
\#O$_3$ & $\psi(M_{\rm 1,i})$ \citep{Chabrier03} & 0.0 & 0.083 & 0.38 & 0.034 & 0.0 & 0.047 & 0.0078 & 0.0 & 0.15 & 0.054 & 0.0 & 0.15 & 0.18 & 0.39 & 0.034\\
\#O$_4$ & $\psi(M_{\rm 1,i})$ \citep{Bell03} & 0.0 & 0.079 & 0.37 & 0.033 & 0.0 & 0.044 & 0.0074 & 0.0 & 0.14 & 0.051 & 0.0 & 0.14 & 0.17 & 0.37 & 0.033\\

\hline \noalign{\smallskip}
\#P & $Q_{\rm crit,He-HG} = Q_{\rm crit,He-GB}$, $\alpha_{\rm ce}=0.2$ & 0.0 & 0.012 & 0.028 & 0.0015 & 0.018 & 0.30 & 0.038 & 1.4e-4 & 0.17 & 0.069 & 0.0 & 0.19 & 0.38 & 0.066 & 0.0016\\
\#Q & $\gamma$ ($J_{\rm orb}$, wind) = 2, $\alpha_{\rm BH}$ = 5 & 0.0 & 0.076 & 0.31 & 0.025 & 0.0 & 0.028 & 0.0046 & 1.1e-4 & 0.10 & 0.036 & 0.0 & 0.10 & 0.14 & 0.32 & 0.025\\
\#R & $\psi(M_{1})$ \citep{Chabrier03}, $\phi(q) \propto$ $q^{1}$ & 0.0 & 0.11 & 0.53 & 0.043 & 0.0 & 0.040 & 0.0051 & 0.0 & 0.17 & 0.051 & 0.0 & 0.17 & 0.20 & 0.53 & 0.043\\
\hline

\end{tabular}
\tablefoot{\tablefoottext{a}{The rate of the SD$_{\rm He}$ channel between 1 and 10~Gyr is not given because it is always 0.}}
\end{center}
\end{table}

\end{landscape}

\normalsize

\begin{appendix}

\section{Mass distribution of the envelope\label{app::lambda}}

A method similar to \cite{DewiTauris00} is used to calculate $\lambda_{\rm ce}$, by fitting to detailed models from the STARS code \citep[]{Eggleton71,Pols95}. The fitting formulae we use are also given and discussed in \cite{IzzardThesis04}.

For hydrogen-rich stars we distinguish between stars with radiative envelopes and stars with deep convective envelopes. The expression for
$\lambda_{\rm ce}$ thus depends on the mass of the convective envelope, $M_{\rm env}$, expressed in solar units: 
\begin{equation} \label{eq:lambda-ce}
\lambda_{\rm ce} = 2 \cdot \begin{cases}
\lambda_{2} & M_{\rm env} = 0, \\
\lambda_{2}+ M_{\rm env}^{0.5}(\lambda_{1}-\lambda_{2})  &  0 < M_{\rm env} < 1,\\
\lambda_{1} & M_{\rm env} \geqslant 1,
\end{cases}
\end{equation}
where
\begin{equation}
\lambda_{2} = 0.42 \left(\frac{R_{\rm zams}}{R}\right)^{0.4}.
\end{equation}
The expression for $\lambda_{1}$ is more complicated and depends on the type of star. For HG and GB stars we have
\begin{equation}
\lambda_{1} = \min \left(0.80, \; \frac{3}{2.4 + M^{-3/2}} - 0.15 \log_{10} L \right)
\end{equation}
where $M$ and $L$ are the mass and luminosity of the star in solar units.
For more evolved stars we define
\begin{equation}
\lambda_{3} =  \min (-0.9, \; 0.58 + 0.75\log_{10} M) - 0.08\log_{10} L
\end{equation}
and
\begin{equation}
\lambda_{1} = \begin{cases}
\min(0.8, \; 1.25 - 0.15 \log_{10} L, \; \lambda_{3}) & \text{CHeB, E-AGB,} \\
\max(-3.5 - 0.75\log_{10}M + \log_{10}L, \; \lambda_{3}) & \text{TP-AGB,}
\end{cases}
\end{equation}
with the latter expression for TP-AGB stars capped at a maximum $\lambda_{1}$ of 1.0.
This results in typical values of $\lambda_{\rm ce} \approx 1.0-2.0$ for stars on the GB or AGB, and $\lambda_{\rm ce} \approx 0.25-0.75$ for HG stars. For helium stars no fit is available and we take $\lambda_{\rm ce} = 0.5$.

The loss of the envelope can be enhanced by using a fraction $\lambda_{\rm ion}$ of the ionization energy. In the code this is expressed by replacing $\lambda_{\rm 1}$ in Eq.~\ref{eq:lambda-ce} by
\begin{equation}
\lambda_{1} \rightarrow \lambda_{1} + \lambda_{\rm ion}(\lambda_{4}-\lambda_{1}),
\end{equation}
where $0 \leq \lambda_{\rm ion} \leq 1$ is a free parameter and
\begin{equation}
\lambda_{4} = \max (\min[\lambda_{5}, \; 100], \; \lambda_{1}),
\end{equation}
with
\begin{equation}
\lambda_{5} = \frac{1}{a + \arctan(b[c - \log_{10} L]) + d(\log_{10} L - 2)}.
\end{equation}
The coefficients $a$ -- $d$ depend on stellar type and mass as follows:
\begin{equation}
a = \begin{cases}
\min(-0.5, \; 1.2\,[\log_{10} M - 0.25]^{2} - 0.7) & \text{HG, GB,} \\
\max(-0.5, \; -0.2 - \log_{10} M)  & \text{CHeB, AGB,}
\end{cases}
\end{equation}
\begin{equation}
b = \max(1.5, \; 3 - 5\log_{10} M),
\end{equation}
\begin{equation}
c = \max(3.7 + 1.6\log_{10} M, \; 3.3 + 2.1\log_{10} M)
\end{equation}
and
\begin{equation}
d = \begin{cases}
\max(0, \; \min[0.15, \; 0.15 - 0.25\log_{10} M])  & \text{HG, GB,} \\
0  & \text{CHeB, AGB.}
\end{cases}
\end{equation}

\section{Accretion efficiency of WDs\label{app::efficiencyWD}}

Material that is transferred to a WD, $\dot{M}_{\rm tr}$, can only be burnt by the WD at a specific rate \cite[]{Nomoto82}. If the mass transfer rate is lower than this specific rate the material is added onto the surface and later on ejected in a nova explosion. When the material is transferred to the WD at a higher rate, we assume that the material which is not burnt is blown away through a optically thick wind from the accreting WD \cite[]{Hachisu96}.

When He-rich material is accreted, it burns into carbon and oxygen. The net accretion efficiency ($\eta_{\rm He}$) is written
\begin{equation}
\label{eq::HachisuHe}
\dot{M}_{\rm WD} = \eta_{\rm He} \dot{M}_{\rm tr}\,,
\end{equation}
where $\dot{M}_{\rm tr}$ is the mass transfer rate and $\dot{M}_{\rm WD}$ the net mass growth of the WD.

When hydrogen-rich material is transferred to the WD the net accretion efficiency ($\eta_{\rm He} \eta_{\rm H}$) is written
\begin{equation}
\label{eq::HachisuH}
\dot{M}_{\rm WD} = \eta_{\rm He}  \eta_{\rm H} \dot{M}_{\rm tr}\,,
\end{equation}
because first hydrogen is burned into helium, and subsequently helium is burnt into carbon and oxygen. 

The accretion efficiencies for hydrogen and helium burning, $\eta_{\rm H}$ and $\eta_{\rm He}$, are calculated following \cite{Hachisu99}
\begin{equation}
\label{eq::etaH}
\eta_{\rm H} = \begin{cases}
\dot{M}_{\rm cr,H}/ \dot{\rm M}_{\rm tr}  & \dot{M}_{\rm tr} > \dot{M}_{\rm cr,H}\,, \\
1  & \dot{M}_{\rm cr,H} > \dot{M}_{\rm tr} > \dot{M}_{\rm cr,H}/8\,,  \\
0  & \dot{M}_{\rm tr} < \dot{M}_{\rm cr,H}/8\,,
\end{cases}
\end{equation}
where
\begin{equation}
\label{eq::McrH}
\dot{M}_{\rm cr,H} = 5.3 \cdot 10^{-7} \left(\frac{1.7-X}{X}\right) \left(\frac{M_{\rm WD}}{M_{\odot}} - 0.4 \right) { \Msun\, {\rm yr}^{-1}}\,,
\end{equation}
where $X$ is the hydrogen abundance and

\begin{equation}
\label{eq::etaHe}
\eta_{\rm He} = \begin{cases}
\dot{M}_{\rm up}/ \dot{M}_{\rm tr}  & \dot{M}_{\rm tr} > \dot{M}_{\rm up}\,, \\
1  & \dot{M}_{\rm up} > \dot{M}_{\rm tr} > \dot{M}_{\rm cr,He}\,, \\
\eta_{\rm KH04}  & \dot{M}_{\rm cr,He} > \dot{M}_{\rm tr} > \dot{M}_{\rm low}\,, \\
0  & \dot{M}_{\rm tr} < \dot{M}_{\rm low}\,,
\end{cases}
\end{equation}
where
\begin{align}
\label{eq::McrHe}
\dot{M}_{\rm up} & = 7.2 \cdot 10^{-6} \left(\frac{M_{\rm WD}}{\Msun} - 0.6\right) \rm{ \Msun\ \rm{yr}^{-1}}\,, \\
\dot{M}_{\rm cr,He} & = 10^{-5.8} \rm{\Msun\ \rm{yr}^{-1}}\,,\\
\dot{M}_{\rm low} & = 10^{-7.4} \rm{\Msun \ \rm{yr}^{-1}}\,.
\end{align}
The expression for $\dot{M}_{\rm up}$ is based on \cite{Nomoto82} and $\dot{M}_{\rm cr,He}$ and $\dot{M}_{\rm low}$ are based on the models of \cite{KatoHachisu04}, more specifically $\dot{M}_{\rm low}$ is the lower limit of the models that have He-shell flashes. The accretion efficiency $\eta_{\rm KH04}$ is based on the models for He-shell flashes of \cite{KatoHachisu04}, implemented in a similar way as in \cite{Meng09}.
This process is limited by the Eddington limit for accretion.

\end{appendix}

\end{document}